\documentclass[twocolumn]{openjournal}

\usepackage{bm}
\usepackage{xspace}
\usepackage{xcolor}
\usepackage{amsmath}
\usepackage{graphicx}
\usepackage[hidelinks]{hyperref}
\usepackage[caption=false]{subfig}
\usepackage{booktabs}
\usepackage[capitalise]{cleveref}
\creflabelformat{equation}{#2\textup{#1}#3}
\newcommand{\iap}{CNRS \& Sorbonne Universit\'e, Institut d’Astrophysique de Paris (IAP), UMR 7095, 98 bis bd Arago, F-75014 Paris, France}
\newcommand{\ccm}{Center for Computational Mathematics, Flatiron Institute, 162 5th Avenue, New York, NY 10010, USA}
\newcommand{\cca}{Center for Computational Astrophysics, Flatiron Institute, 162 5th Avenue, New York, NY 10010, USA}
\newcommand{\cfa}{Center for Astrophysics | Harvard \& Smithsonian, 60 Garden St, Cambridge, MA 02138, USA}
\newcommand{\nfsai}{The NSF AI Institute for Artificial Intelligence and Fundamental Interactions}
\newcommand{\mitphys}{Department of Physics, Massachusetts Institute of Technology, Cambridge, MA 02139, USA}
\newcommand{\montreal}{Department of Physics, Université de Montréal, Montréal, Canada}
\newcommand{\mila}{Mila - Quebec Artificial Intelligence Institute, Montréal, Canada}
\newcommand{\ciela}{Ciela - Montreal Institute for Astrophysical Data Analysis and Machine Learning, Montréal, Canada}
\newcommand{\seoul}{Center for the Gravitational-Wave Universe, Astronomy Program Department of Physics and Astronomy, Seoul National University, Seoul 08826, Korea}
\newcommand{\imperial}{Imperial Centre for Inference and Cosmology (ICIC) \& Astrophysics Group, Imperial College London, Blackett Laboratory, Prince Consort Road, London SW7 2AZ, United Kingdom}
\newcommand{\portsmouth}{Institute of Cosmology and Gravitation, University of Portsmouth, Burnaby Road, Portsmouth, PO1 3FX, United Kingdom}
\newcommand{\columbia}{Columbia Astrophysics Laboratory, Columbia University, 550 West 120th Street, New York, NY 10027, USA}
\newcommand{\hubble}{Hubble Fellow}
\newcommand{\lpnhe}{Sorbonne Universit\'e, Universit\'e Paris Diderot, Sorbonne Paris Cit\'e, CNRS,
Laboratoire de Physique Nucléaire et de Hautes Energies (LPNHE). 4 place Jussieu, F-75252, Paris Cedex 5, France}

\newcommand{\rev}[1]{{#1}}  

\newcommand{\ili}{\textsc{LtU-ILI}\xspace}
\newcommand{\pydelfi}{\code{pydelfi}\xspace}
\newcommand{\sbi}{\code{sbi}\xspace}
\newcommand{\lampe}{\code{lampe}\xspace}

\newcommand{\bmt}{\bm{\theta}}
\newcommand{\bmx}{\bm x}
\newcommand{\bmxo}{\bmx_{\rm o}}
\newcommand{\bmw}{\bm w}
\newcommand{\bmu}{\bm u}
\newcommand{\bmv}{\bm v}

\newcommand{\calP}{\mathcal{P}}
\newcommand{\calD}{\mathcal{D}}
\newcommand{\calL}{\mathcal{L}}
\newcommand{\bbR}{\mathbb{R}}
\newcommand{\bbE}{\mathbb{E}}

\newcommand{\eg}{\emph{e.g.,}\xspace}
\newcommand{\ie}{\emph{i.e.,}\xspace}

\newcommand{\code}[1]{\texttt{#1}}

\shorttitle{\ili}
\shortauthors{Ho et al.}

\graphicspath{{./}{figures/}}

\begin{document}

\title{\ili: An All-in-One Framework for Implicit Inference in Astrophysics and Cosmology}

\author{Matthew Ho$^{1*}$}
\email{$^*$matthew.ho@iap.fr}
\author{Deaglan J. Bartlett$^1$}
\author{Nicolas Chartier$^2$}
\author{Carolina Cuesta-Lazaro$^{3,4,5}$}
\author{Simon Ding$^1$}
\author{Axel Lapel$^{1,14}$}
\author{Pablo Lemos$^{6,7,8,9}$}
\author{Christopher C. Lovell$^{10}$}
\author{T. Lucas Makinen$^{11}$}
\author{Chirag Modi$^{9,12}$}
\author{Viraj Pandya$^{13,15}$}  
\author{Shivam Pandey$^{13}$}  
\author{Lucia A. Perez$^9$}
\author{Benjamin Wandelt$^{1,9}$}
\author{Greg L. Bryan$^{13}$}

\affiliation{$^1$ \iap}
\affiliation{$^2$ \seoul}
\affiliation{$^3$ \cfa}
\affiliation{$^4$ \nfsai}
\affiliation{$^5$ \mitphys}
\affiliation{$^6$ \montreal}
\affiliation{$^7$ \mila}
\affiliation{$^8$ \ciela}
\affiliation{$^9$ \cca}
\affiliation{$^{10}$ \portsmouth}
\affiliation{$^{11}$\imperial}
\affiliation{$^{12}$ \ccm}
\affiliation{$^{13}$ \columbia}
\affiliation{$^{14}$ \lpnhe}
\altaffiliation{$^{15}$ \hubble}


\begin{abstract}
This paper presents the Learning the Universe Implicit Likelihood Inference (\ili) pipeline, a codebase for rapid, user-friendly, and cutting-edge machine learning (ML) inference in astrophysics and cosmology. The pipeline includes software for implementing various neural architectures, training schemata, priors, and density estimators in a manner easily adaptable to any research workflow. It includes comprehensive validation metrics to assess posterior estimate coverage, enhancing the reliability of inferred results. Additionally, the pipeline is easily parallelizable and is designed for efficient exploration of modeling hyperparameters. To demonstrate its capabilities, we present real applications across a range of astrophysics and cosmology problems, such as: estimating galaxy cluster masses from X-ray photometry; inferring cosmology from matter power spectra and halo point clouds; characterizing progenitors in gravitational wave signals; capturing physical dust parameters from galaxy colors and luminosities; and establishing properties of semi-analytic models of galaxy formation. We also include exhaustive benchmarking and comparisons of all implemented methods as well as discussions about the challenges and pitfalls of ML inference in astronomical sciences. All code and examples are made publicly available at \url{https://github.com/maho3/ltu-ili}.
\end{abstract}


\section{Introduction} \label{sec:intro}

Statistical inference of unknown quantities is a fundamental problem in science. In the astronomical sciences, we largely rely on Bayesian inference to test new physical models \citep{Feigelson2012,dodelson2020modern, eadie2023practical}, wherein we assume some prior hypothesis of an observed system and calculate constraints on model parameters which would be consistent with our observations. 
For nearly a century, the practice of building linear or perturbative physical models from first-principles allowed us to make substantial progress towards understanding the universe. Yet, as highlighted in the 2020 Decadal Survey \citep{national2021pathways}, exploring the complex, nonlinear regime of physical phenomena through data-driven inference promises significant gains in constraining power. The large data volume of next-generation surveys, the improvements in high-resolution simulations, and the rapid rise of machine learning (ML) techniques have driven an explosion of inquiry into how one can automatically and rapidly learn complex physical phenomena \citep{carleo2019machine}. Despite the growing enthusiasm, a major hurdle to the widespread adoption of data science methods in astronomy is in their limited accessibility to the general community, which when compounded by the rapid pace of developments in the field, makes it challenging for individuals to stay well-informed on best practices \citep{huppenkothen2023constructing}. There is currently no widespread consensus within the field regarding the approach to building a framework for implicit inference problems that is both reliable, robust, and accessible.

Implicit Likelihood Inference \citep[ILI,][]{cranmer2020frontier,marin2012approximate}, also known as simulation-based inference (SBI) and likelihood-free inference (LFI), is an approach for learning the statistical relationship between parameters and data. ILI is a mathematically rigorous way of framing ML under the umbrella of Bayesian statistics, a language common to astrophysicists and cosmologists since the turn of the century \citep{christensen2001bayesian}. \rev{Given a parameterized model and observed data,} ILI estimates posterior distributions over model parameters, \rev{while accounting for} predictive uncertainties.
It is best explained in contrast to traditional explicit likelihood analyses, wherein one might write down an analytic likelihood to represent the probability of observations given a model and its parameters, and then use sampling methods such as Markov Chain Monte Carlo \citep[MCMC;][]{brooks2011handbook} to sample from the posterior. In ILI, one seeks to automatically learn the likelihood, i.e. to model the relationship between model parameters and observations through training on simulations. As such, ILI can be applied to infer parameters for any phenomenon that can be simulated, without the need for assumptions about the analytical form of the likelihood. 
Furthermore, ILI can accommodate complex data cuts (provided that consistent filters are applied to both the data and the forward model) something that can be hard to model in likelihood-based approaches. 

Given a prescribed training set of model parameters and observations or a fast physical simulator, ILI can produce robust, tightly-constraining inference without analytic likelihoods. In astronomy and cosmology, ILI has consistently been shown to accelerate and improve upon traditional analyses such as cosmological galaxy lensing and clustering \citep[\eg][]{jeffrey2021likelihood, Makinen_2021, Makinen_2022, de2023robust, hahn2023rm},  gravitational waves \citep[\eg][]{daxGW2021, cheung2022testing}, galaxy cluster mass estimation \citep[\eg][]{ho2022dynamical, de2022deep}, galaxy morphology \citep[\eg][]{walmsley2020galaxy, ghosh2022gampen}, stellar streams \citep[\eg][]{hermans2021towards, alvey2023albatross}, and exoplanets \citep[\eg][]{rogers2023exoplanet, aubin2023simulation}.

Despite its popularity, robust applications of ILI have their challenges. First, fully Bayesian models include capturing epistemic (modeling) uncertainty, which requires marginalizing over model parameter uncertainty (\ie network weights and biases) and is intractable for large architectures. However, approximations to this method using dropout marginalization \citep{gal2016dropout}, model ensembling \citep{lakshminarayanan2017simple} or stochastic weight averaging \citep{maddox2019simple,Lemos2023}
prove to be empirically sufficient surrogates. Second, choices of modeling hyperparameters (\eg preprocessing, architecture, learning procedures, etc.) can greatly affect the quality of a trained ILI model. This is generally addressed through model ensembling or hyperparameter searches \citep[\eg][]{jin2019auto, white2021bananas}. Lastly, a pervasive problem in ILI is model misspecification, wherein the simulator or training catalog used to learn likelihoods is not representative of reality. Methods such as \code{SBI++} \citep{sbipp2023} deal with missing data and outliers and present applications to galactic photometric redshift inference \citep{modi2023sensitivity}. For a thorough discussion of caveats, see Section \ref{sec:discussion}.
We argue that addressing these common problems is paramount to responsible application of ILI, and testing them exhaustively necessitates a unified, accessible framework.

We introduce the first version of the Learning the Universe Implicit Likelihood Inference code (\ili), a pipeline for training ML models for regressive parameter estimation. Given a labeled training set of observed data and parameters, \ili{} trains neural networks to emulate posterior probability distributions. The code also provides scientists with a diverse toolkit for testing and validation. It was built under the following design principles:
\begin{itemize}
    \item The code includes state-of-the-art neural architectures, validation tools, and samplers for enabling ILI.
    \item The pipeline is modular, easily customizable, and parallelizable for rapid testing of design choices and hyperparameters.
    \item The interface is accessible to non-specialists while being practical for high-level production.
    \item The methodologies automatically implement standard best practices for machine learning \citep[See][ for a review]{huppenkothen2023constructing}.
\end{itemize}
These attributes are paramount to make ILI a relevant tool for a wide range of inference problems, whether in the context of exploratory analysis or exhaustive empirical testing.

The novel contributions of this work are as follows: 
\begin{itemize}
    \item We unify several leading implicit inference codes--\code{sbi} \citep{tejero-cantero2020sbi}, \code{pydelfi} \citep{alsing2019fast}, and \code{lampe} \citep{rozet2021lampe}--under a single common interface, for the first time allowing rigorous apples-to-apples comparisons of their performance.
    \item We supply extensions to these codes, enabling them to utilize new types of embedding networks (\eg for image- and graph-like datasets) and providing state-of-the-art validation metrics for enabling robust, maximally-informative inference.
    \item We provide an accessible, comprehensive, dual Jupyter and command-line interface, for rapid research and development.
    \item We demonstrate the abilities of ILI in an extensive selection of inference problems in astrophysics and cosmology, and we provide these as public benchmarks for future works.
\end{itemize}
At the time of publishing, \ili{} offers the broadest and most complete toolbox of features among ILI software. Over \code{sbi}, we include the capacity for custom data-loading procedures, additional neural density estimators (NDEs), and adaptability to exotic embedding architectures (including sequential and graph encoders). Over \code{pydelfi}, we allow the capacity for neural posterior and ratio estimation. Over \code{lampe}, we provide a configuration-based interface to training procedures, the ability to specify separate proposal and prior distributions, and automated validation tests. We also include new multivariate coverage tests such as TARP \citep{tarp} and adaptive, automatic integration of data-loading, inference, and validation.

In this software release paper, we provide background, details, and example applications of the \ili{} pipeline. The code is publicly available on Github\footnote{\url{https://github.com/maho3/ltu-ili}}.
In Section \ref{sec:theory}, we review the theoretical foundations of ILI. In Section \ref{sec:structure}, we describe the structure of our pipeline and the various configurations that users can interact with. In Section \ref{sec:synth}, we apply \ili{} on several synthetic problems where the true solution is known, to demonstrate and benchmark its performance relative to traditional methods. In Section \ref{sec:examples}, we provide numerous diverse examples of parameter inference in science using \ili{}. In \rev{Section} \ref{sec:discussion}, we discuss the important considerations to take into account when designing an ILI pipeline, as well as possible failure modes and relevant solutions. Lastly, we provide a summary of our findings in Section \ref{sec:conclusion}.

\vspace{3em}

\section{Implicit Likelihood Inference}\label{sec:theory}

\begin{figure}
    \centering
    \includegraphics[width=\linewidth]{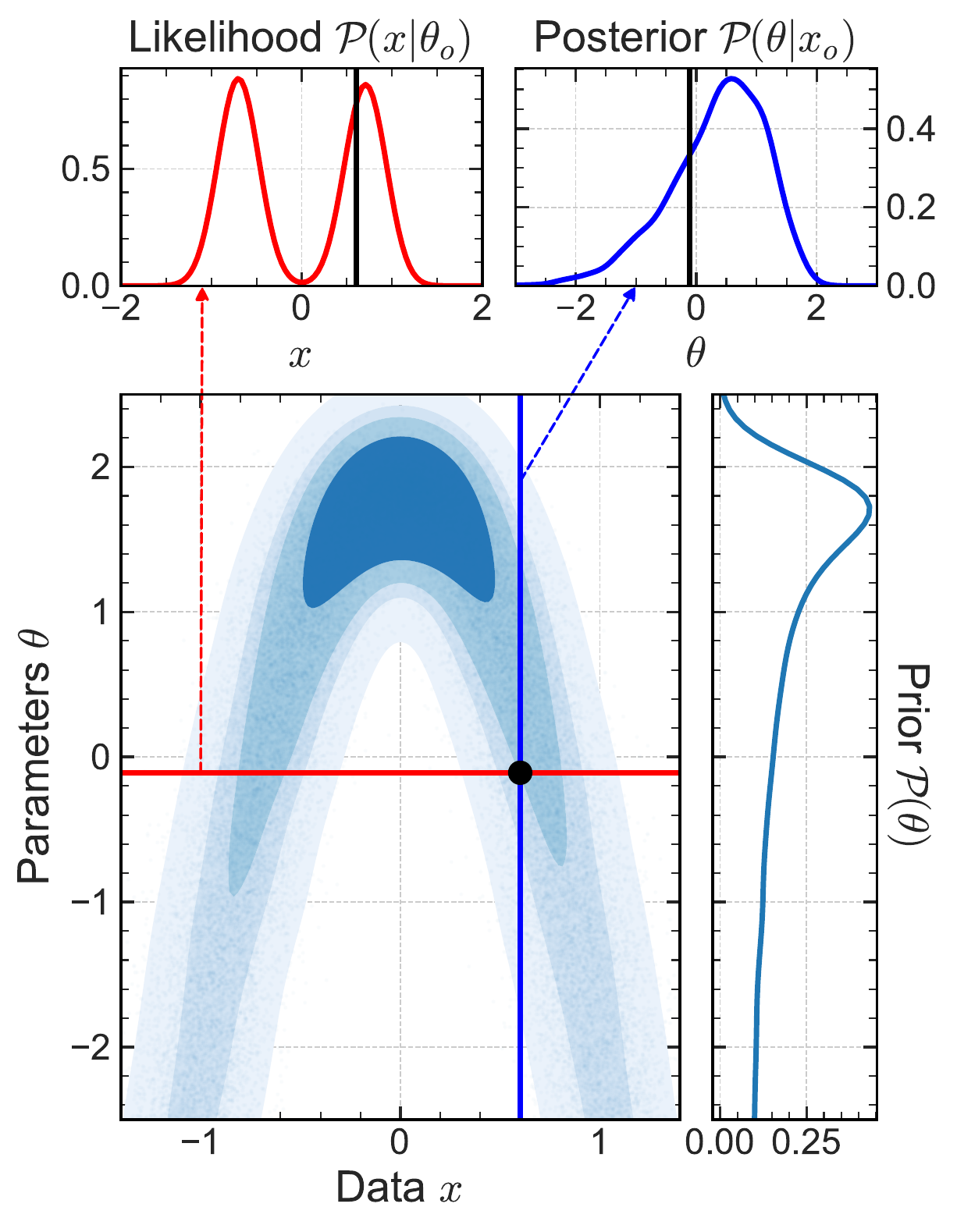}
    \caption{Demonstration of the differences between a likelihood (top-left), posterior (top-right), prior (bottom-right), and joint distribution (bottom-left) for an arbitrary one-dimensional inference problem. Here, the univariate data $x$ and parameters $\theta$ are associated by a quadratic relationship with Gaussian scatter. Given the observed data point $x_o$ (in black), our goal is to recover posterior constraints on $\theta$. 
    In an ILI framework, we seek to capture knowledge of the joint distribution, $P(x,\theta)$, by building neural-network emulators for the likelihood (in red) or the posterior (in blue).}
    \label{fig:example}
\end{figure}

The goal of ILI is the same as any Bayesian inference problem: given a model and observed data $\bmxo$, we wish to know the distribution of model parameters $\bmt$ which are consistent with our data, \ie the \textit{posterior distribution} $\calP \left( \bmt \vert \bmxo \right)$. The data and parameters can take many forms. In astronomy, we might consider the observed data $\bmxo$ to be, for example, the observed X-ray photon count of a galaxy cluster, the gravitational wave signal from a collapsing black hole binary, or the full 2D sky projection of galaxy weak lensing measurements. From this data, we might wish to infer parameters $\bmt$ such as, respectively, the mass of a galaxy cluster, the mass ratio and ellipticity of the black hole binary, or the cosmological parameters governing dark matter and dark energy. For Bayesians, the posterior $\calP \left( \bmt \vert \bmxo \right)$ is the key to constraining our belief in physical laws after observing new data $\bmxo$.

According to Bayes' theorem, the posterior is composed of
\begin{equation}\label{eqn:bayes}
    \calP \left( \bmt \vert \bmx \right) \propto \calP \left( \bmx \vert \bmt \right) \ \calP \left( \bmt \right),
\end{equation}
where $\calP \left( \bmx \vert\bmt \right)$ is the \textit{likelihood}, which defines the probability of the data given the model and some parameters, and $\calP \left( \bmt \right)$ is the \textit{prior}, which specifies our prior belief on the true distribution of the parameters. A demonstration of the differences of the prior, posterior, and likelihood in an inference problem is given in Figure~\ref{fig:example}.

There are two approaches to modeling the terms in Equation \ref{eqn:bayes}: explicit (theory-driven) analyses or implicit (data-driven) analyses. In a classical \textit{explicit} likelihood-based analysis, we assume an analytic form for the likelihood, $\calP\left( \bmx \vert \bmt \right)$, and substitute it into Equation \ref{eqn:bayes}. Typically, this expression is derived from analytic theory or simple simulations, coupled with strong assumptions on the statistical distribution of noise or unknown physics. Sometimes, these statistical assumptions are relatively simple, \eg a Gaussian distribution for continuous data \citep[\eg][]{sdssPk2004, christensen2022parameter, porqueres2022lifting} or a Poisson distribution for discrete counts \citep[\eg][]{boese2001maximum, braun2008methods, tsaprazi2023higher}. Although these may be appropriate distributions for many situations, for a general inference problem we may not know which distribution to choose or whether this choice is valid. 
Also, MCMC-like samplers scale poorly with dimensionality and have bad mode coverage. This can cause problems for high-dimensional inference problems, especially if the likelihood is not differentiable with respect to the parameters.
However, if we have access to a set of examples of the data-parameter relationship (\ie samples from the joint distribution, $(\bmx, \bmt) \sim \calP(\bmx, \bmt)$), we can use \textit{implicit} likelihood analyses to learn the shape of a likelihood or posterior automatically with full differentiability. This flexibility makes ILI an extremely powerful tool for modeling complex or otherwise poorly-understood physical phenomena. 

ILI learns approximate posteriors $\hat{\calP}(\bmt\vert\bmx)$ or likelihoods $\hat{\calP}(\bmx\vert\bmt)$ from the distribution of example data-parameter pairs in a training catalog, $\calD_{\rm train} := \{(\bmx_i, \bmt_i)\}_{i=1}^{N_{\rm train}}$. 
These data can take any form; they can be sensor readings in an experiment \citep[\eg][]{dingeldein2023simulation}, images of the night sky \citep[\eg][]{ntampaka2019deep}, or even compressed summaries of high-dimensional stochastic properties \citep[\eg][]{modi2023sensitivity}. They can be gathered from a pre-run simulation \citep[\eg][]{fluriWL2022}, be simulated on-the-fly \citep[\eg][]{alsing2019fast}, or curated from manually-labeled observations \citep[\eg][]{lanusse2018cmu}, but they must be representative of the true underlying problem.  Given these data-parameter pairs, we wish to construct a model capable of doing inference on real data, \ie to evaluate the posterior at some observed value $\bmx=\bmxo$.

In this section, we describe the theoretical foundations of \ili{}. We begin by providing a simple introductory example for posterior inference with Approximate Bayesian Computation (ABC; Section \ref{subsec:abc}). We then explain how one can vastly accelerate this process using neural networks to emulate probability distributions (Section \ref{subsec:nde}). We then discuss how to practically frame machine learning in the context of Bayesian inference (Sections \ref{subsec:npe}, \ref{subsec:nle} \& \ref{subsec:nre}). Lastly, we discuss auxiliary topics on utilizing active learning to accelerate training (Section \ref{subsec:sequentialL}) and validating black-box models to ensure robustness in real-world applications (Section \ref{subsec:valid}).

\subsection{Approximate Bayesian Computation}\label{subsec:abc}
Approximate Bayesian Computation \citep[ABC;][]{rubinABC} is the original basis of ILI. Given a prior $\calP(\bmt)$, a simulator ($\bmt\rightarrow\bmx$), and an observation $\bmxo$, ABC can construct a posterior. 
The general procedure of Rejection ABC is straightforward: (1) first, draw candidate parameter values from the prior $\bmt\sim \calP(\bmt)$, (2) forward-simulate these parameter values to observations $\bmx\sim \calP(\bmx|\bmt)$, (3) measure a distance metric between the candidate observations and the real observation, $d(\bmx, \bmxo)$, and (4) if the distance is sufficiently close to the real observation (\ie $d(\bmx, \bmxo)<\epsilon$ for small $\epsilon$), accept the candidate parameters $\bmt$ as samples from the inferred posterior. ABC has been widely adopted in statistical inference because it does not require explicit knowledge of the likelihood and the accepted samples are guaranteed to converge to the true posterior as $\epsilon\rightarrow 0$. When $\epsilon$ is non-zero, the inferred posterior is guaranteed to be broader than the true posterior, enabling conservative inference. Despite successful applications in astrophysics and cosmology \citep[\eg][]{schafer2012likelihood, cameron2012approximate, hahn2017approximate}, the ultimate problem with ABC methods is that the acceptance rate vanishes exponentially as the dimensionality of $\bmt$ increases, requiring thus significantly more simulations, which can be alleviated with a larger $\epsilon$ at the cost of targeting a broader posterior than the true posterior \citep{Alsing_2018}. This limitation makes ABC highly intractable in regimes where running simulations is expensive.
Occasionally ABC is used as a synonym for LFI/SBI/ILI, however throughout the paper we exclusively use ABC to refer to the algorithm discussed above.

\subsection{Neural Density Estimation}\label{subsec:nde}
In modern applications, implicit inference is made tractable by using machine learning models to emulate conditional probability distributions, a procedure called Neural Density Estimation \citep{papamakarios2019neural}. As an example, consider a situation wherein we seek to train a mixture density network \citep[MDN;][]{bishop1994mixture} to directly model the conditional distribution $\calP(\bmv\vert\bmu)$ using our training set of data-parameter pairs $(\bmx, \bmt)\sim\calD_{\rm train}$. For this subsection only, we have $(\bmu, \bmv) = (\bmx, \bmt) $ or $(\bmu, \bmv) = (\bmx, \bmt)$ interchangeably. The goal is to build a neural architecture $q_{\bmw}(\bmv|\bmu)$ with weights $\bmw$ that outputs a probability distribution over $\bmv$ which targets the conditional probability $\calP(\bmv\vert\bmu)$. An MDN outputs probability distributions by allowing the neural network to specify the statistical parameters of a mixture density distribution. In the independent Gaussian case, this means the neural network outputs a mean and a variance for each component of the mixture density distribution, \ie
\begin{equation}\label{eqn:Pmdn}
    q_{\bmw}(\bmv|\bmu) = \frac{1}{N_{\rm c}\sqrt{2\pi}} \sum_{i=1}^{N_{\rm c}} 
    \frac{1}{\sigma_i(\bmu;\bmw)}e^{-\frac{(\bmv - \mu_i(\bmu;\bmw))^2}{2\sigma_i^2(\bmu;\bmw)}},
\end{equation}
where $N_c$ is the chosen number of components of the mixture density distribution, $\bmw$ are the learned weights and biases of the neural network, and $\mu_i(\bmu;\bmw)$ and $\sigma_i(\bmu;\bmw)$ are the neural network outputs representing the mean and standard deviation (or \textit{scale}) of the $i$-th Gaussian component as a function of the input $\bmu$ and the weights. 

To train such a network to properly emulate the conditional probability, all we need to do is maximize the joint likelihood of our training data $\calD_{\rm train}$. This is equivalently and often more conveniently expressed in the form of minimizing the negative log-probability loss (typically log-likelihood or log-posterior, see sections \ref{subsec:npe} to \ref{subsec:nre})
\begin{equation}\label{eqn:Lmdn}
    \calL_{\rm MDN} := -\bbE_{\calD_{\rm train}}\left[\log q_{\bmw}(\bmv|\bmu)\right],
\end{equation}
where the expectation $\bbE_{\calD_{\rm train}}$ is taken over all data-parameter pairs in the training set $\sim\calD_{\rm train}$. 
For the case of the the posterior distribution ($\bmv = \bmt, \bmu = \bmx$), if the training data is sufficiently diverse and if the MDN is sufficiently flexible, then the minimization of $\calL_{\rm MDN}$ over the network weights $\bmw$ will converge $q_{\bmw}(\bmt|\bmx)$ to the true posterior $\calP(\bmt|\bmx)$ \citep{papamakarios2016fast}.  

The MDN is just one example of an ever-increasing variety of neural density estimators in the literature. Another very popular class of methods is normalizing flows \citep{papamakarios2021normalizing}, which output a flexible conditional distribution defined via learnable, invertible transformations of a base Gaussian distribution. Normalizing flows are often better equipped for handling cases wherein the target distribution is non-Gaussian, but low-dimensional and uni-modal. For any type of neural density estimator, the training procedure always involves minimizing the negative log-probability of training data, as in Equation~\ref{eqn:Lmdn}.


Now we have established the tools for neural density estimation, we can apply them to fitting components of Bayes' theorem (Equation~\ref{eqn:bayes}).

\subsubsection{Neural Posterior Estimation (NPE)} \label{subsec:npe}
The most straightforward method to do this is to train neural networks to directly emulate the posterior distribution, also known as Neural Posterior Estimation \citep[NPE;][]{papamakarios2016fast, greenberg2019automatic}. As in the MDN example in the previous section, the loss function for NPE is simply the negative log-posterior of our learned neural density estimator $\hat{\calP}(\bmt|\bmx)$, and can be trained using a loss function like that of \citet{papamakarios2016fast}:
\begin{equation}\label{eqn:Lnde}
    \begin{aligned}
    \calL_{\rm NPE} &:= -\bbE_{\calD_{\rm train}}\log \hat{\calP}(\bmt_i|\bmx_i)\\
    &= -\bbE_{\calD_{\rm train}}\log \left[\frac{p(\bmt)}{\tilde{p}(\bmt)}q_{\bmw}(\bmt|\bmx)\right],
    \end{aligned}
\end{equation}
where our neural posterior $\hat{\calP}(\bmt_i|\bmx_i)$ is decomposed into a neural network output $q_{\bmw}(\bmt | \bmx)$ and a weighting factor taken as the ratio of the assumed prior $p(\bmt)$ to the proposal prior $\Tilde{p}(\bmt)$.
\rev{The proposal prior is defined as the distribution of $\bmt$ present in the training dataset $\calD_{\rm train}$, while the assumed prior is an experimental design choice representing the assumed knowledge of the global distribution of $\bmt$. In many cases, the assumed and proposal priors are identical, however this distinction is particularly relevant in the case of sequential learning (see section \ref{subsec:sequentialL})}.
Note, that the normalization factor \rev{$p(\bmt)/\tilde{p}(\bmt)$} cancels out when optimizing $\calL_{\rm NPE}$ over network weights $\bmw$, but is still required for constructing the posterior estimator $\hat{\calP}(\bmt|\bmx)$.

This method has the advantage that it allows for quick evaluation and sampling of the full posterior at inference time. \rev{This accelerated emulation of the posterior is often called \textit{amortized} inference.}  However, the disadvantage is that it requires knowledge of the analytic form of the proposal prior $\tilde{p}(\bmt)$ from which training data was sampled, which may not be accessible for astrophysical training data (\eg cluster properties in cosmological simulations).
Recently, \cite{vasist2023} used an NPE strategy to constrain exoplanetary atmospheric models. NPE has also become a known tool in Gravitational-Wave astronomy where analytical Compact Binary Coalescent (CBC) waveforms play the role of the forward model (\eg with \code{DINGO} from \citealt{daxGW2021}, and the constraints on GW150914 by \citealt{cristosomi2023}, among many others).

\subsubsection{Neural Likelihood Estimation (NLE)} \label{subsec:nle}
An alternative method which circumvents the issue of fixed priors is to only fit for the likelihood, $\calP \left(\bmx \vert \bmt \right)$, via Neural Likelihood Estimation \citep[NLE;][]{Alsing_2018, Alsing_2019, papamakarios2019sequential}. The loss function for NLE notably does not include the assumed or proposal prior, and simply maximizes the global log-likelihood,
\begin{equation}\label{eqn:Lnle}
    \calL_{\rm NLE} := -\bbE_{\calD_{\rm train}}\log q_{\bmw}(\bmx|\bmt),
\end{equation}
wherein we emphasize the swapped arguments of $q$ as compared to Equation \ref{eqn:Lnde}.
Multiplying a learned likelihood with an assumed prior results in a proxy which is proportional to the posterior, \ie $\hat{\calP}(\bmt|\bmx) \propto q_{\bmw}(\bmx|\bmt) p(\bmt)$. Generating samples from the posterior is then simply a matter of using this proxy in running Markov chain Monte Carlo \citep[MCMC;][]{robert1999monte} or Variational Inference \citep{blei2017variational}. For example, in Metropolis-Hastings MCMC \citep{robert2004metropolis}, the learned likelihood and assumed prior can be used to calculate the acceptance probability of a transition $\bmt_t\rightarrow\bmt'$ without knowledge of the full normalized posterior, \ie
\begin{equation}\label{eqn:mcmc}
    \alpha := \min\left(1, \frac{\calP(\bmt')\calP(\bmx\vert\bmt')q(\bmt'\vert\bmt_t)}{\calP(\bmt_t)\calP(\bmx\vert\bmt_t)q(\bmt_t\vert\bmt')}\right),
\end{equation}
where $q(\bmt'\vert\bmt_t)$ is the proposal mechanism. Note, the process of obtaining a posterior is identical to the explicit likelihood-based approach, but instead of an analytic likelihood function, NLE has essentially a ``black box'' function instead. The advantage of this method is that one only needs to perform training once for a given model, at the cost of additional sampling once data becomes available.
For instance, \cite{jeffreyDES2021} fit a density model for the likelihood of compressed summaries from \code{DES SV} weak lensing maps. The \code{Madminer} tool \citep{Brehmer2019xox} for particle physics notably allows for NLE using ILI tools.

\subsubsection{Neural Ratio Estimation (NRE)} \label{subsec:nre}
As in NLE, Neural Ratio Estimation \citep[NRE;][]{hermans2020likelihood} targets a proxy that is proportional to the posterior. Instead of outputting a conditional density estimate for the likelihood, NRE models instead target the likelihood ratio,
\begin{equation}
    r(\bmx,\bmt) := \frac{\calP(\bmx\vert\bmt)}{\calP(\bmx\vert\bmt_0)},
\end{equation}
equal to the ratio of the likelihood at a given point in parameter space to the likelihood at a reference point, $\bmt_0$. This formulation is convenient because it can be folded into an acceptance ratio like Equation~\ref{eqn:mcmc} and also be cast as a classification problem \citep{cranmer2015approximating}. The output of a classification network, $d_{\bmw}(\bmx, \bmt)\in [0,1]$, can be considered equivalent to a likelihood ratio as
\begin{equation}
    \hat{r}(\bmx,\bmt) = \frac{d_{\bmw}(\bmx, \bmt)}{1-d_{\bmw}(\bmx, \bmt)}.
\end{equation}
The classification problem can be interpreted as quantifying whether observed data $\bmx$ is consistent with $\bmt$. We can then train this neural network model using the loss:
\begin{equation}\label{eqn:Lnre}
    \calL_{\rm NRE} := \bbE_{\calD_{\rm train}}\left[d_{\bmw}(\bmx, \bmt) + \bbE_{\bmt'\sim p(\bmt)} \left[1-d_{\bmw}(\bmx, \bmt')\right]\right],
\end{equation}
wherein we associate a positive classification with assigning a data $\bmx$ to the correct $\bmt$, and a negative classification with those assigning data to a $\bmt'$ sampled from the prior.
The main benefit of NRE over NPE and NLE is that one need not specify an approximating distribution such as MDNs or normalizing flows to emulate a posterior. The complexity is purely limited in terms of the depth and design of the neural architecture. Among others, \cite{coleTMNRE} and \cite{karchevTMNRE} applied a variation of NRE (namely Truncated Marginal NRE, \citealt{millerTMNRE}) to CMB data and Supernova data, respectively. Additionally, \cite{delaunoy2020} and \cite{peregrine2023} use NRE for Gravitational Wave parameter inference.

\subsection{Choosing an Estimator} \label{subsec:choose_est}
The optimal choice of NPE, NLE, or NRE is highly dependent on the properties on the underlying problem, the dimensionalities of $\bmx$ and $\bmt$, and the intended application. First, the ease of learning the shape of a likelihood or posterior distribution can vary dramatically between different physical problems. For example, the shape of the unimodal posterior in Figure \ref{fig:example} is considerably simpler than that of the bimodal likelihood, and thus NPE will be easier to converge than NLE. However, this is the opposite for some applications presented in Section \ref{sec:synth}, where the likelihood is simpler and thus NLE converges much faster than NPE or NRE. In cases where both the posterior and likelihood exhibit complicated shapes, NRE models are a strong option, as they do not require an explicit choice of NDE.

\rev{Another} strong rule of thumb is that neural networks are easier to train when using high-dimensional input to produce low-dimensional output than vice versa. Thus, if $\operatorname{dim}(\bmx)$ is large (\eg for image- or graph-like data) then NPE may work better than NLE models. \rev{Alternatively, for high-dimensional posterior inference, \ie when $\operatorname{dim}(\bmt)$ is large, NLE often performs better than NPE, especially when strong degeneracies exist between parameters. The NRE implementations in \ili also struggle in this regime, as evidence suggests that NRE methods require truncation or marginalization to resolve regions of high posterior density \citep{millerTMNRE, Miller:2022shs}. This heuristic may change with the advent of extremely flexible diffusion architectures \citep[\eg][]{song2020score}, but \rev{such models} are not yet integrated into \ili.}

Lastly, it is imperative to consider the downstream applications of the inference. NPE models are much better for cases in which the inference must be repeated for many tests $\bmxo$ (\eg estimating morphology for many JWST galaxies), as it can be prohibitively expensive to run MCMC chains for NLE and NRE for many test points. However, if the learned posteriors are going to be injected into a broader hierarchical likelihood sampling (\eg using cluster masses to constrain cosmology), NLE or NRE will be a more natural choice.

For a quantitative comparison of NPE, NLE, and NRE on machine learning benchmarks, see \citet{lueckmann2021benchmarking}. In any case, the best way to choose a method for a new problem is to try each and perform quantitative comparisons, tests which are made considerably easier with the configuration interface in \ili{}.

\subsection{Sequential Learning}\label{subsec:sequentialL}

Thus far, we have phrased our problem in such a way that we first have a set of pre-run simulations, and only later do we obtain the posterior distribution given our observations. This can be the case when the simulations are particularly expensive (\eg $N$-body simulations in cosmology, or Earth systems models for climate science \eg from which \citealt{climateILI} built emulators). In these cases, we must be careful about any difference between the proposal prior for the parameters from which the simulations were run (\eg a Latin hypercube) and the chosen prior for the parameters used in the inference (\eg a multivariate Gaussian). It may be the case that this proposal prior is significantly wider than the posterior distribution inferred once the real data have been observed, meaning that many of the simulations used in training for NPE/NLE/NRE have been ``wasted'' as the posterior has almost no support in the region of parameter space occupied by a (potentially significant) fraction of the simulations. 

One can instead adopt a multi-round inference approach to improve simulation efficiency, referred to as Sequential NPE/NLE/NRE or \rev{SNPE/SNLE/SNRE}. In this method, we begin by running a fraction of the total desired simulations with parameters sampled from the broad prior $p(\bmt)=\Tilde{p}(\bmt)$, conducting the first round of training for the NPE/NLE/NRE model. This weak inference is then applied to $\bm{x_{\rm o}}$ to derive a first estimate for the posterior. \rev{For instance, in the case of NPE, this is equivalent to minimizing Equation \ref{eqn:Lnde} with $p(\bmt)= \Tilde{p}(\bmt)$ during the first round, which gives a model $q_{\bmw}(\bmt|\bmx)$, and then retraining the model with new simulations drawn from $\Tilde{p}(\bmt)=q_{\bmw}(\bmt|\bmx = \bmxo) $ for the second round, and so on.} The algorithm \citep[\eg][]{greenberg2019automatic} then identifies regions of high posterior density around the observation $\bmxo$ and conducts additional simulations within these areas to produce a refined posterior estimate. This can be repeated until a convergence criterion is met or the total simulation budget is exhausted. This iterative refinement of the posterior around the region of interest can lead to improved posterior estimates for a given experiment. 
However, it is important to note that these results are deliberately optimized for a singular observed data point. The sampled simulations may not be appropriate for a different experiment, potentially resulting in a less accurate estimate of the posterior distribution if no further simulations are run.

Note that an alternative method for generating parameters of interest using acquisition functions that minimize the expected uncertainty in the posterior density approximation exists. This is based on the
Bayesian Optimization for Likelihood-Free Inference (BOLFI) framework \citep{gutmann2016bayesian} has also been suggested and applied to cosmological data \citep{Leclercq2018}, although this is not yet part of \ili{}.

\subsection{Validation}\label{subsec:valid}
The goal of model validation is to assess whether the posteriors $\hat{\calP}(\bmt|\bmxo)$ learned in Section \ref{subsec:nde} will be accurate and reliable when applied to new data. Explicitly, we want to evaluate: (1) whether the learned posteriors are maximally constraining of $\bmt$ given the observed data $\bmxo$, and (2) whether the predictive uncertainties quoted by our learned model are accurately calibrated to our training data. These two criteria are naturally adversarial, and is often referred to as the bias-variance tradeoff. For example, a model can satisfy condition (1) by reporting tight error bars, but it will then fail condition (2) when its error bars are smaller than the true distribution of training data. Similarly, a model that produces a posterior equal to the prior will easily satisfy condition (2) but will ultimately be uninformative for solving condition (1). 
Despite the `black box' nature of neural networks, these criteria 
control the quality of the learned implicit inference posteriors, mitigating the chance that the learned model is not representative of the data used to train it.

We evaluate the constraints and coverage of our learned posteriors on a labeled test set of data-parameter pairs $\calD_{\rm test} := \{(\bmx_i, \bmt_i)\}_{i=1}^{N_{\rm test}}$. This test set is meant to represent a fully independent sampling of the real data distribution, unseen by the model prior to validation. This test set may originate from the same source as our labeled training set $\calD_{\rm train}$ but must be held independent to avoid overfitting and underestimation of the uncertainty (see \citealt{huppenkothen2023constructing} for a review of best practices). If the distribution of data-parameter pairs in the test set matches that of the predictions of our learned posterior, then our model is calibrated and can be reliably extended to new, unlabeled data.

We build tests by comparing the true parameters from the test set to posterior samples derived from our models. Sampling-based comparisons provide a unified framework for all NPE/NLE/NRE learning strategies since they do not require normalization of the posterior PDF. The technical details for sampling from neural networks depend on the implemented learning strategy and are discussed in more detail in Section \ref{subsec:structure_validation}. 
Here, we will assume that, for a given input $\bmx$, we can draw independent and identically distributed (i.i.d.) samples from the learned posterior distributions $\hat{\bmt}\sim \hat{\calP}(\bmt\vert\bmx)$. These samples can then be used to compare the level of constraint on $\bmt$ and the consistency with the true values. 



To test precision, which reflects the ability of a learned posterior to constrain parameter values, we examine the distribution and shape of posterior samples $\hat{\bmt}$ around the true value $\bmt$. These distributions are commonly visualized using multivariate corner plots 
or marginal true vs. predicted plots. 
The contours in these plots help qualitatively assess constraining power, identify posterior degeneracies, and detect systematic biases.

A quantitative measure of precision is the cumulative likelihood of the test dataset, denoted as $\hat{\calP}(\calD_{\rm test}) = \prod_{i=1}^{N_{\rm test}} \hat{\calP}(\bmt_i|\bmx_i)$. Larger test likelihoods indicate a model posterior that concentrates more probability mass around the true value, implying greater constraining power. The test likelihood is simple to calculate for NPE methods, which directly estimate $\calP(\bmt|\bmx)$. However, for NLE and NRE methods, evaluating the log-likelihood involves an additional step of training a generative model $\mathcal{G}(\bmt)$ from the inferred posterior samples $\hat{\bmt}$. This process is computationally expensive, especially for large datasets, and not sensitive to overconfidence.

Lastly, given samples from a reference posterior, we can evaluate the discrepancy between a model's prediction and its optimum. Reference posteriors can often be attained via long-run MCMC chains for explicit likelihood approaches, and then used to compare with ILI methods. \citet{lueckmann2021benchmarking} performed an extensive quantitative analysis on various sample-wise distances for benchmarking ILI methods and found that Classifier 2 Sample Tests \citep[C2ST;][]{lopez2016revisiting} were the most constraining measure of accuracy. C2ST is defined as the accuracy of a trained classifier (often a neural network itself) to distinguish between the true and inferred posterior samples. High C2ST suggests that the true and inferred posterior are very different, whereas a C2ST of $\sim 0.5$ suggests they are nearly indistinguishable. C2ST is later used to benchmark \ili{} in Section \ref{subsec:toy}.


Posterior samples are also useful for quantifying the calibration of our model uncertainty in parameter space. We can construct a direct comparison of the predicted percentiles by our inference engine to the true error observed in our validation dataset. We first define the Probability Integral Transform \citep[PIT;][]{cook2006} as the \textit{cumulative density function} (CDF) of our model posterior given an input $\bmxo$,
\begin{equation}\label{eqn:pit}
    {\rm PIT}(\bmt;\bmxo) = \int_{-\infty}^{\bmt}\ d\bmt\ \hat{\calP}(\bmt\vert\bmxo).
\end{equation}
Given i.i.d. samples from the posterior, we can construct an estimator for the PIT value as
\begin{equation} \label{eqn:pit_est}
    \hat{\rm PIT}(\bmt;\bmxo) = \bbE_{\hat{\bmt}\sim \hat{\calP}(\bmt|\bmxo)}\left[\Theta\left(\hat{\bmt}-\bmt\right)\right],
\end{equation}
where $\Theta$ is the Heaviside step function. Stated plainly, the PIT value counts the number of times that our posterior samples $\hat{\bmt}$ fall below the true parameter value $\bmt$. If our model posteriors are globally consistent with the truth, i.e. we match the true posterior everywhere in data-parameter space, then the distribution of PIT values evaluated on data-parameter pairs from our test set must be uniformly distributed \citep{zhao2021diagnostics}: 
\begin{equation}
    {\rm PIT}(\bmt_i;\bmx_i) \sim U(0,1),\ \forall\ (\bmx_i, \bmt_i)\in\calD_{\rm test}\,.
\end{equation}
As a result of this property, the distribution of PIT values is a common goodness-of-fit metric for conditional density models \citep[e.g.][]{cook2006, bordoloi2010photo, tanaka2018photometric}. This is also known as Simulation-Based Calibration \citep{talts2018}.

To better interpret errors in the PIT distribution, we can use a percentile-percentile (P-P) plot, explicitly comparing the CDF of PIT values to the CDF of a uniform random variable. Intuitively, the P-P plot measures: `What percentile level is my posterior model assigning to the true value, and with what frequency does this occur in the test set?' If the predictive posterior is well-calibrated, these should be the same, \eg we should predict the true value below the 50-th percentile 50\% of the time. If it is not, the shape of the P-P plot acts as a sensitive probe of global bias or over-/under-dispersion of predictive uncertainty (see Figure 1 of \citealt{zhao2021diagnostics} for interpretations of error with P-P plots). This is an extremely practical tool for understanding and correcting biases and over/under-dispersion in complex ILI posteriors. 

As the dimensionality of $\bmt$ increases, proper coverage of the posterior distribution requires exponentially more samples. As a result, the PIT estimator (Equation \ref{eqn:pit_est}) becomes intractably difficult to measure with low variance in the high-dimensionality regime. In this regime, we use approximate methods to bound the constraints of posterior coverage tests. Firstly, we can construct a PIT value from the marginal posterior over each $i$-th component of $\bmt$, \ie 
\begin{equation}
    \hat{\rm PIT}(\theta_i;\bmx) = \bbE_{\hat{\bmt}\sim \hat{\calP}(\bmt|\bmx)}\left[\Theta\left(\hat{\theta}_i-\theta_i\right)\right],
\end{equation}
where $\hat{\theta}_i$ and $\theta_i$ are the $i$-th components of $\hat{\bmt}$ and $\bmt$ respectively. If our model posterior is globally consistent with the true posterior, then this marginal PIT value has the same properties as Equation \ref{eqn:pit_est}, \ie when evaluated on the test set they should be uniformly distributed. 


It is also sufficient to use approximations to check multivariate posterior coverage. The Tests of Accuracy with Random Points \citep[TARP;][]{tarp} method uses samples in high-dimensional parameter space to estimate expected coverage probabilities. By examining the density of posterior samples within regions of parameter space near true values, TARP constructs estimates of posterior coverage which are guaranteed to converge to the true posterior coverage with sufficient samples. With enough test samples, the TARP method is necessary and sufficient to \rev{test the accuracy of posterior estimators}.
\section{Code Structure}\label{sec:structure}

\begin{figure*}
    \centering
    \includegraphics[width=\linewidth]{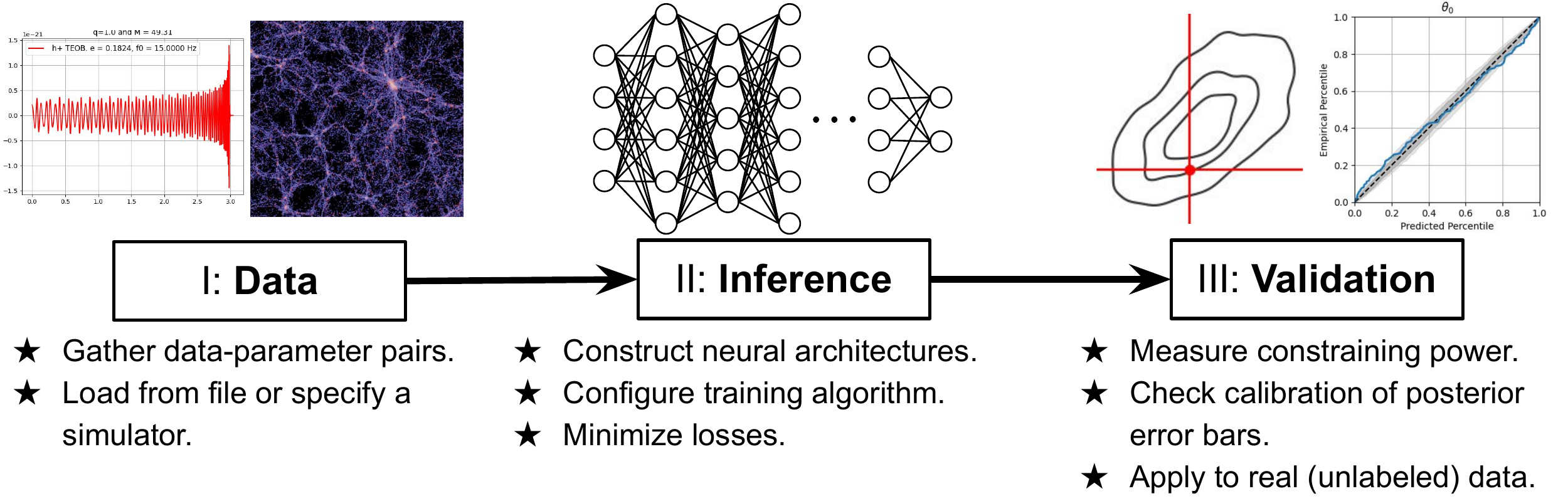}
    \caption{Overview of the \ili{} pipeline, displaying the procedural processes when running each stage. Each of the three stages (Data, Inference, Validation) is independently configurable, meaning any setup of one stage will automatically link to the others.}
    \label{fig:pipeline}
\end{figure*}

The \ili{} software is a pipeline for training implicit inference models to infer scalar parameters from observational data. A typical inference pipeline consists of three main stages: Data, Inference, and Validation. The Data stage involves loading datasets from file or setting up on-the-fly simulators (e.g. for the methodologies in Section \ref{subsec:sequentialL}) in order to provide data-parameter pairs for implicit inference. The Inference stage then takes these data-parameter pairs, trains neural networks to connect them with posterior or likelihood representations (as in Section \ref{subsec:nde}), and saves the fitted models to file. Finally, the Validation stage loads a test dataset as well as the fitted models from Inference in order to evaluate goodness-of-fit metrics (e.g. constraint level, posterior coverage, etc.) and produces inference on unlabeled data. A schematic of this pipeline is shown in Figure~\ref{fig:pipeline}.

Each stage is independent and adaptive, such that the configuration of one stage can be changed while keeping the others the same and the pipeline will still run. For example, if you had a fixed dataset but wanted to try out many different neural architectures, you would only need to change the Inference configuration. Similarly, if you wanted to see the gain in constraining power of using a short, conservative data vector versus a longer, more informative data vector, you could simply change the Data configuration to load the different datasets and the Inference and Validation would proceed exactly the same way. 
This design principle is extremely practical for doing apples-to-apples comparisons of the many design choices and hyperparameters associated with ILI, a key tenet of robust and responsible machine learning \citep{huppenkothen2023constructing}.

The pipeline is designed to promote exploratory research while also supporting efficient production-level testing. We provide a fully functional Jupyter interface and detailed examples to access each stage of \ili{}. The Jupyter interface is a widely-used tool for examining new problems, building educational material, and facilitating collaborative efforts \citep{wang2019data}. The Jupyter interface of \ili{} provides step-by-step guidance for establishing and testing a new inference pipeline that is familiar to novices and experts alike. This allows \ili{} to act both as a user-friendly toolkit to explore inference on a new problem as well as carry out thorough testing and hyperparameter tuning through its \code{yaml}-based configuration API. Users can generate many configurations of the pipeline and run multiple fittings in parallel on large-scale GPU compute clusters.

In this section, we discuss the design of the Data, Inference, and Validation stages of the \ili{} pipeline, including each of their capabilities and relevant reference works. For specifics on using the API, we refer the reader to our full code documentation\footnote{\url{https://ltu-ili.readthedocs.io}}.

\subsection{Data}

The objective of the Data stage is to gather data-parameter pairs and present them in an amenable format to the Inference stage. For standard inference, with fixed training and test catalogs, the Data stage simply loads data-parameter pairs from in-memory or on-disk storage and provides \code{get\_all\_data()} and \code{get\_all\_parameters()} helper functions to pass along data and parameters in the correct format. These functions are called during the Inference stage for later pre-processing within the \sbi{}, \pydelfi{}, and \lampe{} backends. Within \ili{}, we provide several convenience classes to prepare data from either \code{numpy}-, \code{xarray}-, or \code{torch}-like data formats. Users are also can create their own data-loading objects, so long as they adhere to the aforementioned conventions.

For multi-round inference (Section \ref{subsec:sequentialL}), the Data stage also provides a \code{simulate()} function, wherein the loader will execute an on-the-fly simulation for given parameters to generate data for efficient sequential learning. In the example class provided in \ili{}, users can initialize this stage with an arbitrary simulation function that intakes parameters $\bmt$ and outputs data $\bmx$, and the sequential Inference stage will handle any calls to the function. Also, users can load a pre-run initial catalog from file and write any newly-generated simulations to disk during the training procedure.

\subsection{Inference}\label{subsec:training}

The \ili{} code encompasses a broad diversity of methodologies and architectures for performing implicit inference. However, the core configurations for all types of inference are universal. 
Every Inference stage requires specifying: a posterior-, likelihood-, or ratio-estimation; one or multiple neural architectures; a prior distribution; and training hyperparameters (e.g. learning rate, batch size, etc.).

Once these are provided, the Inference stage knows how to pull data-parameter pairs from the Data stage, perform data preprocessing, construct each training algorithm, minimize losses (see Equations \ref{eqn:Lnde}, \ref{eqn:Lnle}, and \ref{eqn:Lnre}), and produce a learned posterior model, $\hat{\calP}(\bmt|\bmx)$. Each inference algorithm automatically implements best practices for training, including data normalization \citep{singh2020investigating}, adaptive stochastic gradient descent \citep{kingma2014adam}, and validation-guided early stopping \citep{bai2021understanding}. The remainder of this section is dedicated to the core considerations of configuring the Inference stage.

The Inference stage is built on three distinct computational backends: \pydelfi{} \citep{alsing2019fast}, \sbi{} \citep{tejero-cantero2020sbi}, and \lampe{} \citep{rozet2021lampe}. \pydelfi{} is a Tensorflow package \citep{abadi2016tensorflow} for performing NLE using active learning and data compression techniques. \sbi{} is a PyTorch package \citep{paszke2019pytorch} for performing NPE, NLE, and NRE, as well as their sequential analogs.  \lampe{} is also a PyTorch package focused on NPE methods, with substantial support for diverse embedding networks and normalizing flow models. The implementations of NLE in \pydelfi{} and \sbi{} are similar in theory \citep{papamakarios2019sequential} but differ in applications of embedding networks. \pydelfi{} has the option to train in a two-step process, wherein first it trains an embedding network to optimally compress data via Fisher information maximization \citep{charnock2018automatic, alsing2018generalized} and then attaches it to the head of an NLE model, whereas \sbi{} exclusively uses NLE embedding networks to compress parameters. \pydelfi's two-step compression-estimation process has been shown to significantly improve simulation-efficiency during sequential training for applications in cosmology \citep{alsing2018massive}. The \lampe{} NPE in \ili{} is a custom implementation of \citet{lueckmann2017flexible}, similar to SNPE-B of \sbi{} but allowing for greater variety of configuration, embedding networks, and NDEs.
\ili{} is a unifying framework for all of these mutually-exclusive backends, allowing, for the first time, convenient apples-to-apples comparison of each implementation of ILI. 

The Tensorflow and PyTorch backends have a variety of pre-implemented NDEs and ratio classifiers. The NDEs are either Mixture Density Networks (MDNs; See Section \ref{subsec:nde}) or normalizing flows \citep[See][for a review]{papamakarios2021normalizing}. 
\rev{
The \pydelfi{} backend possesses its own neural architectures as originally implemented from scratch in \citet{alsing2019fast}, whereas the \sbi{} and \lampe{} models have software dependencies, respectively \code{nflows} \citep{nflows} and \code{zuko} \citep{rozet2022zuko}.}
Both backends also accept various standard and customizable prior distributions. The primary requirement of a custom prior is the ability to evaluate $\log p(\bmt)$ (for NDE models) or a proportional proxy (for NLE/NRE models).  If the assumed prior's support is bounded (\eg a top-hat prior), the code internally implements an affine parameter transformation to avoid assigning a nonzero probability to regions of parameter space beyond prior bounds.

A key component of \ili{} is model ensembling, wherein we independently train multiple neural networks on the same dataset and combine their predictions to improve robustness to overfitting and model uncertainty. Singular NDEs are prone to overconfidence \citep{hermans2022trust}, \ie they have a tendency to underestimate the predictive uncertainty. Deep ensembling \citep{lakshminarayanan2017simple} is a well-tested schema for correcting overconfidence by averaging NDE predictions from multiple models to inflate error bars and correctly capture model uncertainty. It is an alternative to the computationally-expensive Bayesian neural networks \citep{blundell2015weight, gal2016dropout, cobb2021scaling}, in which the full weight posterior is learned during training and sampled during inference. We recommend that users utilize deep ensembles to ensure robust inference within \ili{}.

The adaptability of neural networks to various data inputs has greatly contributed to their growth in astronomy and cosmology. The convenient implementations of the various NDEs and classifiers in \ili{} can be further augmented with customizable embedding networks. These embedding networks can use classic \code{Keras}-like neural layers \citep{chollet2015keras} in both Tensorflow and PyTorch, greatly simplifying implementation. The integration with embedding networks in the \lampe{} backend allow for exotic inputs, like graphs and sequences. In the code, we provide several examples of embedding networks, including connections to convolutional and graph neural networks \citep{lecun1998gradient, kipf2016semi}. We also note that all backends can be run on GPUs, to take advantage of computational speed-ups when using complex embedding networks.

\subsection{Validation}\label{subsec:structure_validation}
The Validation stage provides modular functions to implement the validation metrics described in Section \ref{subsec:valid} on the learned posteriors $\hat{\calP}(\bmt|\bmx)$. The execution of the Validation stage follows: 
\begin{enumerate}
    \item Load the learned neural posteriors from the Inference Stage.
    \item Sample the posterior $\hat{\bmt}\sim \hat{\calP}(\bmt|\bmx)$ at test data points, $\calD_{\rm test}$, or at an unlabeled observational point, $\bmx_{\rm obs}$.
    \item Construct the validation metric and return or save it to file.
\end{enumerate}

The most impactful configuration of the Validation stage is the choice of sampling methods. \ili{} provides three classes of samplers: MCMC, variational inference (VI), and direct sampling.
All models (NPE/NLE/NRE) in both backends are natively integrated with the \code{emcee} \citep{foreman2013emcee} sampling package. \code{emcee} implements the \citet{goodman2010ensemble} MCMC sampler and improves efficiency through parallelization and affine transformations but does not take advantage of probability gradients. The \sbi{} models also have access to \code{PyTorch}'s \code{pyro} samplers \citep{bingham2019pyro}, including slice samplers \citep{neal2003slice}, Hamiltonian Monte Carlo \citep[HMC;][]{neal2011mcmc}, and the No-U-Turn sampler \citep[NUTS;][]{hoffman2014no}. All \code{emcee} and \code{pyro} samplers have a unified interface within \ili{}, wherein users can simply specify their choice of sampler as well as the number, length, and thinning of MCMC chains. Also, we provide the functionality for \sbi{} models to fit and sample from NDEs through neural VI \citep{graves2011practical}. VI is a fast, approximate alternative to MCMC sampling, particularly useful in the high-dimensional parameter regime. Lastly, solely for the NPE models in \sbi{}, users can utilize direct sampling, which is faster than both MCMC and VI. We remark that the choice of sampling configuration might change according to the validation metric in question, so \ili{} allows for the flexibility to specify this on a case-by-case basis.

\rev{\subsection{Choosing a Backend}\label{subsec:choose_back}
As described above and later tested in Section \ref*{sec:synth}, the algorithmic behavior of the \sbi, \pydelfi, and \lampe backends are nearly identical for the same experimental configurations (\ie the same training data, hyperparameters, neural architectures, etc.). However, each backend has unique strengths and weaknesses and may be more appropriate for different problems. For example, \sbi is the most commonly-used backend, with a set of standardized training procedures and architectures which have been validated across many problems \citep{lueckmann2021benchmarking}. It also has custom samplers for NLE and NRE methods which are faster than the generic \code{emcee} samplers used in the \pydelfi backend. For new users, \sbi is a reliable choice for exploring new datasets or setting up standard ILI applications. In contrast, \lampe's routines have access to a broad scope of exotic NDEs and allow for much greater flexibility when designing training procedures and custom embedding architectures. In our experience, using the flow models unique to \lampe often produced tighter and better-calibrated posterior estimates relative to similar \sbi applications. In addition, graph and recurrent embedding networks are currently only possible through the \lampe interface, because of its capacity to handle complex data modalities. As a result, \lampe is recommended for more advanced use of ILI in complex datasets. Lastly, our \pydelfi tests in Section \ref{subsec:benchmarking} appear to indicate stronger performance in sequential learning tasks than similar approaches with \sbi. Also, \pydelfi's \code{Tensorflow} implementation may make it a more attractive option to users already working in \code{Tensorflow}, rather than the other options in \code{PyTorch}.
Despite these rules of thumb, \ili is a fully unified interface incorporating all these backends. Our code allows users to easily explore the strengths of each backend for their specific applications.
}

\section{Synthetic Experiments}\label{sec:synth}
In this section, we demonstrate the capabilities of \ili{} on a number of synthetic experiments for which an analytic form for the posterior or likelihood is known. We use these traditional experiments to measure benchmark performance the variety of inference engines available in the code against each other and classical methods (\eg ABC, HMC).

\subsection{Toy Problem} \label{subsec:toy}

\begin{figure}
    \centering
    \includegraphics[width=\linewidth]{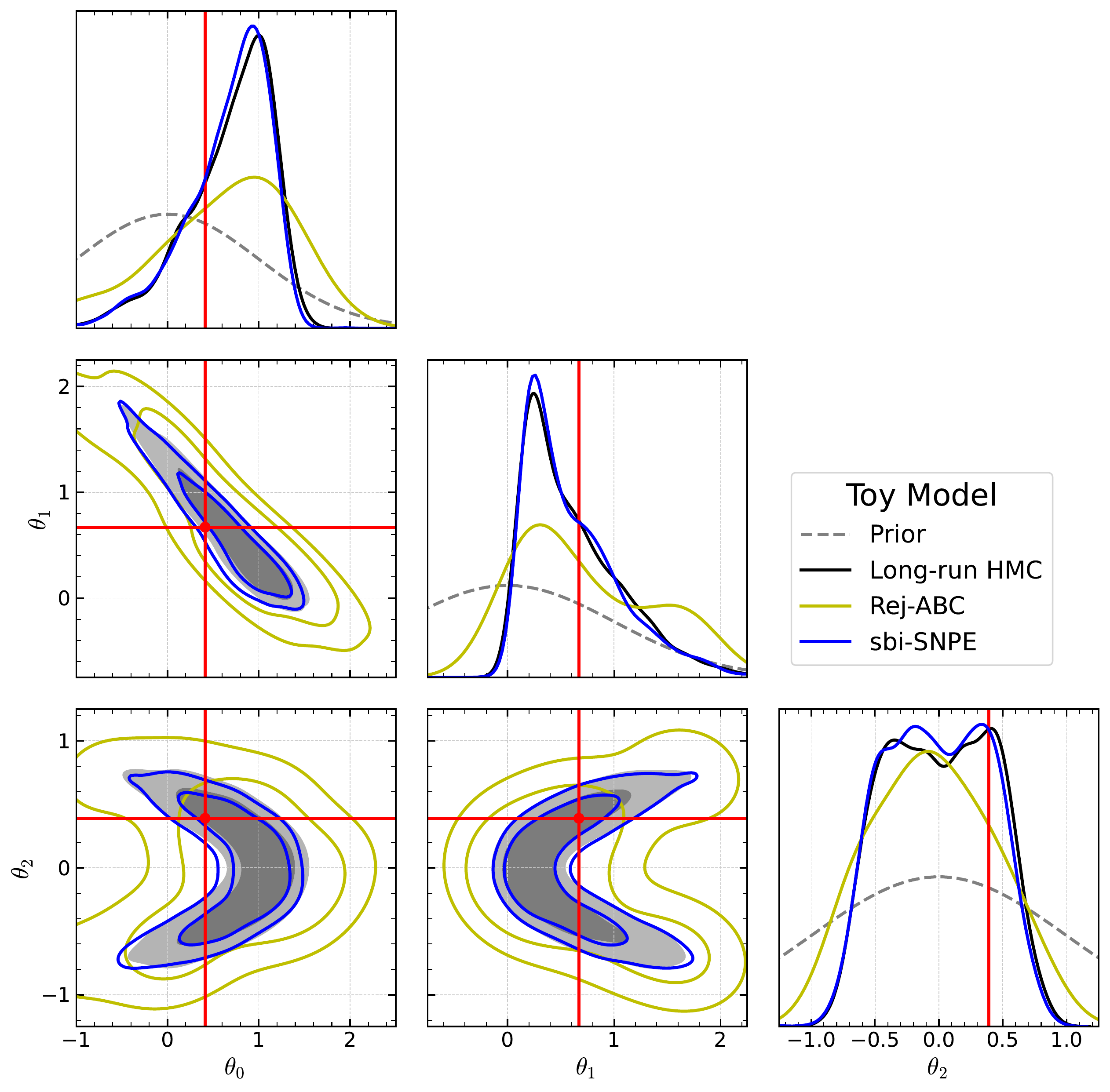}
    \caption{Example SNPE posterior inference in comparison to classical implicit (Rejection ABC) and explicit (HMC) likelihood inference for the known toy simulator in Equation \ref{eqn:toy}. The true value for each parameter is shown in red and contours are shown at the central $68\%$ and $95\%$ confidence intervals. Note that the prior for each $\theta_i$ is a standard normal, as $p(\theta_i) = \mathcal{N}(0,1)$. 
    }\label{fig:toy_corner}
\end{figure}

We first build a toy model to demonstrate \ili{}'s ability to learn non-linear posteriors. We consider a 10-dimensional data vector $\bmx$ constructed element-wise from the following stochastic simulator:
\begin{equation}\label{eqn:toy}
    x_i = 3\sin (k_i+\phi_0) + \phi_1 k_i^2 + \epsilon_i,
\end{equation}
where $k_i = (2i/3)-3$ and $i\in [0,9]$.
Here, the data vector $\bmx := \{x_i\}_{i=0}^{9}\in\bbR^{10}$ is composed of a sinusoidal signal with phase $\phi_0$, a quadratic signal of amplitude $\phi_1$, and an i.i.d. noise component $\epsilon_i \sim \mathcal{N}(0,1)$. We further decompose ${\bm \phi}$ into our target parameters $\bmt\in\bbR^3$ via $\phi_0 = \theta_0+\theta_1$ and $\phi_1 = \theta_1 - 3\theta_2^2$. We expect our inference engine to forward constraints on latent variables $\bm\phi$ onto target variables $\bmt$. We assume a standard normal prior on each of our parameters of interest, $p(\theta_i)=\mathcal{N}(0,1)$. This is a challenging inference task because the likelihood is highly non-linear and the posteriors exhibit strong degeneracies. However, learning parameters from such complex, correlated data vectors is a common problem in astronomy and cosmology \citep[\eg][]{daxGW2021, modi2023sensitivity, bartlett2023precise}.

Given this simulator, we train \sbi-SNPE using an ensemble of two normalizing flows (MAFs) to produce the example $\bmt$ posterior in Figure \ref{fig:toy_corner}, shown relative to equivalent constraints with Rejection ABC and long-run HMC. We train the SNPE sequentially using 10 rounds of 2000 simulations each, following the procedure described in Section \ref{subsec:sequentialL} and using Equation \ref{eqn:toy}. For ABC, we run the same number of simulations (20,000) and take the closest $0.5\%$ to our data vector, using an L2 distance. For HMC, we run eight chains each with 10,000 tuning steps and 10,000 sampling steps, for a total of 160,000 simulations. This run has an effective sample size of $~3,900-8,000$ and a Gelman-Rubin $\hat{R}$ statistic of $1.001$ for all parameters. We consider these long-run HMC samples to be equivalent to a reference `ground truth' posterior.

Despite the same simulation budget as ABC, the SNPE posterior constraints have converged to the reference posterior and are strongly more constraining than the ABC constraints. In both the SNPE and HMC posteriors, we clearly observe the expected linear degeneracy between $\theta_0$ and $\theta_1$ as well as the quadratic degeneracy between $\theta_1$ and $\theta_2$, direct results from the aforementioned definitions of $\phi_0$ and $\phi_1$. These two degeneracies imply another quadratic degeneracy between $\theta_0$ and $\theta_2$. Despite this non-linearity, we see the multi-dimensional 68\% and 95\% confidence intervals are entirely consistent between SNPE and HMC. However, ABC methods struggle in this regime. Because the data dimensionality is ten, ABC requires orders-of-magnitude more samples to fully capture the tails of the posterior distribution. This regime demonstrates the utility of having a low-cost amortized posterior model like SNPE, especially in further cases where the simulators are not as simple.

\subsection{Benchmarking} \label{subsec:benchmarking}
\begin{figure*}
    \centering
    \includegraphics[width=\linewidth]{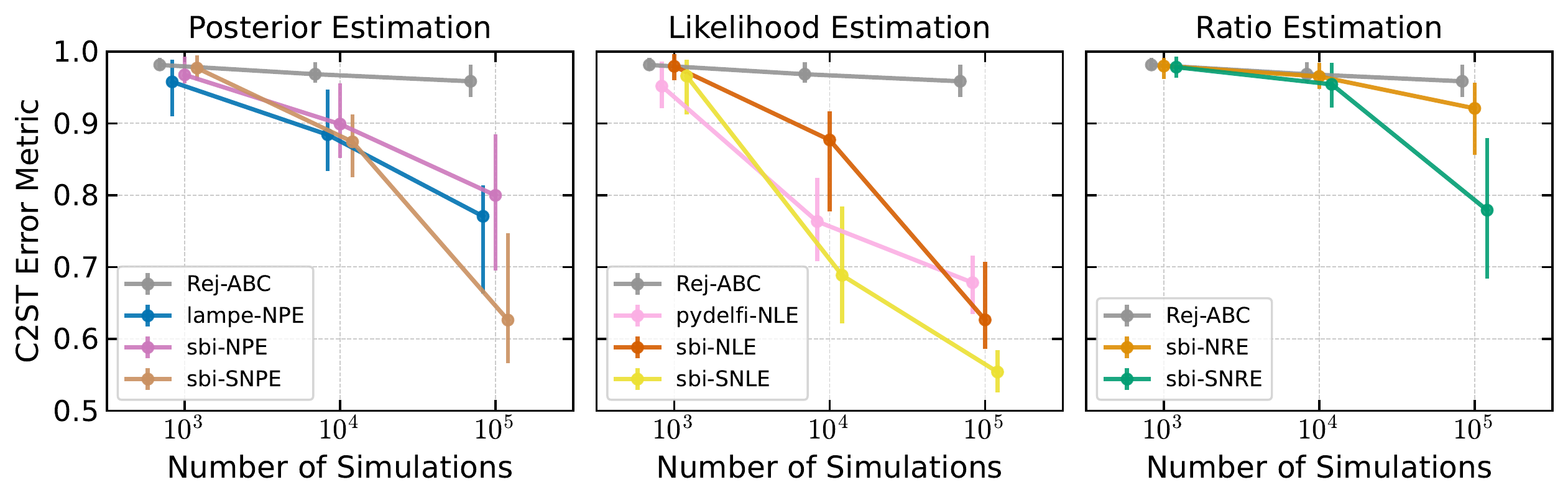}
    \caption{Prediction error as a function of simulation budget for various \ili{} training methodologies on the benchmark SLCP inference problem \citep{papamakarios2019sequential}. For clarity, plotted points for different models at the same simulation budget are slightly offset horizontally. The Rejection-ABC method is shown on all subplots as a baseline. Error is defined in terms of the Classifier 2-sample Test \citep[C2ST;][]{lopez2016revisiting} metric. A lower C2ST indicates a more accurate inferred posterior, with 0.5 being the optimal score. C2ST values are shown at their median and central 95\% confidence interval, calculated over ten independent runs.}
    \label{fig:benchmark}
\end{figure*}

We benchmark all models in \ili{} on the standard ``Simple Likelihood Complex Posterior" problem \citep[SLCP;][]{papamakarios2019sequential} common in simulation-based inference experiments \citep[\eg][]{Greenberg_2019, lueckmann2021benchmarking, ramesh2022gatsbi}. SLCP has a fairly straightforward, unimodal likelihood but a very complex, multimodal posterior distribution, making it a difficult, but attainable benchmark for implicit inference. We follow the implementation of SLCP in \citet{lueckmann2021benchmarking}, wherein we train various NPE/NLE/NRE methods and compare their resultant posteriors to reference posteriors determined with long-run HMC chains. We implement this for all available NPE/NLE/NRE methods and their sequential analogs in each of the \pydelfi, \sbi, and \lampe{} backends. We also include benchmarking of Rejection ABC as a baseline. Throughout all backends and engines, we use the same architectures as in \citet{lueckmann2021benchmarking}, which were determined through exhaustive hyperparameter searches.

For each trial, we measure the error of an inference engine with respect to a reference posterior using C2ST (see Section \ref{subsec:valid}).
For each backend, engine, and simulation budget, we perform independent training and inference on ten separate reference posteriors to aggregate test statistics.

In Figure \ref{fig:benchmark}, we show the performance of each \ili{} engine on the SLCP benchmark as a function of the simulation budget used for generating the training set.  As we would expect, the accuracy of inferred posteriors increases with the number of given simulations across all models. We observe that all models strongly outperform the Rejection ABC baseline due to their efficient use of simulation data. For the SLCP benchmark, we observe that NLE methods perform the best, NPE methods are a close second, and NRE methods have the worst performance. However, we note that the SLCP by definition has a simple, easy-to-learn likelihood, encouraging strong NLE performance. This hierarchy of performance could be entirely different for other problems. Sequential methods (SNPE/SNLE/SNRE) show very little improvement for small simulation budgets ($10^3$), but show massive gains for large budgets ($10^5$), especially for the NLE methods. This is likely because, for small training datasets, the posteriors are poorly constrained and barely deviate from the prior. However, for the tight posterior constraints in large budget runs, sequential methods are able to more easily focus simulation resources on small regions of parameter space. Next, we note that the \lampe-NPE implementation largely mirrors that of \sbi-NPE, and \pydelfi-NLE slightly outperforms \sbi-NLE for a medium simulation budget ($10^4$). Although hyperparameters (\ie architectures, learning rate, validation schema) are fixed for these experiments, slight differences in backend implementations might lead to better or worse performance. Lastly, as a check, the performances of all \sbi{} engines closely match those found in \citet{lueckmann2021benchmarking}, suggesting our implementation of SLCP is correct.

\section{Science Experiments} \label{sec:examples}

In this section, we demonstrate various applications of \ili{} for common astrophysics and cosmology problems. These experiments are designed to display the diverse functionality of the code, including using CNN embedding networks (Section \ref{subsec:clusters}), applying posterior coverage validation (Section \ref{subsec:quijote}), incorporating graph neural network layers (Section \ref{subsec:graph}), using multiround inference (Section \ref{subsec:gravwave}), quantifying information in different data sources (Section \ref{subsec:Dust_example}), and integrating information-optimal embeddings (Section \ref{subsec:sapphire}). We note that these experiments are intended to illuminate  interesting pathways for the scientific application of \ili{} and not necessarily make significant scientific conclusions on their subject material. For readability, experiment descriptions are kept concise, but the code and data are made publicly available through the codebase.




\subsection{X-ray Mass Estimation of Galaxy Clusters} \label{subsec:clusters}

\begin{figure*}
    \centering
    \includegraphics[width=0.95\linewidth]{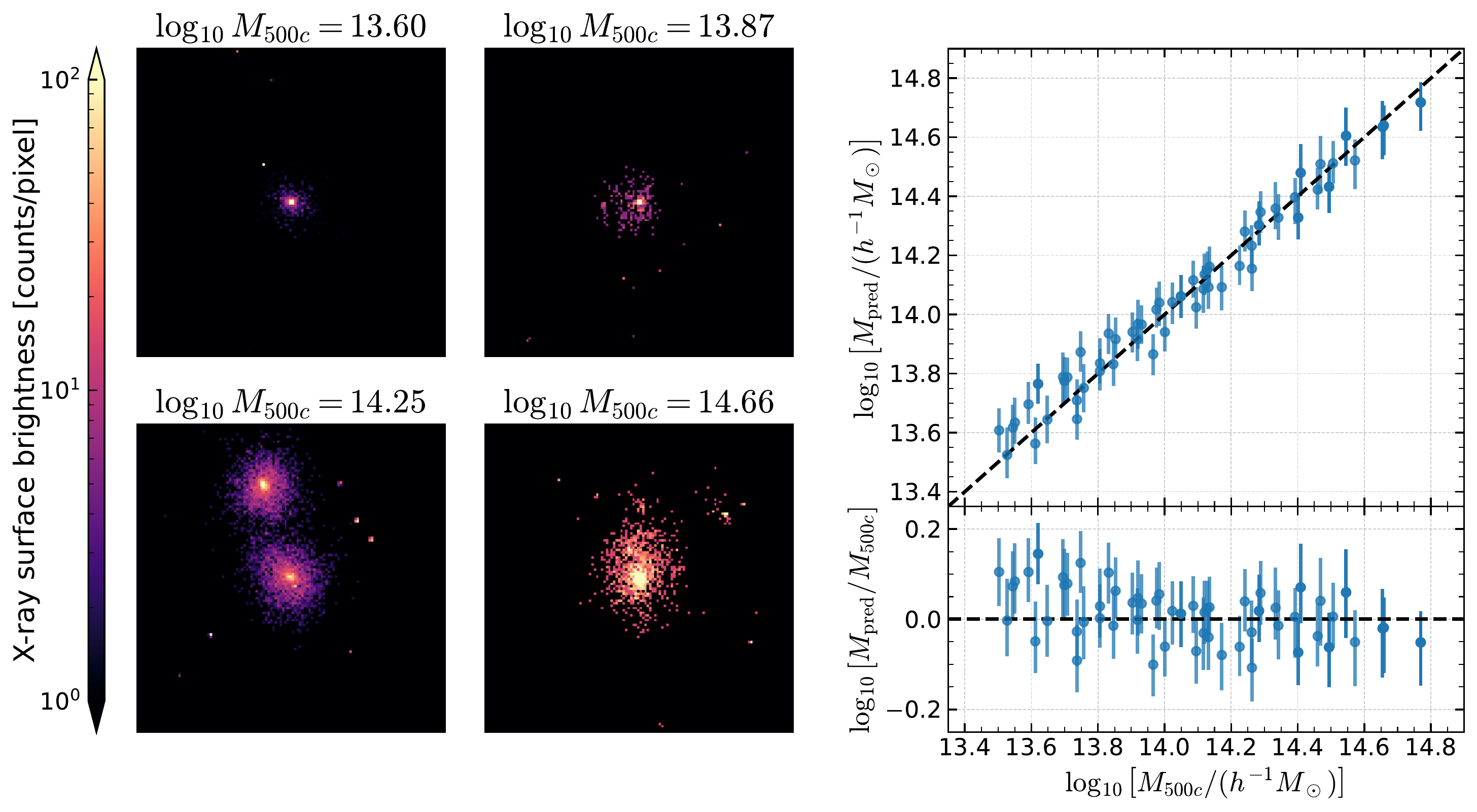}
    \caption{Inference of galaxy cluster $M_{500c}$ mass from eROSITA-like mock X-ray observations, as in \citet{ho2023benchmarks}. The left four subplots show random examples of single-band X-ray observables in terms of their sky-projected surface brightness and target mass (in $h^{-1}M_\odot$). The right plot is the true vs. predicted log-mass estimates on the test set, where we show the median and central 68\% confidence interval of our neural posteriors. For readability, test points in the right plot have been randomly subsampled.}\label{fig:xray}
\end{figure*}

A prime functionality of \ili{} is its flexibility to adapt to many modalities of data. Through the use of a customizable embedding network, we process image data to estimate the mass of galaxy clusters from X-ray observations. Our inference problem mirrors that of \citet{ho2023benchmarks}, wherein we try to estimate the $M_{\rm 500c}$ mass of a galaxy cluster from $128\times128$ images of sky-projected X-ray photon counts. To train and test our model, we use a dataset of $3,285$ mock X-ray observations of clusters derived from the Magneticum hydrodynamical simulation \citep{dolag2016sz} designed to mimic those of the eROSITA telescope \citep{soltis2022machine}. These mock observations are single-band images of the X-ray photon emission from the hot intra-cluster medium (ICM) and include realistic simulation of the contaminating systematics which afflict real X-ray measurements, including cluster morphology, background emission, telescope response, and AGN sources. Examples of these observables are shown in Figure \ref{fig:xray}. A good estimator of cluster mass needs to be able to disentangle the source X-ray signal from the noise and peripheral emission, while also understanding the physical connection between X-ray emission, ICM gas content, and system mass.

To recreate this example, we perform NPE using an embedding network of the same convolutional architecture as \citet{ho2023benchmarks}. That is, instead of the last dense layer mapping to a single point estimate, we forward the final embedding into an NDE. We construct an ensemble of four models, each with the same embedding architecture, but two with Masked Autoregressive Flow \citep[MAF; ][]{papamakarios2017masked} NDEs and two with MDNs. We then train the models from scratch using the NDE losses, instead of the mean-squared-error (MSE) loss of \citet{ho2023benchmarks}. As in the reference, we assume a uniform prior on cluster mass. We split our mock catalog into 90\% training data and 10\% test data, and present our performance results on the latter. The entire training and testing procedure takes about fifteen minutes on an Nvidia V100 GPU.

The true vs. predicted mass estimates for this model are shown in Figure \ref{fig:xray}. Using the original architecture, we achieve a test scatter on the mean prediction of $0.0782\ {\rm dex}$. When compared to the $0.0773\ {\rm dex}$ scatter of the single-band X-ray images in \citet{ho2023benchmarks}, we achieve a very similar level of information extraction, though our uncertainty is slightly higher than the original. We attribute this to the fact that model ensembling slightly inflates predictive uncertainty \citep{hermans2022trust}. However, we remark that by slightly increasing the number of convolutional filters in the first two layers, we observed mean scatters as low as $0.071\ {\rm dex}$, suggesting that the NPEs are capable of state-of-the-art constraints on physical problems when given the right architecture. We also note that hyperparameter search through neural architectures is made efficient and accessible through \ili{}'s configuration-based interface.

\begin{figure*}
    \centering
    \begin{minipage}{.63\linewidth}
        \begin{subfloat}[True vs. Predicted\label{fig:truevpred}]{
        \centering
        \includegraphics[width=\linewidth]{
        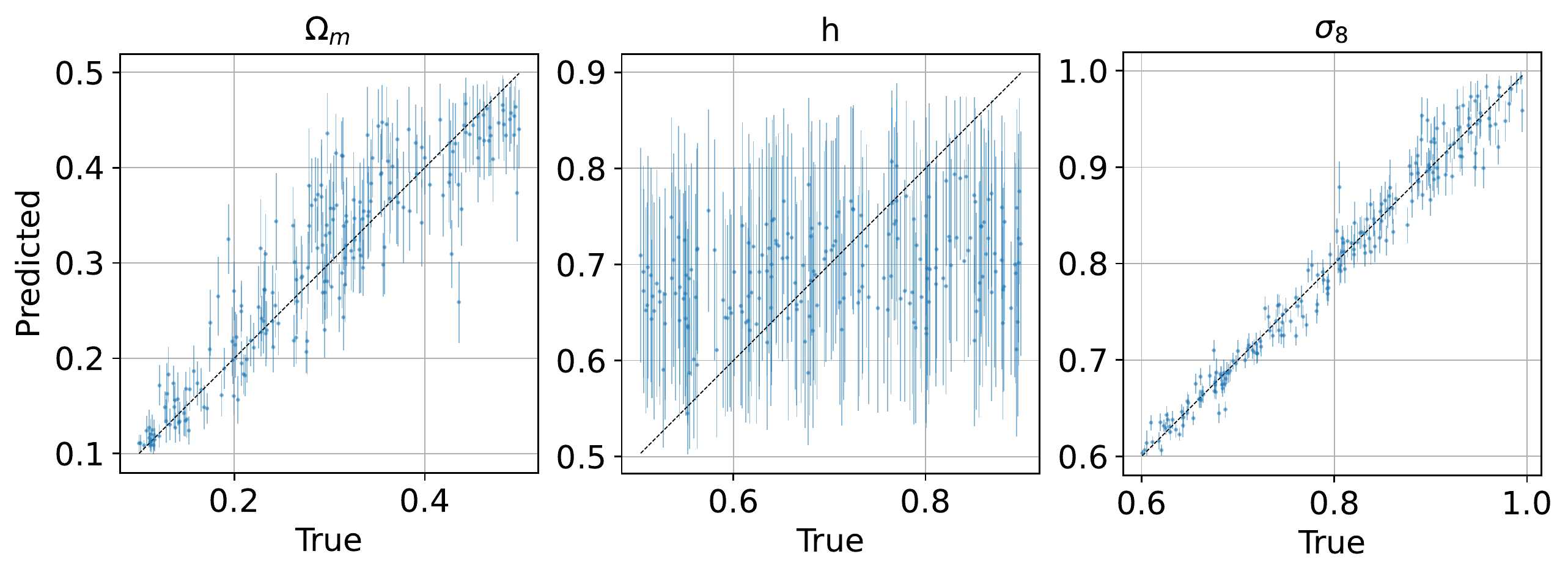}}
        \end{subfloat}\\
        \begin{subfloat}[Marginal Percentile Coverage Test\label{fig:Quijote_MPCT}]{
        \centering
        \includegraphics[width=\linewidth]{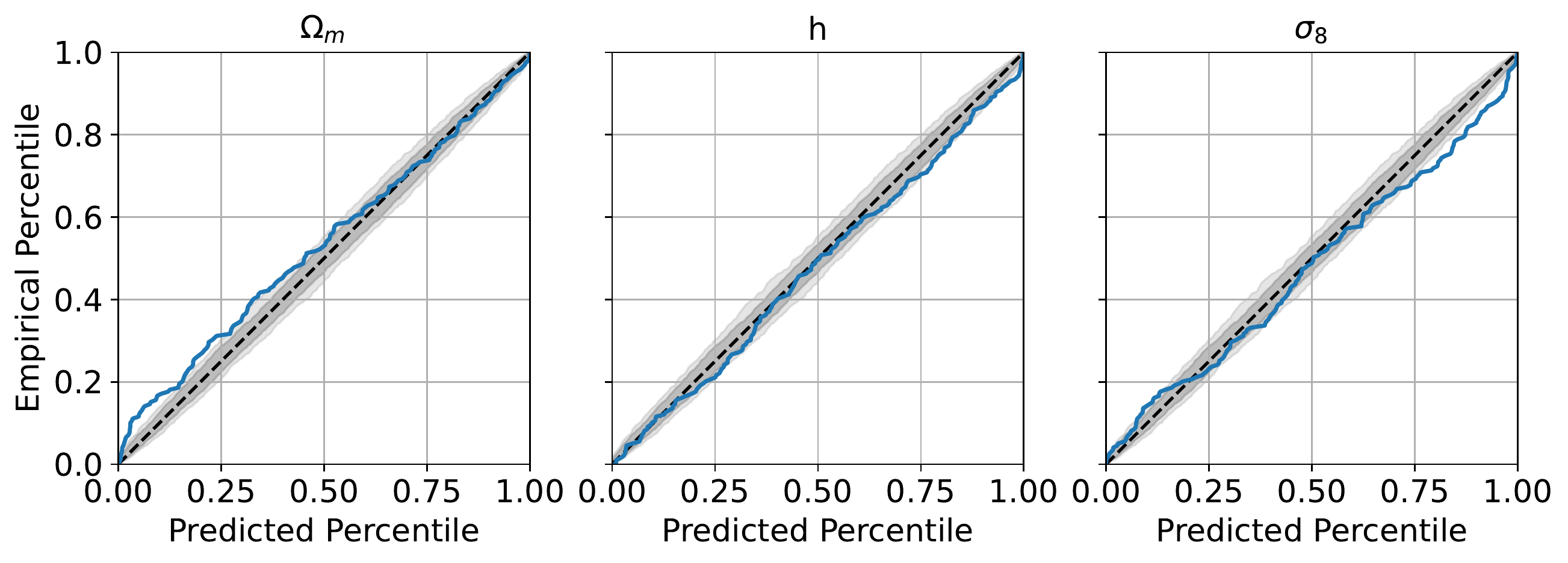}}
        \end{subfloat}\\
    \end{minipage}\hfill
    \begin{minipage}{.35\linewidth} 
        \begin{subfloat}[TARP Multivariate Coverage Test
        \label{fig:tarp}]{
        \centering
        \includegraphics[width=\linewidth]
        {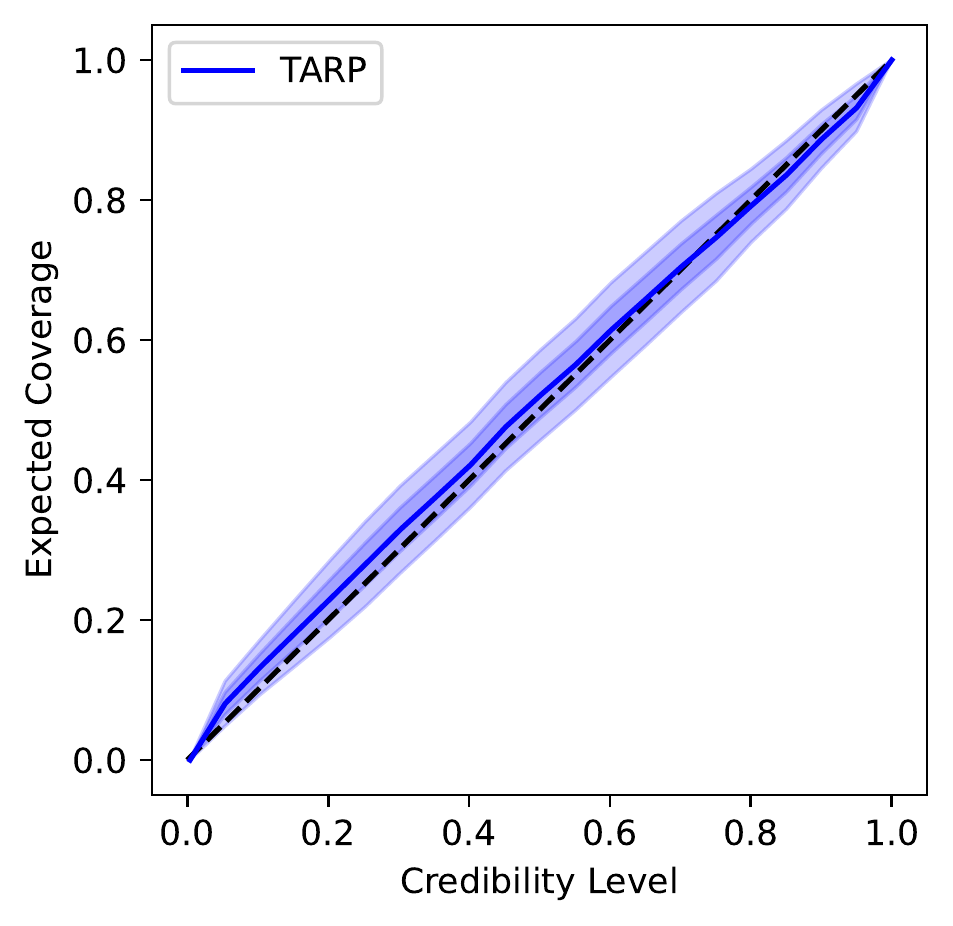}}
        \end{subfloat}
    \end{minipage}
\caption{\label{fig:quijote} Inference of three cosmological parameters $\Omega_m$, $h$, and $\sigma_8$ from halo power spectrum multipoles in Quijote simulations using \rev{\ili}. In (a) we show the comparison between true and predicted value of the parameters in the test simulations. Marginal error bars are shown at the central $68\%$ confidence intervals. In (b) and (c) we show the coverage tests using the P-P plots as well as using TARP method. These show bootstrapping error bars at $68\%$ and $95\%$, indicative of typical variations in an ideal estimator for a finite test set.}
\end{figure*}

\subsection{Quijote Dark Matter Power Spectrum} \label{subsec:quijote}

\rev{Another prime utility of \ili is its capacity to seamlessly integrate ILI training procedures with sampling and validation metrics. A common inference benchmark in the field of cosmology is inferring cosmological parameters from the matter power spectrum \citep[\eg][]{cooray2002halo, spurio2022cosmopower, hahn2023simbig}. In this problem, we run simulations of the matter evolution in the universe for various cosmological models, measure observational summary statistics such as the matter power spectrum, and then try to connect these observations back to constraints on cosmology.}
Here, we demonstrate this application by predicting \rev{the posterior distribution of three cosmological parameters} using the power spectrum multipoles measured from the halo catalog of Quijote simulation \cite[see][for a description of the simulations]{Quijote_sims}. We use the power spectrum multipoles corresponding to $\ell=0,1,2$ in 23 linearly spaced wavenumber ($k$) bins ranging from \rev{$0.08\ h^{-1}{\rm Mpc}$ to $2.8\ h^{-1}{\rm Mpc}$} as our observations, $x$. We use the \rev{NLE} method to learn the likelihood $p(x|\theta)$, where the parameters $\theta$ correspond to three cosmological parameters: the total matter density $\Omega_m$, amplitude of the matter fluctuations $\sigma_8$, and the Hubble constant $h$ measuring the rate of the expansion of the Universe. 

Using the \sbi backend, we train a NLE model using an ensemble of six NDEs, each using a thin MAF architecture, with ten hidden layers and three transformations. We use 1800 simulations from Quijote to train the network, retaining the remaining 200 to test its performance as shown in Figure~\ref{fig:quijote}. We then use VI sampling to obtain the posterior of the parameters from this trained likelihood ensemble, weighted by their validation losses. This greatly accelerates the sampling process on the full test set when compared to traditional MCMC sampling.

We show the true vs predicted value of the three parameters on the top-left panels (Figure~\ref{fig:truevpred}), finding good predictive performance for $\Omega_m$ and $\sigma_8$. The lack of predictability for $h$ is expected physically, as the power spectrum multipoles alone are significantly less sensitive to expansion rate without some external prior information. We also perform coverage tests on the posteriors, finding uniformly distributed rank histograms as given by the coverage test in the bottom-left panels (Figure~\ref{fig:Quijote_MPCT}). Finally, we show the coverage test using the TARP method in the center-right panel (Figure~\ref{fig:tarp}), finding that the posterior is correctly calibrated.

\subsection{Field-level Inference with Graph Neural Networks} \label{subsec:graph}

\begin{figure*}[htbp]
    \centering
    \begin{minipage}{0.55\textwidth}
        \centering
        \includegraphics[width=\textwidth]{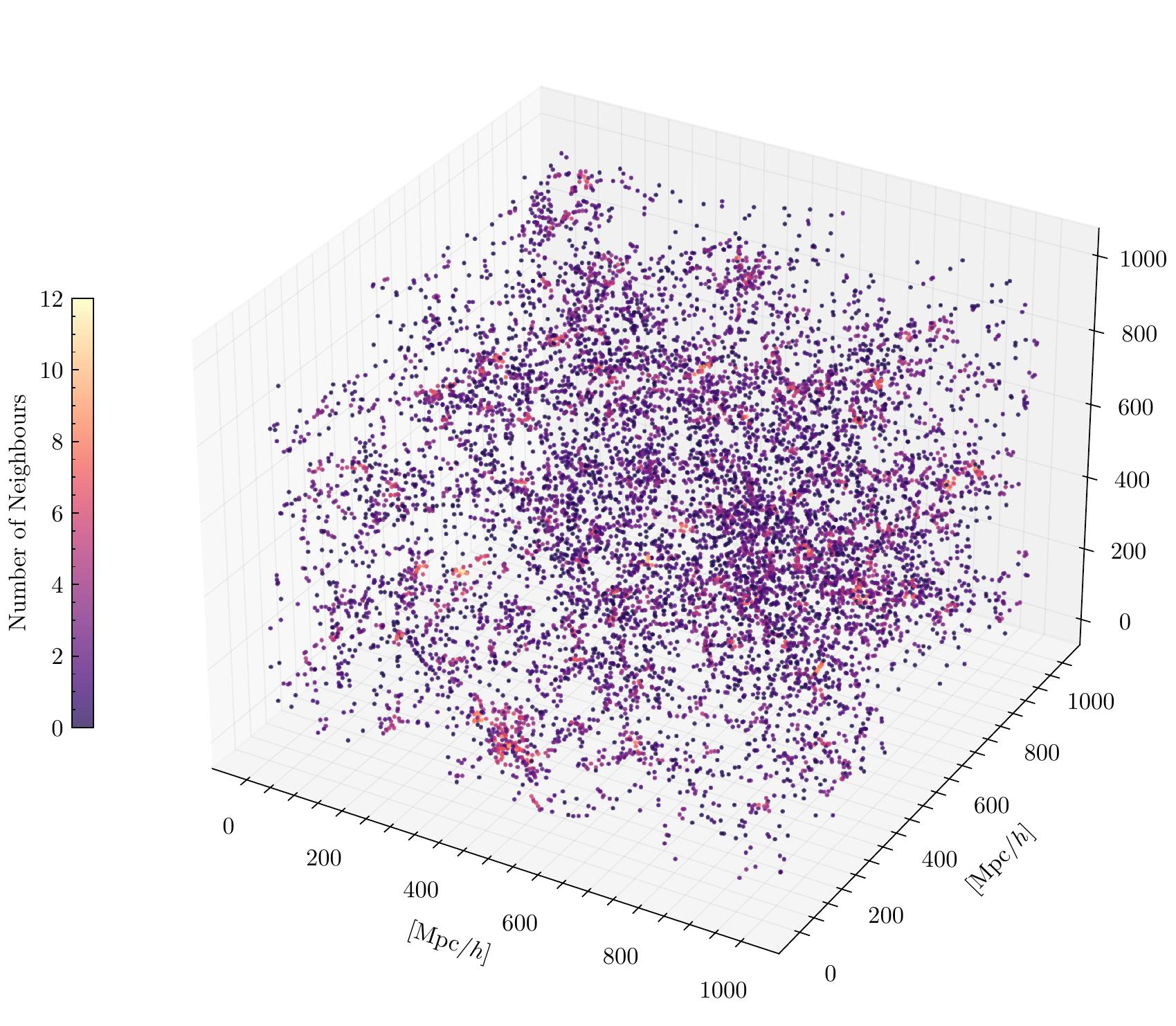} 
    \end{minipage}
    \hfill
    \begin{minipage}{0.43\textwidth}
        \centering
        \includegraphics[width=\textwidth]{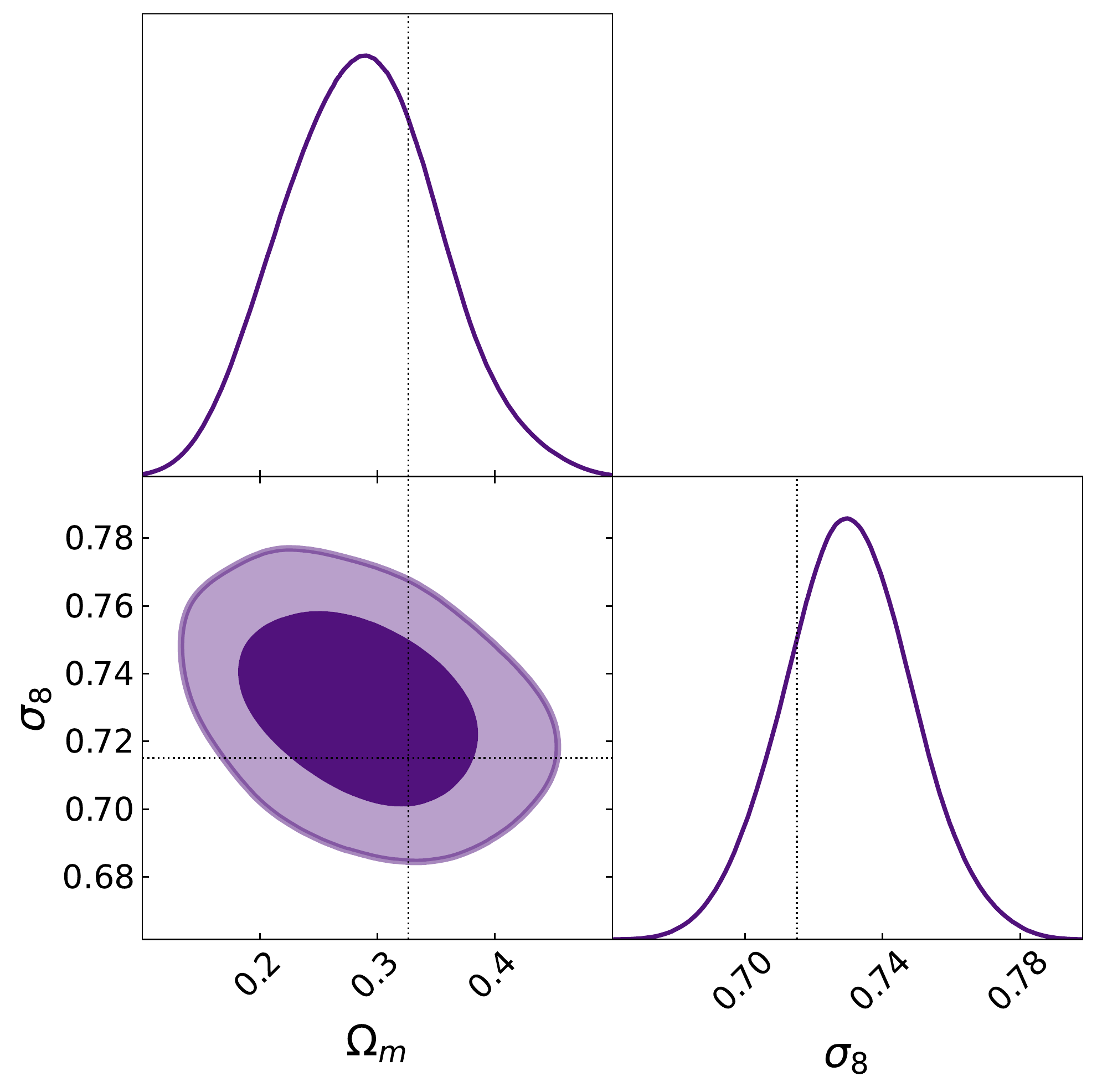} 
    \end{minipage}

    \caption{Cosmological inference from dark matter halo point clouds. Left: Example point cloud from the Quijote simulations. The graph neural network takes as an input the distances between halos separated by less than $20 \, {\rm Mpc}/ h$. Right: Inferred posterior over $\Omega_m$ and $\sigma_8$ on a test set simulation. True values are shown in gray. Contours are shown at the central $68\%$ and $95\%$ confidence intervals.}  
    \label{fig:gnn}
\end{figure*}

\rev{We next demonstrate \ili{}'s ability to handle complex data modalities by addressing the same cosmological inference problem using a point cloud dataset. Instead of power spectrum summaries, we instead use discrete catalogs of the $10,000$ heaviest dark matter halos in each of the Quijote simulations \citep{Quijote_sims}. The high resolution information present in the discrete halo positions has been shown to constrain cosmological parameters to extremely high precision \citep[\eg][]{Makinen2022Cosmic,cuesta2023point, deSanti_2023}. We ingest this catalog into our neural architectures using a message-passing graph neural network \citep{gilmer2020message}, which treats dark matter halos as nodes on a graph and filters down field-level information into highly-informative neural representations. This graph information is then fed into flow-based NDEs for performing NPE. We note that usage of graph-type inputs is only possible through the \lampe backend in \ili, as was employed here.

We train this graph-enhanced NPE model to recover the cosmological parameters $\Omega_m$ and $\sigma_8$ from the positions of Quijote's heaviest dark matter halos. The graph neural network is a message-passing neural network with node and edge neural networks being two multi-layer perceptrons with $3$ layers of $128$ hidden units each. The network takes as inputs 3D distances between halos, and their corresponding modulus. Nearby halos are connected within the graph if their comoving distance smaller than $20 \ h^{-1} {\rm Mpc}$. For NDEs, we train an ensemble of four networks, each one using a MAF architecture with five transformations and $50$ hidden features.}

In Figure \ref{fig:gnn}, we show an example point cloud together with the resulting graph neural network inference. The graph neural network obtains informative posteriors, consistent with true values. We note that the constraints are not as tight as those produced for the Quijote dark matter power spectrum in Figure \ref{fig:quijote}. This is likely because the graph only targets the most massive halos, \ie a more physically realistic probe, while the results in Figure \ref{fig:quijote} access the full, high-resolution dark matter distribution. \rev{In any case, \ili with the \lampe backend provides an easily accessible pathway for new users to integrate exotic architectures such as graph architectures into their development pipeline.}

\subsection{Multi-round Inference for Gravitational Waves} \label{subsec:gravwave}

\begin{figure*}
    \centering
    \begin{minipage}{.43\linewidth}  
        \begin{subfloat}[Example of GW signal in source and detector frame.\label{fig:gw_model}]{
        \centering
        \includegraphics[width=\linewidth]
        {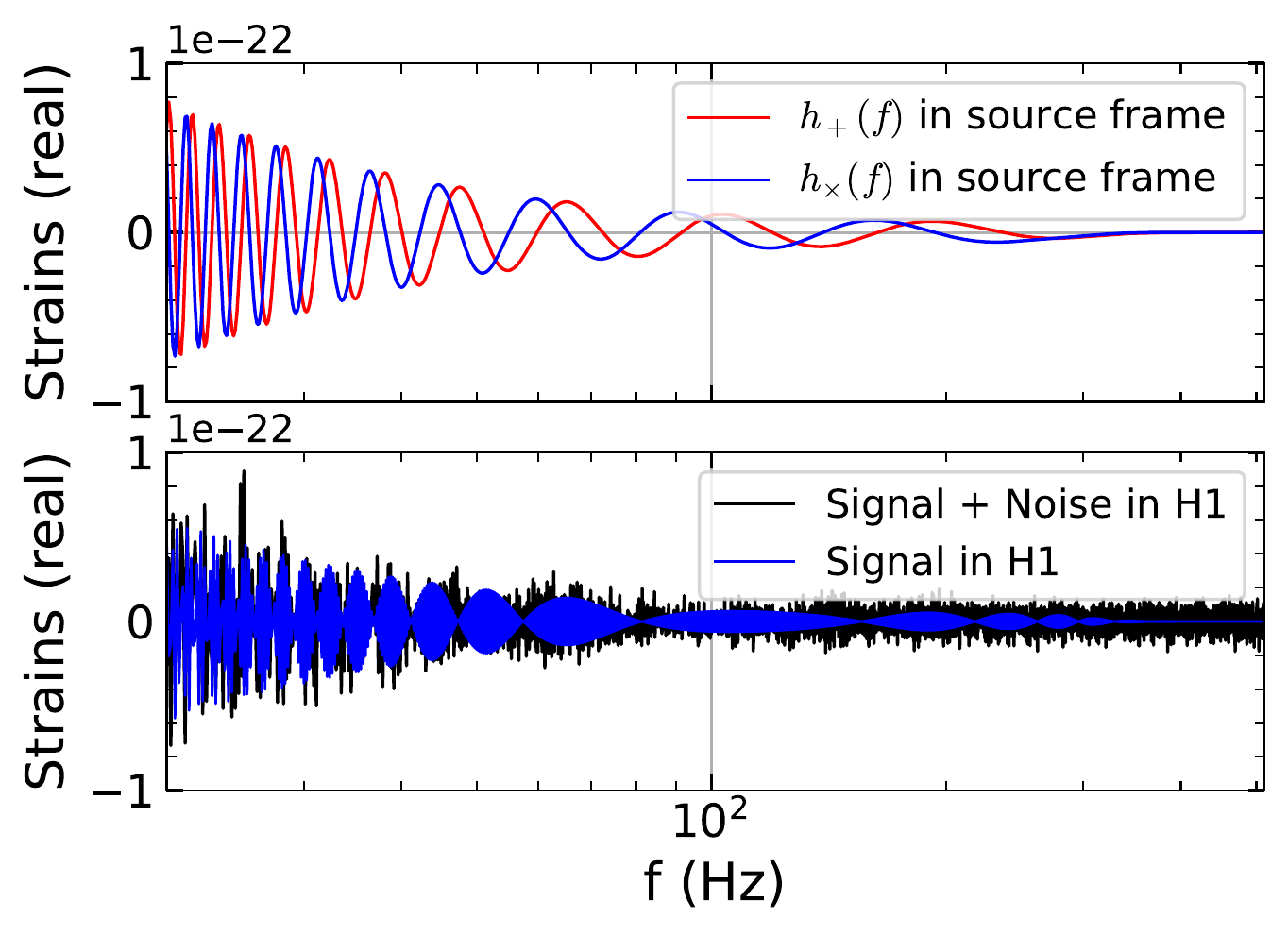}}
        \end{subfloat}\\
        \begin{subfloat}[Multi-round training and validation posterior probability.\label{fig:gw_logprob}]{
        \centering
        \includegraphics[width=\linewidth]{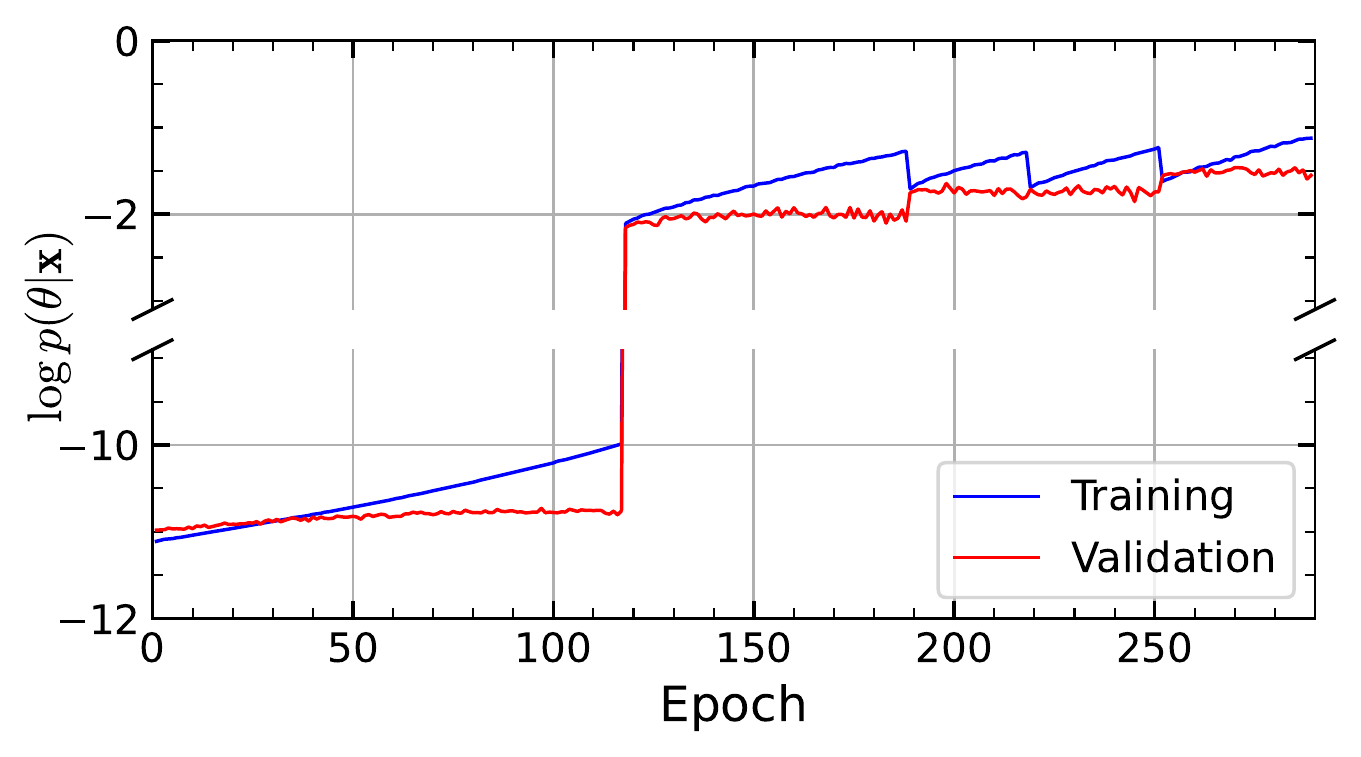}}
        \end{subfloat}
    \end{minipage}\hfill
    \begin{minipage}{.55\linewidth} 
        \begin{subfloat}[Posterior corner plot for SNPE inference.\label{fig:gw_posterior}]{
        \centering
        \includegraphics[width=\linewidth]{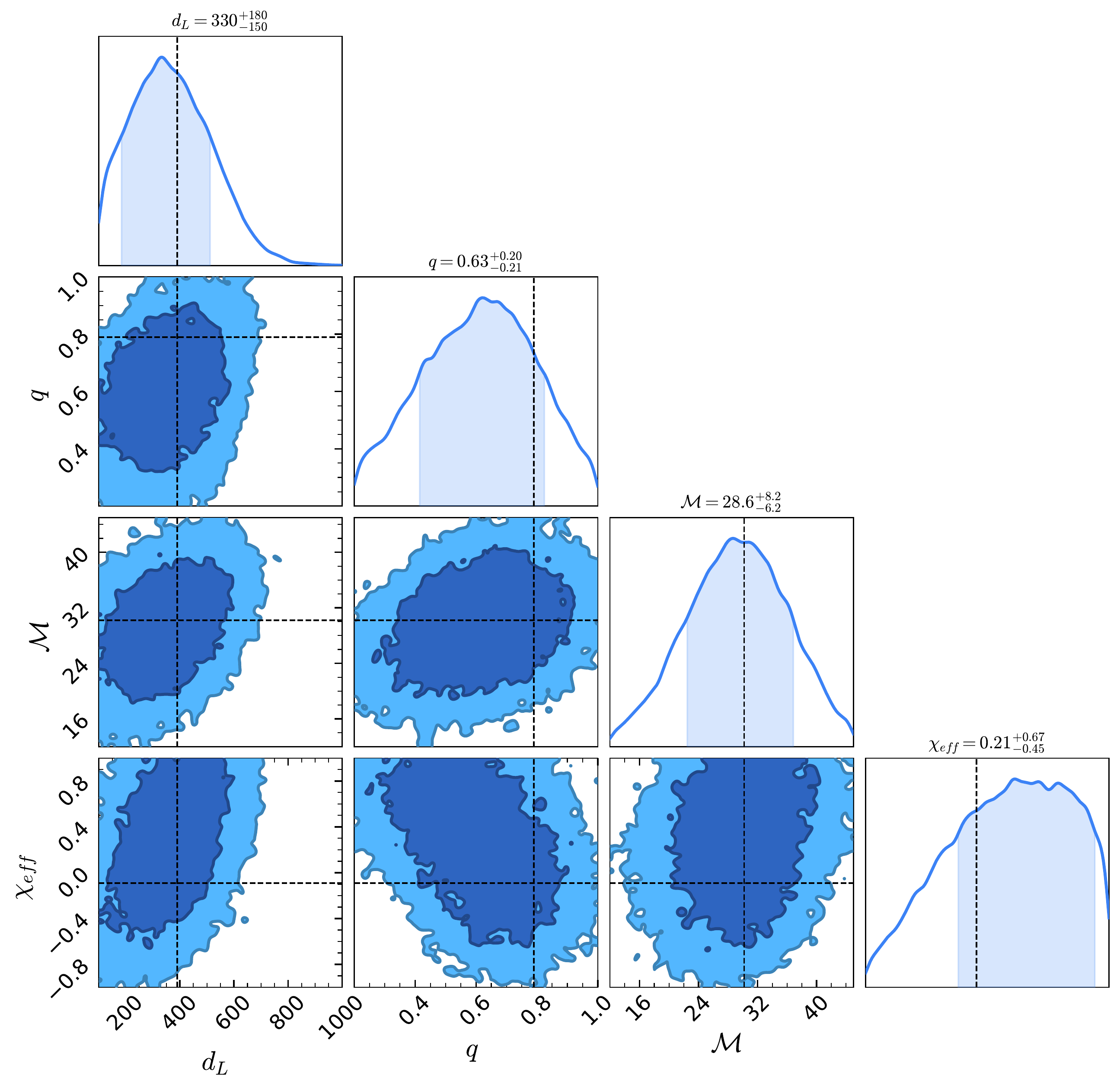}}
        \end{subfloat}
    \end{minipage}
\caption{Example parameter inference of a GW150914-like event as observed by the gravitational wave Hanford (H1) and Livingstone detectors. (a) Example simulated gravitational signal. Top: Radiation polarizations $h_{+}$ and $h_{\times}$ in source frame. Bottom: Real part of the strain and noise realization on Hanford (H1) detector frame. (b) Evolution of the log-probability for the multi-round GW inference example using SNPE, \ie the negative of the loss function in equation \eqref{eqn:Lnde} for $n_r = 5$ rounds. 
 (c)  Resultant posterior for a single run of the multi-round SNPE inference on GW strain data. Contours are shown at $68\%$ and $95\%$ confidence intervals, and dashed lines indicate the fiducial parameter values. The fake observation $\bmx_0$ was generated with the same simulator function as the training forward model.}
\end{figure*}


\rev{We next test the performance of \ili when addressing sequential learning tasks, specifically focusing on gravitational wave reconstruction.}. Over recent years, several studies based on ILI methods have addressed the problem of reconstructing parameters of merger events from gravitational wave (GW) signals \citep{cutler1994, gw_inference}. Initial advancements were marked by significant contributions to posterior estimation using normalizing flows, as highlighted in \cite{green_gwflow, green_gw150914}. Subsequent research \citep[DINGO:][]{daxGW2021} introduced variations of NPE taking advantage of Group Equivariant Neural Posterior Estimation \citep[GNPE:][]{dax_gnpe} or Flow Matching Posterior Estimation  \citep{fmpe_gw}. More recently, \cite{cristosomi2023} showcased a similar application of multi-round NPE to estimate ringdown parameters of GW150914. \cite{delaunoy2020} and \cite{peregrine2023} also serve as recent illustrations of applying NRE to GW analyses -- the latter incorporating multi-round inference as well.

In this example, we explore the application of multi-round inference to constrain a reduced set of gravitational wave (GW) parameters. Our forward-modeled data vector is constructed by merging the simulated frequency-domain strain signals from each of the Hanford (H1) and Livingstone (L1) detectors. We employ the \code{IMRPhenomPv2} algorithm \citep{khan_imr} with a frequency resolution of $\Delta f = 0.125 \textrm{ Hz}$, resulting in an input data vector of $\textrm{dim}(\bm{x}) = 2 \times 7872$. An illustration of this data vector is presented in Figure \ref{fig:gw_model} for clarity. We stress that, in contrast to likelihood-based sampling methods implemented in \code{Bilby} \citep{bibly_ashton, bilby_shaw}, which only necessitate the signal component of the forward model for the likelihood computation, our ILI approach requires the inclusion of noise realizations in the training data vector.

We illustrate a simplified application of LtU-ILI for SNPE on GW signals, fixing parameters related to the time and geometry of the detection and focusing on a set of four parameters: chirp mass ($\mathcal{M}$), mass ratio ($q$), effective aligned spin ($\chi_{\textrm{eff}}$) and luminosity distance ($d_L$). These parameters, chosen for their reduced correlation and significance in GW parameter estimation \citep[\eg][]{veitch_2015, Vitale:2016avz}, are derived from the masses of the coalescing objects -- with $\mathcal{M}\equiv (m_1m_2)^{\frac{3}{5}}/(m_1 + m_2)^{\frac{1}{5}}$ and $q \equiv m_2/m_1 \leq 1$ -- and spin characteristics \citep{racine_spins}.

We established baseline values for our simulated event $\bm{x_0}$ at \{$d_L = 390\ \textrm{Mpc}, q = 0.79, \mathcal{M} = 30.2 \ M_\odot, \chi_{\textrm{eff}} = -0.09$)\}, akin to the GW150914 event as detailed in the first column of Table 1 by \cite{LIGOScientific:2016vlm}. Following the procedure described in \citet{gwsim}, the defined priors are uniform ranging from [$100.0 \textrm{ Mpc}, 0.2, 12.0 \, M_\odot, -1.0$] to [$1000.0 \textrm{ Mpc}, 1.0, 45.0 \, M_\odot, 1.0$], with narrower margins for chirp mass and luminosity distance to facilitate a straightforward example. For density estimation, we utilize the \sbi{} backend with a MAF and a 1D convolutional embedding network, undergoing five rounds of training with 1000 simulations per round.

Figure \ref{fig:gw_logprob} illustrates the progression of the cumulative posterior probability, as defined in Equation \ref{eqn:Lnde}, as a function of training time. The noticeable increase in probability \rev{density} for both training and validation datasets is attributed to the model's focus on learning the posterior density near the target observation $\mathcal{P}(\bmt | \bmx = \bmx_0)$. However, an early increase in validation loss during subsequent rounds suggests the potential for overfitting, indicating the model's limitations in gleaning further insights from the simulation inputs. Sampling the posterior at $\bm{x}=\bm{x_0}$ depicted in Figure \ref{fig:gw_posterior}, reveals that we are able to constrain each of the four parameters around the fiducial point. \rev{For this particular GW example, using \ili allows us to parameterize the simulator (waveform approximant, number of detectors considered, geometry of the detection, duration of the signal\footnote{\rev{Note}, the frequency resolution is only determined by the time duration of the signal, whereas the sampling rate only gives the Nyquist frequency}, ...), the architectures to train and the metrics to compute with one single set of parameter files, thus making the organization and replication of unified inference tests easier for this particularly difficult problem.}

\subsection{Inferring Dust Parameters} \label{subsec:Dust_example}

\rev{\ili is a practical tool for exploratory discovery, especially useful for investigating novel datasets within which likelihoods are not well-defined. In this application, we study how different observables can be used to constrain different degeneracies within galactic dust models, and how their observational combinations can be used to maximally extract information  within an ILI model.}
Dust makes up approximately 1\% of the interstellar medium, but reprocesses around 30\% of stellar emission \citep{silva1998}.
Understanding the properties of dust and its impact on observables is, therefore, a key goal of galactic and extragalactic astronomy, and a fundamental uncertainty in downstream cosmological inference analyses \citep{crocce2016galaxy}.
Dust leads to an overall reddening of the emission, parameterized by the optical depth and attenuation law; these are analogous to the normalization and slope, particularly where the latter is parameterized as a power law.
Below we show how these parameters can be inferred from observational properties using a simple dust model \citep{trayford2015} applied to the Simba cosmological hydrodynamic simulation \citep{dave_simba_2019} within an ILI framework.

We first briefly describe our forward model for the stellar emission and dust attenuation.
We generated spectra for all galaxies with stellar mass $> 10^{10} \; \mathrm{M_{\odot}}$ in the Simba simulation using \code{Synthesizer} (Lovell et al. \textit{in prep.})\footnote{\url{https://flaresimulations.github.io/synthesizer}}.
The integrated intrinsic stellar emission of each galaxy was obtained by linking each stellar particle to the BC03 stellar population synthesis models \citep{bruzual_charlot_2003} based on their age and metallicity.
We then modelled the dust attenuation as a wavelength-dependent power law, with slope $\alpha$,
\begin{align}
    T(\lambda, t) = \mathrm{exp} \left[ \tau(t) \left( \frac{\lambda}{\lambda_{\mathrm{V}}} \right)^{-\alpha} \right] \;\;,
\end{align}
where $\lambda_{\mathrm{V}}$ is the wavelength of the V band ($550 \; \mathrm{nm}$).
Here the optical depth $\tau$ is dependent on the age of the stellar population $t$; star particles younger than 10 Myr are assumed to still reside within their birth clouds (BC), and therefore experience an extra source of attenuation,
\begin{align}
    \tau(t) = \begin{cases} \tau_{\mathrm{BC}} + \tau_{\mathrm{ISM}} &; \quad t \leqslant 10 {\rm Myr} \\ \tau_{\mathrm{ISM}} &; \quad t > 10 {\rm Myr} \end{cases} \;\;.
\end{align}

Assuming the same underlying stellar emission model, we generated rest frame optical photometry in the $g$ and $r$ bands for all galaxies in Simba for 1000 simulations on a Latin hypercube, with flat priors on the parameters in the range $\alpha \in [0.5, 2.0]$, $\tau_{\mathrm{ISM}} \in [0.01, 0.5]$ and $\tau_{\mathrm{BC}} \in [0.3, 1.5]$.
We then take these simulations and measure the $r$-band luminosity function and the $g-r$ color distribution, and use these as our summary statistics.
\rev{We perform NPE using an ensemble of a MAF with 50 hidden features and five neural transformations and an MDN with 50 hidden features and five mixture components.}

\begin{figure}
    \centering
    \includegraphics[width=\linewidth]{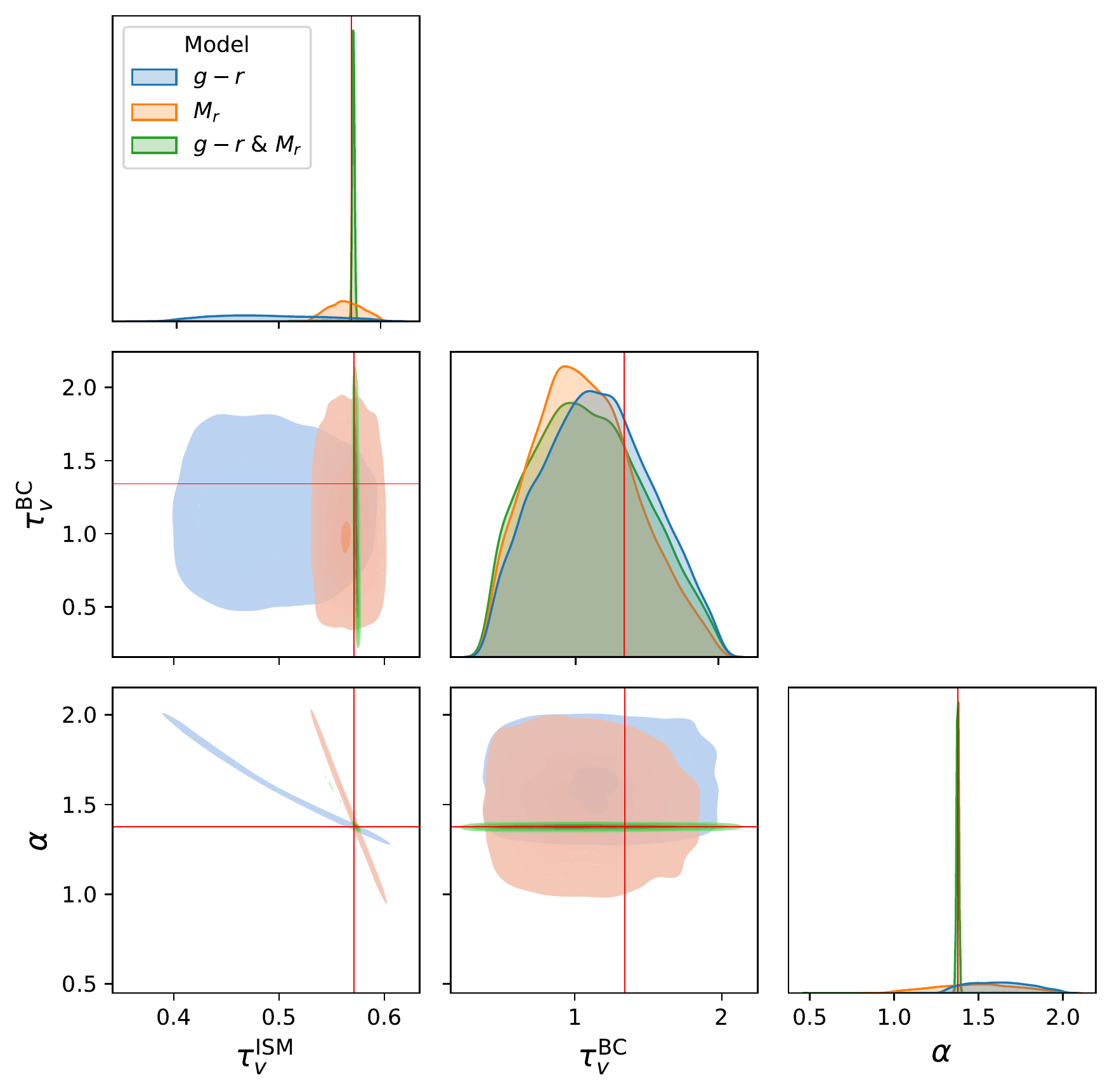}
    \caption{Dust parameter inference example on the Simba simulation. Example posteriors on the ISM and birth cloud optical depth as well as the slope of the attenuation law, from the distribution of $g-r$ colors (blue), the $r$-band magnitude luminosity function (orange), and the combination of both (green).
    }
    \label{fig:dust}
\end{figure}

Figure \ref{fig:dust} shows the posteriors on a test parameter -- data pair.
We show results from training on just the luminosity function, just the color, and the two combined.
It is clear that the overall normalisation of the luminosity function constrains the ISM optical depth, $\tau_{\mathrm{ISM}}$, whereas the colour constrains the slope of the attenuation law, $\alpha$.
When combined, these parameters are simultaneously tightly constrained.
The birth cloud attenuation has a more subtle effect on the luminosity function and colour distribution, but the combination leads to tighter constraints. \rev{\ili greatly simplifies the setup of the inference pipelines used here, for rapid testing and comparison of combinations of data sources.}

\subsection{The Mass and Energy Loading of Galactic Winds} \label{subsec:sapphire}

\begin{figure*}
    \centering
    \includegraphics[width=0.49\linewidth]{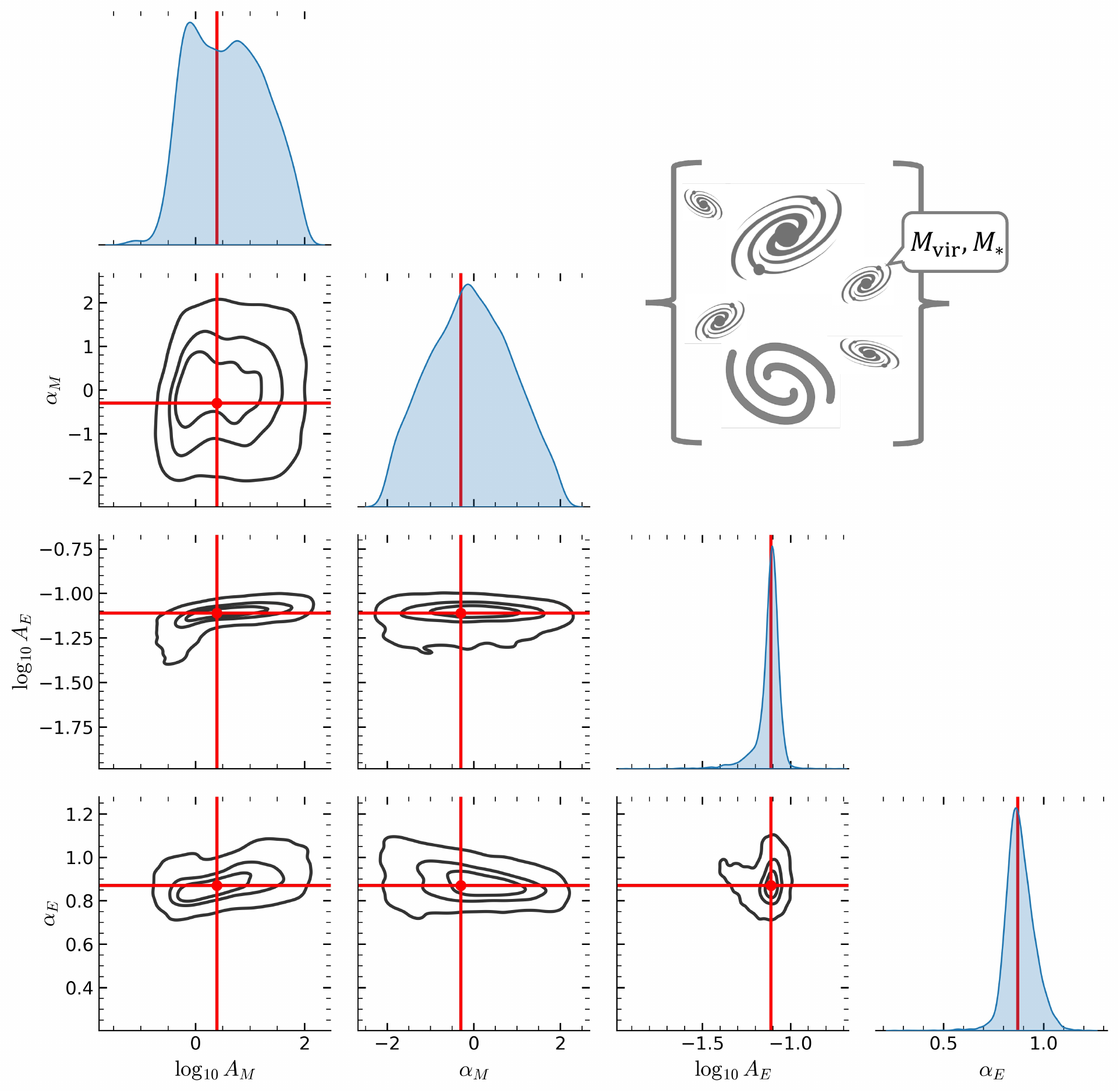}
    \includegraphics[width=0.49\linewidth]{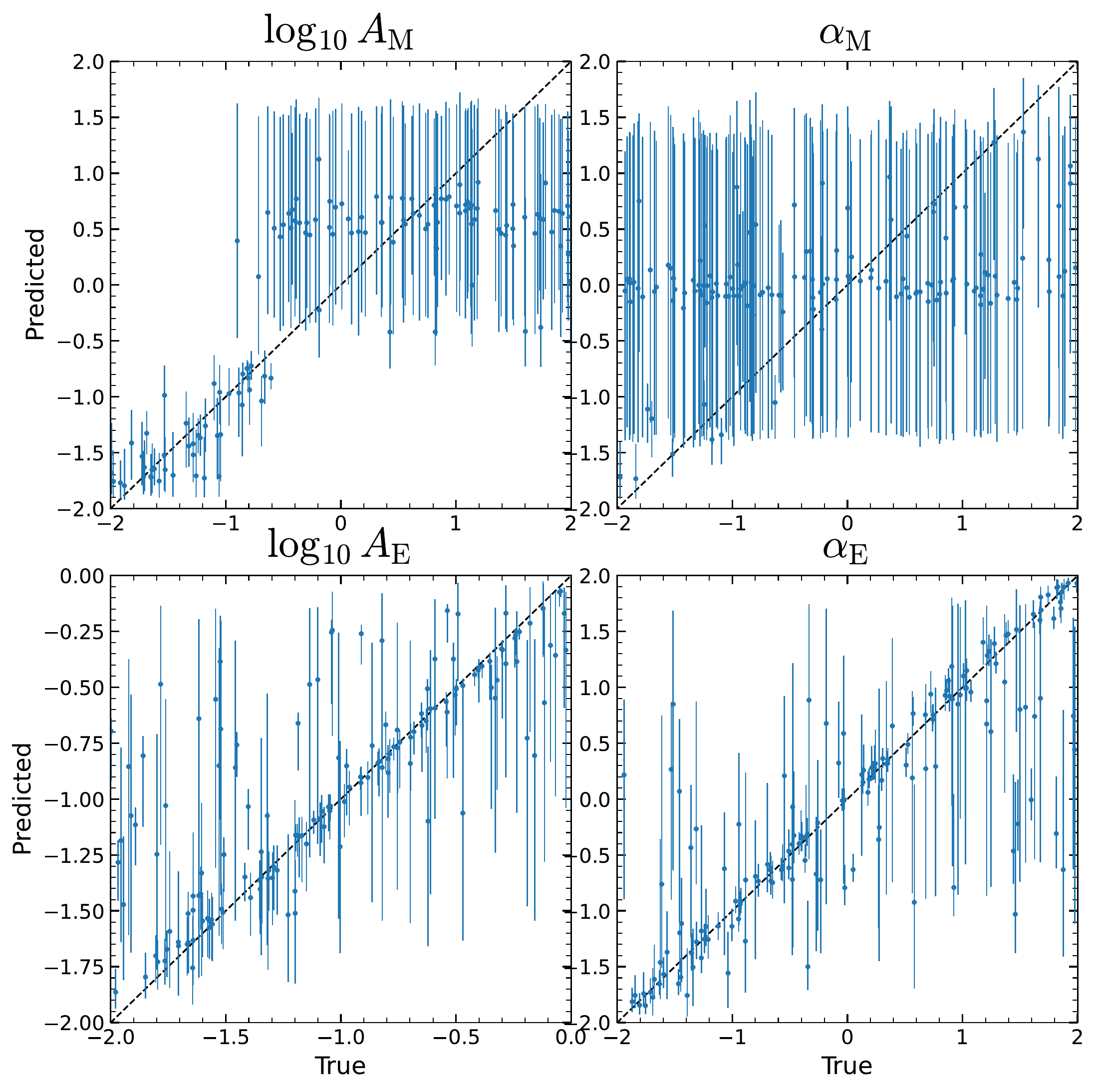}
    \caption{\rev{Inference of mass and energy loading parameters of \code{sapphire} semi-analytic model from sets of galaxy virial and stellar masses, \ie $\textbf{x} = \{ (M_{\rm vir }, M_*,\frac{M_{\rm vir}}{M_*}) \}_{i=1}^{n_x}$. The left plots show contours of an example inferred posterior, while the right plots show the true values vs. predictions across the full test set. Here, we utilize the \texttt{fishnets} architecture to aggregate sets of galaxy properties before performing NPE. Posterior predictive plots show our NPE network is sensitive to both energy loading hyper-parameters but cannot learn the mass loading hyper-parameters with the $M_*-M_{\rm vir}$ relation alone.}}\label{fig:sapphire2}
\end{figure*}

\rev{Lastly, we demonstrate how using custom embedding networks within LtU-ILI can aid posterior recovery for low signal-to-noise problems such as inferring parameters of semi-analytic models of galaxy formation. One such example is studying the poorly constrained role of feedback in regulating star formation and shaping the stellar-to-halo-mass relation.}
Recently, \citet{carr2023} and \cite{pandya23_unified} showed that supernova-driven galactic winds can heat and stir turbulence in the circumgalactic medium of dwarf galaxies and Milky Way-mass halos. This can suppress the cooling and accretion of gas onto galaxies thereby helping to regulate star formation. The strength of this ``preventative feedback'' is controlled by two wind parameters: the mass loading factor ($\eta_{\rm M}$) and the energy loading factor ($\eta_{\rm E}$). Together, these describe the specific energy of winds and summarize the fraction of mass and energy produced by supernovae that leave the galaxy and become available for large-scale heating.

Here we want to explore the connection between $\eta_{\rm M}$, $\eta_{\rm E})$, and the stellar-to-halo-mass relation predicted by numerous realizations of the next-generation semi-analytic model \texttt{sapphire} (Pandya et al \textit{in prep.}). The model used here is more similar to \citet{carr2023} than \cite{pandya23_unified} in that it neglects turbulence and assumes purely thermal heating of the circumgalactic medium (CGM) for simplicity. We assume that $\eta_{\rm M}$ and $\eta_{\rm E}$ both follow power law scalings with halo virial velocity $V_{\rm vir}$ (a proxy for halo mass):

\rev{
\begin{equation}
\eta_{\rm M} = A_{\rm M} \left(\frac{V_{\rm vir}}{125 \rm{km\;s^{-1}}}\right)^{\alpha_{\rm M}}
\end{equation}

\begin{equation}
\eta_{\rm E} = A_{\rm E} \left(\frac{V_{\rm vir}}{125 \rm{km\;s^{-1}}}\right)^{\alpha_{\rm E}}
\end{equation}
}

Here, $A_{\rm M}$, $\alpha_{\rm M}$, $A_{\rm E}$ and $\alpha_{\rm E}$ are hyperparameters that govern these power law scalings, and we have ignored any redshift dependence. We generate 5,000 model realizations on a Latin hypercube by uniformly sampling combinations of $A_{\rm M}\in[0.01,100]$, $\alpha_{\rm M}\in[-2,2]$, $A_{\rm E}\in[0.01,1]$ and $\alpha_{\rm E}\in[-2,2]$. All other parameters of the model were fixed to reasonable values following \cite{pandya23_unified}. With these parameters, the system of ordinary differential equations describing the model (Equations 1, 2, and 4-8 of \citealt{pandya23_unified}) is evolved using mass accretion histories of 120 halos randomly selected from five subvolumes of the TNG100 simulation \citep{pillepich2018one}. We find that this is a sufficient number of halos to capture variations in the stellar-to-halo-mass relation with model parameters. We restrict to halos with $\log M_{\rm vir}/M_{\odot}=10-12.4$ at $z=0$, spanning the dwarf and Milky Way regime in which supernova feedback is thought to dominate.


\rev{As an embedding network, we use the \texttt{fishnets} architecture by \cite{makinen2023fishnets} to extract what information the stellar-to-halo-mass relation contains about $\theta = (A_{\rm M}, \alpha_{\rm M}, A_{\rm E}, \alpha_{\rm E})$. The output from each \texttt{sapphire} realization is a set of $n_x=120$ independently-evolved galaxies, \ie $\textbf{x} = \{ (M_{\rm vir }, M_*, \frac{M_{\rm vir}}{M_*}) \}_{i=1}^{n_x}$ where $M_{\rm vir}$ and $M_*$ are the $z=0$ halo mass and stellar mass, respectively.} \rev{The likelihood here is a product of each individual galaxies' likelihoods, $ \mathcal{L}(\{ x_i \} | \theta) = \prod^{n_x}_{i=0} \mathcal{L}(x_i | \theta)$, requiring an aggregation over heterogeneous data distributions.} Following the optimal \texttt{fishnets} formalism presented in \cite{makinen2023fishnets}, we embed data into neural score embeddings and Fisher weights, each parameterized by fully-connected networks of size $[128,128,128]$ with \texttt{LeakyRelu} activations before passing the weighted scores \rev{to two NPE networks, a Neural Spline Flow \citep{durkan2019neural} and a Gaussianization Flow \citep{meng2020gaussianization}. }

We show the resulting true parameter versus predicted posterior plots in Figure \ref{fig:sapphire2}. We find that galaxy energy loading factors are sensitive to halo and stellar mass properties as expected, while mass loading factors prove more difficult to constrain. This is consistent with \citet{carr2023} who showed that the stellar-to-halo-mass relation is largely insensitive to variations in $\eta_{\rm M}$ but very responsive to changes in $\eta_{\rm E}$. The physical reason for this is that increasing $\eta_{\rm M}$ leads to a higher CGM density and enhanced cooling of gas back into the galaxy without suppressing star formation, thus leading to a similar $M_*-M_{\rm vir}$ relation. Instead, increasing $\eta_{\rm E}$ can heat the CGM, prevent gas accretion, reduce star formation and alter the normalization and shape of the $M_*-M_{\rm vir}$ relation. \rev{Interestingly, for \code{sapphire} realizations run at $A_{\rm M}\lesssim 0.25$ there is some information available to constrain mass-loading parameters, which isn't accesible for high $A_{\rm M}$. \ili is able to capture this effect and carefully estimate posterior error bars, demonstrating it's ability to calibrate inference both in low and high signal-to-noise regimes.} In the future, it will be compelling to use \texttt{fishnets} to understand what additional galaxy properties output by \texttt{sapphire} contain the information needed to better learn both $\eta_{\rm M}$ and $\eta_{\rm E}$.

\section{Discussion} \label{sec:discussion}
ILI methods are guaranteed to recover the true posterior under the assumptions that: 
\begin{enumerate}
    \item There is a sufficient amount of training data to cover data and parameter space.
    \item The training data or simulator is reflective of reality.
    \item The chosen neural architecture is sufficiently flexible to capture the data-parameter relationship.
\end{enumerate}
Despite the best practices implemented in \ili{}, biases may occur when these assumptions are violated. Before applying these approaches to real observational data, it is crucial to understand these failure modes, as well as how to build inference pipelines to be robust to them. Below, we review important considerations provide qualitative guidance for using ILI in observational settings.

\rev{Firstly,} model misspecification occurs when the data-generating function for creating training data is poorly representative of physical processes in the real world \citep{cannon2022investigating}. This can cause an implicit model to interpret the wrong data-parameter connection, resulting in biased predictions. In astrophysics and cosmology, combating model misspecification often requires running more realistic, high-resolution simulators which can be more computationally demanding. In turn, this reduces the number of available simulations, damaging assumption (1). Alternative approaches include increasing the noise model of a given simulation \citep[\eg][]{modi2023sensitivity}, training NDEs with a conservative loss to inflate uncertainties and counteract misspecification \citep[\eg][]{kelly2023misspecification, huang2023learning},
or using latent functions present within intermediate layers of the Bayesian hierarchical model as diagnostics \citep[\eg][]{Leclercq2022}. 

\rev{A more direct approach to addressing model misspecification is through Bayesian model comparison \citep{silva1998, handley2015polychord}. Given observational data $\bmx_o$, this technique seeks to quantify whether a proposed data-generating model $\mathcal{M}(\bmt)$ is reflective of reality. This is often done through calculation of the model evidence, $p\left(\bmx_o|\mathcal{M}\right)$, which, given an explicit likelihood model $\calL(\bmx|\bmt, \mathcal{M})$, can be done numerically, though requiring massive simulation budgets. Recent ILI approaches using the harmonic mean estimator \citep{harmonic, spurio2023bayesian} or Evidence Networks \citep{jeffrey2024evidence} have been shown to greatly accelerate these calculations. They also allow us to learn distributions implicitly from simulations, eliminating the need for explicit likelihoods. These novel evidence estimators are a natural addition to \ili and are planned for a future release. When implemented, these techniques will allow users to form quantitative arguments to addressing model misspecification.

Secondly, even for well-specified models, ILI methods may overfit and suffer from poor out-of-sample performance when training data is limited. This behavior has led to recent discussion as to whether ILI methods are systematically overconfident \citep{hermans2022trust}. PIT tests 
(Section\ \ref{subsec:valid}) on independent validation sets are a useful diagnostic tool for detecting overconfidence. For a fixed simulation budget, this problem can be mitigated through the use of ensembling \citep{hermans2022trust}. Ensembling averages the predictions of multiple, independently trained models together, naturally inflating predictive uncertainty to account for epistemic bias \citep{lakshminarayanan2017simple}. \rev{Alternatively, approximate Bayesian neural networks \citep{blundell2015weight, gal2016dropout} seek to constrain posteriors over model parameters during training, which they then subsequently propagate to model outputs to increase posterior variance. Lastly, once a model has been trained, its predictions can be reweighted through frequentist recalibration \citep{masserano2022simulation} which guarantees accurate posterior coverage over held-out test data. While approximate Bayesian networks and frequentist recalibration are not yet implemented in \ili, their integration is considered for future versions of the codebase.}

Overfitting can also be calibrated through careful hyperparameter selection. Choices of input data, neural architectures, and training procedures all play a quantifiable role in the final accuracy and out-of-sample bias measured in an ILI model. Oftentimes, selection of hyperparameters is a trade-off between choosing a flexible, expressive model which produces high-accuracy but is prone to overfitting, against a shallow, weaker model with more generalizable predictions. This is commonly referred to as the `bias-variance tradeoff' \citep{belkin2019reconciling}, and can be interpreted as placing assumptions (1) and (3) in direct tension. To address this, practitioners often perform automated searches over hyperparameters to find configurations which minimize variance within a given bias tolerance. Often, these searches are accelerated through the use of Bayesian optimization tools such as \code{optuna} \citep{akiba2019optuna}. These can be used to motivate ILI design choices directly from data, a necessary step for scientific studies. In the \ili repository, we provide examples of hyperparameter optimization through grid searching, and we plan to link \ili to optimal search tools such as \code{optuna} in future iterations.
}

Lastly, implicit inference can be very challenging in the high-dimensional parameter regime (\ie when $\operatorname{dim}(\bmt) \gg 100$). As the dimensionality of $\bmt$ increases, the number of parameters necessary to specify a highly-correlated posterior increases exponentially. This problem is of particular relevance in astrophysics and cosmology, in which we often would like to infer voxelized 2D or 3D fields \citep[\eg][]{jasche2013bayesian, de2024three}.
Recent empirical evidence suggests that Moment Networks \citep{jeffrey2020solving} and diffusion-based generative models \citep{song2020score, legin2024posterior} are \rev{remarkably} effective neural architectures for these tasks. Though \ili{} does not currently include these models, we intend to implement them in future work.


\section{Conclusion} \label{sec:conclusion}

In conclusion, we have introduced \ili{}, a comprehensive framework designed for implicit inference in scientific applications.
\ili{} provides extensive routines to facilitate the training of neural networks for regression posterior inference, aimed in particular at cases in which the likelihood is intractable. The code encapsulates best practices and state-of-the-art models, ensuring adaptability to a broad spectrum of scientific domains. By consolidating the training, testing, and robustness assessment of ILI models into a unified, modular package, \ili{} streamlines the process of addressing novel inference problems while providing for production-level inference and hyperparameter searches, making it accessible to researchers both experienced and novice. 

\rev{This manuscript is designed to serve as a reference document for future applications of \ili{}. In summary, we record detailed theoretical descriptions of every flavor of ILI (Section \ref*{sec:theory}), describe the important functionalities of the codebase (Section \ref*{sec:structure}), and provide guidance for experimental design amidst a broad landscape of estimators (Section \ref{subsec:choose_est}) and backends (Section \ref{subsec:choose_back}).
We also have provided benchmarks and science examples on a variety of astronomy and cosmology inference problems. Our results demonstrate that \ili{} yields competitive constraints on both synthetic and real science examples, indicating that these methods are ready out-of-the-box for application to new problems.
}

Looking ahead, \rev{we aim to apply this framework to modern analyses of the problems described in Section~\ref{sec:examples}, with an emphasis on observational applications to survey data from surveys such as SDSS, Planck, VRO, Euclid, DESI, and JWST.}
We aim to provide exhaustive benchmarks on these problems, contributing to the formalization of machine learning analyses in astronomy and beyond. 
We also intend to integrate \ili{} with state-of-the-art hyperparameter tuning \citep[\eg][]{akiba2019optuna} to provide an automatic, yet principled way for doing model selection, as in \citet{choustikov2024inferring}. This includes developing and implementing consolidated metrics for calibrating posterior coverage \citep[\eg][]{zhao2021diagnostics, lemos2024pqmass}, to guarantee accurate posterior coverage through hyperparameter search.
An alternative approach is to implement recalibration methods \citep[\eg][]{masserano2022simulation} which correctly calibrate posteriors given an independent validation set. 

\rev{We note that, while the current release of \ili{} is extensive, we hope to further improve the toolkit by incorporating additional Bayesian machine learning methods. For example, we do not currently include newer ILI procedures such as Truncated Marginal Neural Ratio Estimation \citep[TMNRE;][]{Miller:2021hys, Miller:2022shs} or score-based diffusion networks \citep[\eg][]{sharrock2022sequential, dax2023flow, linhart2024diffusion}, each of which has been shown to greatly improve upon the benchmarks that were investigated in Section \ref{sec:synth}. Also, during the development of this work, \code{nbi} \citep{zhang2023nbi} was released as a neural posterior estimation toolkit focused on spectroscopy and light-curve analysis. Though we do not directly integrate \code{nbi}, we remark that the functionality of \code{nbi} is already encompassed and extended by \ili.
Lastly, a major future addition to \ili will involve integration with implicit evidence estimators such as Evidence Networks \citep{jeffrey2024evidence} and \code{harmonic} \citep{harmonic, spurio2023bayesian}. With these features, users will be able to do Bayesian posterior inference and model comparison, all within one interface.
We reserve implementation and analysis of each of these improvements to future work.}


\begin{acknowledgments}
We thank Rosa Malandrino, Niall Jeffrey, and Justin Alsing for useful comments and suggestions on developing the codebase and writing the manuscript.
This project was developed as part of the Simons Collaboration on ``Learning the Universe.''
The Flatiron Institute is supported by the Simons Foundation.
\end{acknowledgments}

\bibliography{references}

\begin{thebibliography}{172}%
\makeatletter
\providecommand \@ifxundefined [1]{%
 \@ifx{#1\undefined}
}%
\providecommand \@ifnum [1]{%
 \ifnum #1\expandafter \@firstoftwo
 \else \expandafter \@secondoftwo
 \fi
}%
\providecommand \@ifx [1]{%
 \ifx #1\expandafter \@firstoftwo
 \else \expandafter \@secondoftwo
 \fi
}%
\providecommand \natexlab [1]{#1}%
\providecommand \enquote  [1]{``#1''}%
\providecommand \bibnamefont  [1]{#1}%
\providecommand \bibfnamefont [1]{#1}%
\providecommand \citenamefont [1]{#1}%
\providecommand \href@noop [0]{\@secondoftwo}%
\providecommand \href [0]{\begingroup \@sanitize@url \@href}%
\providecommand \@href[1]{\@@startlink{#1}\@@href}%
\providecommand \@@href[1]{\endgroup#1\@@endlink}%
\providecommand \@sanitize@url [0]{\catcode `\\12\catcode `\$12\catcode `\&12\catcode `\#12\catcode `\^12\catcode `\_12\catcode `\%12\relax}%
\providecommand \@@startlink[1]{}%
\providecommand \@@endlink[0]{}%
\providecommand \url  [0]{\begingroup\@sanitize@url \@url }%
\providecommand \@url [1]{\endgroup\@href {#1}{\urlprefix }}%
\providecommand \urlprefix  [0]{URL }%
\providecommand \Eprint [0]{\href }%
\providecommand \doibase [0]{http://dx.doi.org/}%
\providecommand \selectlanguage [0]{\@gobble}%
\providecommand \bibinfo  [0]{\@secondoftwo}%
\providecommand \bibfield  [0]{\@secondoftwo}%
\providecommand \translation [1]{[#1]}%
\providecommand \BibitemOpen [0]{}%
\providecommand \bibitemStop [0]{}%
\providecommand \bibitemNoStop [0]{.\EOS\space}%
\providecommand \EOS [0]{\spacefactor3000\relax}%
\providecommand \BibitemShut  [1]{\csname bibitem#1\endcsname}%
\let\auto@bib@innerbib\@empty
\bibitem [{\citenamefont {Feigelson}\ and\ \citenamefont {Babu}(2012)}]{Feigelson2012}%
  \BibitemOpen
  \bibfield  {author} {\bibinfo {author} {\bibfnamefont {E.~D.}\ \bibnamefont {Feigelson}}\ and\ \bibinfo {author} {\bibfnamefont {G.~J.}\ \bibnamefont {Babu}},\ }\href@noop {} {\emph {\bibinfo {title} {Modern statistical methods for astronomy: with R applications}}}\ (\bibinfo  {publisher} {Cambridge University Press},\ \bibinfo {year} {2012})\BibitemShut {NoStop}%
\bibitem [{\citenamefont {Dodelson}\ and\ \citenamefont {Schmidt}(2020)}]{dodelson2020modern}%
  \BibitemOpen
  \bibfield  {author} {\bibinfo {author} {\bibfnamefont {S.}~\bibnamefont {Dodelson}}\ and\ \bibinfo {author} {\bibfnamefont {F.}~\bibnamefont {Schmidt}},\ }\href@noop {} {\emph {\bibinfo {title} {Modern cosmology}}}\ (\bibinfo  {publisher} {Academic press},\ \bibinfo {year} {2020})\BibitemShut {NoStop}%
\bibitem [{\citenamefont {Eadie}\ \emph {et~al.}(2023)\citenamefont {Eadie}, \citenamefont {Speagle}, \citenamefont {Cisewski-Kehe}, \citenamefont {Foreman-Mackey}, \citenamefont {Huppenkothen}, \citenamefont {Jones}, \citenamefont {Springford},\ and\ \citenamefont {Tak}}]{eadie2023practical}%
  \BibitemOpen
  \bibfield  {author} {\bibinfo {author} {\bibfnamefont {G.~M.}\ \bibnamefont {Eadie}}, \bibinfo {author} {\bibfnamefont {J.~S.}\ \bibnamefont {Speagle}}, \bibinfo {author} {\bibfnamefont {J.}~\bibnamefont {Cisewski-Kehe}}, \bibinfo {author} {\bibfnamefont {D.}~\bibnamefont {Foreman-Mackey}}, \bibinfo {author} {\bibfnamefont {D.}~\bibnamefont {Huppenkothen}}, \bibinfo {author} {\bibfnamefont {D.~E.}\ \bibnamefont {Jones}}, \bibinfo {author} {\bibfnamefont {A.}~\bibnamefont {Springford}}, \ and\ \bibinfo {author} {\bibfnamefont {H.}~\bibnamefont {Tak}},\ }\href@noop {} {\bibfield  {journal} {\bibinfo  {journal} {arXiv preprint arXiv:2302.04703}\ } (\bibinfo {year} {2023})}\BibitemShut {NoStop}%
\bibitem [{\citenamefont {{National Academies of Sciences, Engineering, and Medicine}}(2023)}]{national2021pathways}%
  \BibitemOpen
  \bibfield  {author} {\bibinfo {author} {\bibnamefont {{National Academies of Sciences, Engineering, and Medicine}}},\ }\href {\doibase 10.17226/26141} {\emph {\bibinfo {title} {Pathways to Discovery in Astronomy and Astrophysics for the 2020s}}}\ (\bibinfo  {publisher} {The National Academies Press},\ \bibinfo {address} {Washington, DC},\ \bibinfo {year} {2023})\BibitemShut {NoStop}%
\bibitem [{\citenamefont {Carleo}\ \emph {et~al.}(2019)\citenamefont {Carleo}, \citenamefont {Cirac}, \citenamefont {Cranmer}, \citenamefont {Daudet}, \citenamefont {Schuld}, \citenamefont {Tishby}, \citenamefont {Vogt-Maranto},\ and\ \citenamefont {Zdeborov{\'a}}}]{carleo2019machine}%
  \BibitemOpen
  \bibfield  {author} {\bibinfo {author} {\bibfnamefont {G.}~\bibnamefont {Carleo}}, \bibinfo {author} {\bibfnamefont {I.}~\bibnamefont {Cirac}}, \bibinfo {author} {\bibfnamefont {K.}~\bibnamefont {Cranmer}}, \bibinfo {author} {\bibfnamefont {L.}~\bibnamefont {Daudet}}, \bibinfo {author} {\bibfnamefont {M.}~\bibnamefont {Schuld}}, \bibinfo {author} {\bibfnamefont {N.}~\bibnamefont {Tishby}}, \bibinfo {author} {\bibfnamefont {L.}~\bibnamefont {Vogt-Maranto}}, \ and\ \bibinfo {author} {\bibfnamefont {L.}~\bibnamefont {Zdeborov{\'a}}},\ }\href@noop {} {\bibfield  {journal} {\bibinfo  {journal} {Reviews of Modern Physics}\ }\textbf {\bibinfo {volume} {91}},\ \bibinfo {pages} {045002} (\bibinfo {year} {2019})}\BibitemShut {NoStop}%
\bibitem [{\citenamefont {Huppenkothen}\ \emph {et~al.}(2023)\citenamefont {Huppenkothen}, \citenamefont {Ntampaka}, \citenamefont {Ho}, \citenamefont {Fouesneau}, \citenamefont {Nord}, \citenamefont {Peek}, \citenamefont {Walmsley}, \citenamefont {Wu}, \citenamefont {Avestruz}, \citenamefont {Buck} \emph {et~al.}}]{huppenkothen2023constructing}%
  \BibitemOpen
  \bibfield  {author} {\bibinfo {author} {\bibfnamefont {D.}~\bibnamefont {Huppenkothen}}, \bibinfo {author} {\bibfnamefont {M.}~\bibnamefont {Ntampaka}}, \bibinfo {author} {\bibfnamefont {M.}~\bibnamefont {Ho}}, \bibinfo {author} {\bibfnamefont {M.}~\bibnamefont {Fouesneau}}, \bibinfo {author} {\bibfnamefont {B.}~\bibnamefont {Nord}}, \bibinfo {author} {\bibfnamefont {J.}~\bibnamefont {Peek}}, \bibinfo {author} {\bibfnamefont {M.}~\bibnamefont {Walmsley}}, \bibinfo {author} {\bibfnamefont {J.~F.}\ \bibnamefont {Wu}}, \bibinfo {author} {\bibfnamefont {C.}~\bibnamefont {Avestruz}}, \bibinfo {author} {\bibfnamefont {T.}~\bibnamefont {Buck}},  \emph {et~al.},\ }\href@noop {} {\bibfield  {journal} {\bibinfo  {journal} {arXiv preprint arXiv:2310.12528}\ } (\bibinfo {year} {2023})}\BibitemShut {NoStop}%
\bibitem [{\citenamefont {Cranmer}\ \emph {et~al.}(2020)\citenamefont {Cranmer}, \citenamefont {Brehmer},\ and\ \citenamefont {Louppe}}]{cranmer2020frontier}%
  \BibitemOpen
  \bibfield  {author} {\bibinfo {author} {\bibfnamefont {K.}~\bibnamefont {Cranmer}}, \bibinfo {author} {\bibfnamefont {J.}~\bibnamefont {Brehmer}}, \ and\ \bibinfo {author} {\bibfnamefont {G.}~\bibnamefont {Louppe}},\ }\href@noop {} {\bibfield  {journal} {\bibinfo  {journal} {Proceedings of the National Academy of Sciences}\ }\textbf {\bibinfo {volume} {117}},\ \bibinfo {pages} {30055} (\bibinfo {year} {2020})}\BibitemShut {NoStop}%
\bibitem [{\citenamefont {Marin}\ \emph {et~al.}(2012)\citenamefont {Marin}, \citenamefont {Pudlo}, \citenamefont {Robert},\ and\ \citenamefont {Ryder}}]{marin2012approximate}%
  \BibitemOpen
  \bibfield  {author} {\bibinfo {author} {\bibfnamefont {J.-M.}\ \bibnamefont {Marin}}, \bibinfo {author} {\bibfnamefont {P.}~\bibnamefont {Pudlo}}, \bibinfo {author} {\bibfnamefont {C.~P.}\ \bibnamefont {Robert}}, \ and\ \bibinfo {author} {\bibfnamefont {R.~J.}\ \bibnamefont {Ryder}},\ }\href@noop {} {\bibfield  {journal} {\bibinfo  {journal} {Statistics and computing}\ }\textbf {\bibinfo {volume} {22}},\ \bibinfo {pages} {1167} (\bibinfo {year} {2012})}\BibitemShut {NoStop}%
\bibitem [{\citenamefont {Christensen}\ \emph {et~al.}(2001)\citenamefont {Christensen}, \citenamefont {Meyer}, \citenamefont {Knox},\ and\ \citenamefont {Luey}}]{christensen2001bayesian}%
  \BibitemOpen
  \bibfield  {author} {\bibinfo {author} {\bibfnamefont {N.}~\bibnamefont {Christensen}}, \bibinfo {author} {\bibfnamefont {R.}~\bibnamefont {Meyer}}, \bibinfo {author} {\bibfnamefont {L.}~\bibnamefont {Knox}}, \ and\ \bibinfo {author} {\bibfnamefont {B.}~\bibnamefont {Luey}},\ }\href@noop {} {\bibfield  {journal} {\bibinfo  {journal} {Classical and Quantum Gravity}\ }\textbf {\bibinfo {volume} {18}},\ \bibinfo {pages} {2677} (\bibinfo {year} {2001})}\BibitemShut {NoStop}%
\bibitem [{\citenamefont {Brooks}\ \emph {et~al.}(2011)\citenamefont {Brooks}, \citenamefont {Gelman}, \citenamefont {Jones},\ and\ \citenamefont {Meng}}]{brooks2011handbook}%
  \BibitemOpen
  \bibfield  {author} {\bibinfo {author} {\bibfnamefont {S.}~\bibnamefont {Brooks}}, \bibinfo {author} {\bibfnamefont {A.}~\bibnamefont {Gelman}}, \bibinfo {author} {\bibfnamefont {G.}~\bibnamefont {Jones}}, \ and\ \bibinfo {author} {\bibfnamefont {X.-L.}\ \bibnamefont {Meng}},\ }\href@noop {} {\emph {\bibinfo {title} {Handbook of markov chain monte carlo}}}\ (\bibinfo  {publisher} {CRC press},\ \bibinfo {year} {2011})\BibitemShut {NoStop}%
\bibitem [{\citenamefont {Jeffrey}\ \emph {et~al.}(2021)\citenamefont {Jeffrey}, \citenamefont {Alsing},\ and\ \citenamefont {Lanusse}}]{jeffrey2021likelihood}%
  \BibitemOpen
  \bibfield  {author} {\bibinfo {author} {\bibfnamefont {N.}~\bibnamefont {Jeffrey}}, \bibinfo {author} {\bibfnamefont {J.}~\bibnamefont {Alsing}}, \ and\ \bibinfo {author} {\bibfnamefont {F.}~\bibnamefont {Lanusse}},\ }\href@noop {} {\bibfield  {journal} {\bibinfo  {journal} {Monthly Notices of the Royal Astronomical Society}\ }\textbf {\bibinfo {volume} {501}},\ \bibinfo {pages} {954} (\bibinfo {year} {2021})}\BibitemShut {NoStop}%
\bibitem [{\citenamefont {Makinen}\ \emph {et~al.}(2021)\citenamefont {Makinen}, \citenamefont {Charnock}, \citenamefont {Alsing},\ and\ \citenamefont {Wandelt}}]{Makinen_2021}%
  \BibitemOpen
  \bibfield  {author} {\bibinfo {author} {\bibfnamefont {T.~L.}\ \bibnamefont {Makinen}}, \bibinfo {author} {\bibfnamefont {T.}~\bibnamefont {Charnock}}, \bibinfo {author} {\bibfnamefont {J.}~\bibnamefont {Alsing}}, \ and\ \bibinfo {author} {\bibfnamefont {B.~D.}\ \bibnamefont {Wandelt}},\ }\href {\doibase 10.1088/1475-7516/2021/11/049} {\bibfield  {journal} {\bibinfo  {journal} {Journal of Cosmology and Astroparticle Physics}\ }\textbf {\bibinfo {volume} {2021}},\ \bibinfo {pages} {049} (\bibinfo {year} {2021})}\BibitemShut {NoStop}%
\bibitem [{\citenamefont {Makinen}\ \emph {et~al.}(2022{\natexlab{a}})\citenamefont {Makinen}, \citenamefont {Charnock}, \citenamefont {Lemos}, \citenamefont {Porqueres}, \citenamefont {Heavens},\ and\ \citenamefont {Wandelt}}]{Makinen_2022}%
  \BibitemOpen
  \bibfield  {author} {\bibinfo {author} {\bibfnamefont {T.~L.}\ \bibnamefont {Makinen}}, \bibinfo {author} {\bibfnamefont {T.}~\bibnamefont {Charnock}}, \bibinfo {author} {\bibfnamefont {P.}~\bibnamefont {Lemos}}, \bibinfo {author} {\bibfnamefont {N.}~\bibnamefont {Porqueres}}, \bibinfo {author} {\bibfnamefont {A.~F.}\ \bibnamefont {Heavens}}, \ and\ \bibinfo {author} {\bibfnamefont {B.~D.}\ \bibnamefont {Wandelt}},\ }\href {\doibase 10.21105/astro.2207.05202} {\bibfield  {journal} {\bibinfo  {journal} {The Open Journal of Astrophysics}\ }\textbf {\bibinfo {volume} {5}} (\bibinfo {year} {2022}{\natexlab{a}}),\ 10.21105/astro.2207.05202}\BibitemShut {NoStop}%
\bibitem [{\citenamefont {de~Santi}\ \emph {et~al.}(2023{\natexlab{a}})\citenamefont {de~Santi}, \citenamefont {Shao}, \citenamefont {Villaescusa-Navarro}, \citenamefont {Abramo}, \citenamefont {Teyssier}, \citenamefont {Villanueva-Domingo}, \citenamefont {Ni}, \citenamefont {Angl{\'e}s-Alc{\'a}zar}, \citenamefont {Genel}, \citenamefont {Hernandez-Martinez} \emph {et~al.}}]{de2023robust}%
  \BibitemOpen
  \bibfield  {author} {\bibinfo {author} {\bibfnamefont {N.~S.}\ \bibnamefont {de~Santi}}, \bibinfo {author} {\bibfnamefont {H.}~\bibnamefont {Shao}}, \bibinfo {author} {\bibfnamefont {F.}~\bibnamefont {Villaescusa-Navarro}}, \bibinfo {author} {\bibfnamefont {L.~R.}\ \bibnamefont {Abramo}}, \bibinfo {author} {\bibfnamefont {R.}~\bibnamefont {Teyssier}}, \bibinfo {author} {\bibfnamefont {P.}~\bibnamefont {Villanueva-Domingo}}, \bibinfo {author} {\bibfnamefont {Y.}~\bibnamefont {Ni}}, \bibinfo {author} {\bibfnamefont {D.}~\bibnamefont {Angl{\'e}s-Alc{\'a}zar}}, \bibinfo {author} {\bibfnamefont {S.}~\bibnamefont {Genel}}, \bibinfo {author} {\bibfnamefont {E.}~\bibnamefont {Hernandez-Martinez}},  \emph {et~al.},\ }\href@noop {} {\bibfield  {journal} {\bibinfo  {journal} {arXiv preprint arXiv:2302.14101}\ } (\bibinfo {year} {2023}{\natexlab{a}})}\BibitemShut {NoStop}%
\bibitem [{\citenamefont {Hahn}\ \emph {et~al.}(2023{\natexlab{a}})\citenamefont {Hahn}, \citenamefont {Lemos}, \citenamefont {Parker}, \citenamefont {Blancard}, \citenamefont {Eickenberg}, \citenamefont {Ho}, \citenamefont {Hou}, \citenamefont {Massara}, \citenamefont {Modi}, \citenamefont {Dizgah} \emph {et~al.}}]{hahn2023rm}%
  \BibitemOpen
  \bibfield  {author} {\bibinfo {author} {\bibfnamefont {C.}~\bibnamefont {Hahn}}, \bibinfo {author} {\bibfnamefont {P.}~\bibnamefont {Lemos}}, \bibinfo {author} {\bibfnamefont {L.}~\bibnamefont {Parker}}, \bibinfo {author} {\bibfnamefont {B.~R.-S.}\ \bibnamefont {Blancard}}, \bibinfo {author} {\bibfnamefont {M.}~\bibnamefont {Eickenberg}}, \bibinfo {author} {\bibfnamefont {S.}~\bibnamefont {Ho}}, \bibinfo {author} {\bibfnamefont {J.}~\bibnamefont {Hou}}, \bibinfo {author} {\bibfnamefont {E.}~\bibnamefont {Massara}}, \bibinfo {author} {\bibfnamefont {C.}~\bibnamefont {Modi}}, \bibinfo {author} {\bibfnamefont {A.~M.}\ \bibnamefont {Dizgah}},  \emph {et~al.},\ }\href@noop {} {\bibfield  {journal} {\bibinfo  {journal} {arXiv preprint arXiv:2310.15246}\ } (\bibinfo {year} {2023}{\natexlab{a}})}\BibitemShut {NoStop}%
\bibitem [{\citenamefont {{Dax}}\ \emph {et~al.}(2021{\natexlab{a}})\citenamefont {{Dax}}, \citenamefont {{Green}}, \citenamefont {{Gair}}, \citenamefont {{Macke}}, \citenamefont {{Buonanno}},\ and\ \citenamefont {{Sch{\"o}lkopf}}}]{daxGW2021}%
  \BibitemOpen
  \bibfield  {author} {\bibinfo {author} {\bibfnamefont {M.}~\bibnamefont {{Dax}}}, \bibinfo {author} {\bibfnamefont {S.~R.}\ \bibnamefont {{Green}}}, \bibinfo {author} {\bibfnamefont {J.}~\bibnamefont {{Gair}}}, \bibinfo {author} {\bibfnamefont {J.~H.}\ \bibnamefont {{Macke}}}, \bibinfo {author} {\bibfnamefont {A.}~\bibnamefont {{Buonanno}}}, \ and\ \bibinfo {author} {\bibfnamefont {B.}~\bibnamefont {{Sch{\"o}lkopf}}},\ }\href {\doibase 10.1103/PhysRevLett.127.241103} {\bibfield  {journal} {\bibinfo  {journal} {\prl}\ }\textbf {\bibinfo {volume} {127}},\ \bibinfo {eid} {241103} (\bibinfo {year} {2021}{\natexlab{a}})},\ \Eprint {http://arxiv.org/abs/2106.12594} {arXiv:2106.12594 [gr-qc]} \BibitemShut {NoStop}%
\bibitem [{\citenamefont {Cheung}\ \emph {et~al.}(2022)\citenamefont {Cheung}, \citenamefont {Wong}, \citenamefont {Hannuksela}, \citenamefont {Li},\ and\ \citenamefont {Ho}}]{cheung2022testing}%
  \BibitemOpen
  \bibfield  {author} {\bibinfo {author} {\bibfnamefont {D.~H.}\ \bibnamefont {Cheung}}, \bibinfo {author} {\bibfnamefont {K.~W.}\ \bibnamefont {Wong}}, \bibinfo {author} {\bibfnamefont {O.~A.}\ \bibnamefont {Hannuksela}}, \bibinfo {author} {\bibfnamefont {T.~G.}\ \bibnamefont {Li}}, \ and\ \bibinfo {author} {\bibfnamefont {S.}~\bibnamefont {Ho}},\ }\href@noop {} {\bibfield  {journal} {\bibinfo  {journal} {Physical Review D}\ }\textbf {\bibinfo {volume} {106}},\ \bibinfo {pages} {083014} (\bibinfo {year} {2022})}\BibitemShut {NoStop}%
\bibitem [{\citenamefont {Ho}\ \emph {et~al.}(2022)\citenamefont {Ho}, \citenamefont {Ntampaka}, \citenamefont {Rau}, \citenamefont {Chen}, \citenamefont {Lansberry}, \citenamefont {Ruehle},\ and\ \citenamefont {Trac}}]{ho2022dynamical}%
  \BibitemOpen
  \bibfield  {author} {\bibinfo {author} {\bibfnamefont {M.}~\bibnamefont {Ho}}, \bibinfo {author} {\bibfnamefont {M.}~\bibnamefont {Ntampaka}}, \bibinfo {author} {\bibfnamefont {M.~M.}\ \bibnamefont {Rau}}, \bibinfo {author} {\bibfnamefont {M.}~\bibnamefont {Chen}}, \bibinfo {author} {\bibfnamefont {A.}~\bibnamefont {Lansberry}}, \bibinfo {author} {\bibfnamefont {F.}~\bibnamefont {Ruehle}}, \ and\ \bibinfo {author} {\bibfnamefont {H.}~\bibnamefont {Trac}},\ }\href@noop {} {\bibfield  {journal} {\bibinfo  {journal} {Nature Astronomy}\ }\textbf {\bibinfo {volume} {6}},\ \bibinfo {pages} {936} (\bibinfo {year} {2022})}\BibitemShut {NoStop}%
\bibitem [{\citenamefont {de~Andres}\ \emph {et~al.}(2022)\citenamefont {de~Andres}, \citenamefont {Cui}, \citenamefont {Ruppin}, \citenamefont {De~Petris}, \citenamefont {Yepes}, \citenamefont {Gianfagna}, \citenamefont {Lahouli}, \citenamefont {Aversano}, \citenamefont {Dupuis}, \citenamefont {Jarraya} \emph {et~al.}}]{de2022deep}%
  \BibitemOpen
  \bibfield  {author} {\bibinfo {author} {\bibfnamefont {D.}~\bibnamefont {de~Andres}}, \bibinfo {author} {\bibfnamefont {W.}~\bibnamefont {Cui}}, \bibinfo {author} {\bibfnamefont {F.}~\bibnamefont {Ruppin}}, \bibinfo {author} {\bibfnamefont {M.}~\bibnamefont {De~Petris}}, \bibinfo {author} {\bibfnamefont {G.}~\bibnamefont {Yepes}}, \bibinfo {author} {\bibfnamefont {G.}~\bibnamefont {Gianfagna}}, \bibinfo {author} {\bibfnamefont {I.}~\bibnamefont {Lahouli}}, \bibinfo {author} {\bibfnamefont {G.}~\bibnamefont {Aversano}}, \bibinfo {author} {\bibfnamefont {R.}~\bibnamefont {Dupuis}}, \bibinfo {author} {\bibfnamefont {M.}~\bibnamefont {Jarraya}},  \emph {et~al.},\ }\href@noop {} {\bibfield  {journal} {\bibinfo  {journal} {Nature Astronomy}\ }\textbf {\bibinfo {volume} {6}},\ \bibinfo {pages} {1325} (\bibinfo {year} {2022})}\BibitemShut {NoStop}%
\bibitem [{\citenamefont {Walmsley}\ \emph {et~al.}(2020)\citenamefont {Walmsley}, \citenamefont {Smith}, \citenamefont {Lintott}, \citenamefont {Gal}, \citenamefont {Bamford}, \citenamefont {Dickinson}, \citenamefont {Fortson}, \citenamefont {Kruk}, \citenamefont {Masters}, \citenamefont {Scarlata} \emph {et~al.}}]{walmsley2020galaxy}%
  \BibitemOpen
  \bibfield  {author} {\bibinfo {author} {\bibfnamefont {M.}~\bibnamefont {Walmsley}}, \bibinfo {author} {\bibfnamefont {L.}~\bibnamefont {Smith}}, \bibinfo {author} {\bibfnamefont {C.}~\bibnamefont {Lintott}}, \bibinfo {author} {\bibfnamefont {Y.}~\bibnamefont {Gal}}, \bibinfo {author} {\bibfnamefont {S.}~\bibnamefont {Bamford}}, \bibinfo {author} {\bibfnamefont {H.}~\bibnamefont {Dickinson}}, \bibinfo {author} {\bibfnamefont {L.}~\bibnamefont {Fortson}}, \bibinfo {author} {\bibfnamefont {S.}~\bibnamefont {Kruk}}, \bibinfo {author} {\bibfnamefont {K.}~\bibnamefont {Masters}}, \bibinfo {author} {\bibfnamefont {C.}~\bibnamefont {Scarlata}},  \emph {et~al.},\ }\href@noop {} {\bibfield  {journal} {\bibinfo  {journal} {Monthly Notices of the Royal Astronomical Society}\ }\textbf {\bibinfo {volume} {491}},\ \bibinfo {pages} {1554} (\bibinfo {year} {2020})}\BibitemShut {NoStop}%
\bibitem [{\citenamefont {Ghosh}\ \emph {et~al.}(2022)\citenamefont {Ghosh}, \citenamefont {Urry}, \citenamefont {Rau}, \citenamefont {Perreault-Levasseur}, \citenamefont {Cranmer}, \citenamefont {Schawinski}, \citenamefont {Stark}, \citenamefont {Tian}, \citenamefont {Ofman}, \citenamefont {Ananna} \emph {et~al.}}]{ghosh2022gampen}%
  \BibitemOpen
  \bibfield  {author} {\bibinfo {author} {\bibfnamefont {A.}~\bibnamefont {Ghosh}}, \bibinfo {author} {\bibfnamefont {C.~M.}\ \bibnamefont {Urry}}, \bibinfo {author} {\bibfnamefont {A.}~\bibnamefont {Rau}}, \bibinfo {author} {\bibfnamefont {L.}~\bibnamefont {Perreault-Levasseur}}, \bibinfo {author} {\bibfnamefont {M.}~\bibnamefont {Cranmer}}, \bibinfo {author} {\bibfnamefont {K.}~\bibnamefont {Schawinski}}, \bibinfo {author} {\bibfnamefont {D.}~\bibnamefont {Stark}}, \bibinfo {author} {\bibfnamefont {C.}~\bibnamefont {Tian}}, \bibinfo {author} {\bibfnamefont {R.}~\bibnamefont {Ofman}}, \bibinfo {author} {\bibfnamefont {T.~T.}\ \bibnamefont {Ananna}},  \emph {et~al.},\ }\href@noop {} {\bibfield  {journal} {\bibinfo  {journal} {The Astrophysical Journal}\ }\textbf {\bibinfo {volume} {935}},\ \bibinfo {pages} {138} (\bibinfo {year} {2022})}\BibitemShut {NoStop}%
\bibitem [{\citenamefont {Hermans}\ \emph {et~al.}(2021)\citenamefont {Hermans}, \citenamefont {Banik}, \citenamefont {Weniger}, \citenamefont {Bertone},\ and\ \citenamefont {Louppe}}]{hermans2021towards}%
  \BibitemOpen
  \bibfield  {author} {\bibinfo {author} {\bibfnamefont {J.}~\bibnamefont {Hermans}}, \bibinfo {author} {\bibfnamefont {N.}~\bibnamefont {Banik}}, \bibinfo {author} {\bibfnamefont {C.}~\bibnamefont {Weniger}}, \bibinfo {author} {\bibfnamefont {G.}~\bibnamefont {Bertone}}, \ and\ \bibinfo {author} {\bibfnamefont {G.}~\bibnamefont {Louppe}},\ }\href@noop {} {\bibfield  {journal} {\bibinfo  {journal} {Monthly Notices of the Royal Astronomical Society}\ }\textbf {\bibinfo {volume} {507}},\ \bibinfo {pages} {1999} (\bibinfo {year} {2021})}\BibitemShut {NoStop}%
\bibitem [{\citenamefont {Alvey}\ \emph {et~al.}(2023)\citenamefont {Alvey}, \citenamefont {Gerdes},\ and\ \citenamefont {Weniger}}]{alvey2023albatross}%
  \BibitemOpen
  \bibfield  {author} {\bibinfo {author} {\bibfnamefont {J.}~\bibnamefont {Alvey}}, \bibinfo {author} {\bibfnamefont {M.}~\bibnamefont {Gerdes}}, \ and\ \bibinfo {author} {\bibfnamefont {C.}~\bibnamefont {Weniger}},\ }\href@noop {} {\bibfield  {journal} {\bibinfo  {journal} {arXiv preprint arXiv:2304.02032}\ } (\bibinfo {year} {2023})}\BibitemShut {NoStop}%
\bibitem [{\citenamefont {Rogers}\ \emph {et~al.}(2023)\citenamefont {Rogers}, \citenamefont {Jan{\'o}~Mu{\~n}oz}, \citenamefont {Owen},\ and\ \citenamefont {Makinen}}]{rogers2023exoplanet}%
  \BibitemOpen
  \bibfield  {author} {\bibinfo {author} {\bibfnamefont {J.~G.}\ \bibnamefont {Rogers}}, \bibinfo {author} {\bibfnamefont {C.}~\bibnamefont {Jan{\'o}~Mu{\~n}oz}}, \bibinfo {author} {\bibfnamefont {J.~E.}\ \bibnamefont {Owen}}, \ and\ \bibinfo {author} {\bibfnamefont {T.~L.}\ \bibnamefont {Makinen}},\ }\href@noop {} {\bibfield  {journal} {\bibinfo  {journal} {Monthly Notices of the Royal Astronomical Society}\ }\textbf {\bibinfo {volume} {519}},\ \bibinfo {pages} {6028} (\bibinfo {year} {2023})}\BibitemShut {NoStop}%
\bibitem [{\citenamefont {Aubin}\ \emph {et~al.}(2023)\citenamefont {Aubin}, \citenamefont {Cuesta-Lazaro}, \citenamefont {Tregidga}, \citenamefont {Via{\~n}a}, \citenamefont {Garraffo}, \citenamefont {Gordon}, \citenamefont {L{\'o}pez-Morales}, \citenamefont {Hargreaves}, \citenamefont {Makhnev}, \citenamefont {Drake} \emph {et~al.}}]{aubin2023simulation}%
  \BibitemOpen
  \bibfield  {author} {\bibinfo {author} {\bibfnamefont {M.}~\bibnamefont {Aubin}}, \bibinfo {author} {\bibfnamefont {C.}~\bibnamefont {Cuesta-Lazaro}}, \bibinfo {author} {\bibfnamefont {E.}~\bibnamefont {Tregidga}}, \bibinfo {author} {\bibfnamefont {J.}~\bibnamefont {Via{\~n}a}}, \bibinfo {author} {\bibfnamefont {C.}~\bibnamefont {Garraffo}}, \bibinfo {author} {\bibfnamefont {I.~E.}\ \bibnamefont {Gordon}}, \bibinfo {author} {\bibfnamefont {M.}~\bibnamefont {L{\'o}pez-Morales}}, \bibinfo {author} {\bibfnamefont {R.~J.}\ \bibnamefont {Hargreaves}}, \bibinfo {author} {\bibfnamefont {V.~Y.}\ \bibnamefont {Makhnev}}, \bibinfo {author} {\bibfnamefont {J.~J.}\ \bibnamefont {Drake}},  \emph {et~al.},\ }\href@noop {} {\bibfield  {journal} {\bibinfo  {journal} {arXiv preprint arXiv:2309.09337}\ } (\bibinfo {year} {2023})}\BibitemShut {NoStop}%
\bibitem [{\citenamefont {Gal}\ and\ \citenamefont {Ghahramani}(2016)}]{gal2016dropout}%
  \BibitemOpen
  \bibfield  {author} {\bibinfo {author} {\bibfnamefont {Y.}~\bibnamefont {Gal}}\ and\ \bibinfo {author} {\bibfnamefont {Z.}~\bibnamefont {Ghahramani}},\ }in\ \href@noop {} {\emph {\bibinfo {booktitle} {international conference on machine learning}}}\ (\bibinfo {organization} {PMLR},\ \bibinfo {year} {2016})\ pp.\ \bibinfo {pages} {1050--1059}\BibitemShut {NoStop}%
\bibitem [{\citenamefont {Lakshminarayanan}\ \emph {et~al.}(2017)\citenamefont {Lakshminarayanan}, \citenamefont {Pritzel},\ and\ \citenamefont {Blundell}}]{lakshminarayanan2017simple}%
  \BibitemOpen
  \bibfield  {author} {\bibinfo {author} {\bibfnamefont {B.}~\bibnamefont {Lakshminarayanan}}, \bibinfo {author} {\bibfnamefont {A.}~\bibnamefont {Pritzel}}, \ and\ \bibinfo {author} {\bibfnamefont {C.}~\bibnamefont {Blundell}},\ }\href@noop {} {\bibfield  {journal} {\bibinfo  {journal} {Advances in neural information processing systems}\ }\textbf {\bibinfo {volume} {30}} (\bibinfo {year} {2017})}\BibitemShut {NoStop}%
\bibitem [{\citenamefont {Maddox}\ \emph {et~al.}(2019)\citenamefont {Maddox}, \citenamefont {Izmailov}, \citenamefont {Garipov}, \citenamefont {Vetrov},\ and\ \citenamefont {Wilson}}]{maddox2019simple}%
  \BibitemOpen
  \bibfield  {author} {\bibinfo {author} {\bibfnamefont {W.~J.}\ \bibnamefont {Maddox}}, \bibinfo {author} {\bibfnamefont {P.}~\bibnamefont {Izmailov}}, \bibinfo {author} {\bibfnamefont {T.}~\bibnamefont {Garipov}}, \bibinfo {author} {\bibfnamefont {D.~P.}\ \bibnamefont {Vetrov}}, \ and\ \bibinfo {author} {\bibfnamefont {A.~G.}\ \bibnamefont {Wilson}},\ }\href@noop {} {\bibfield  {journal} {\bibinfo  {journal} {Advances in neural information processing systems}\ }\textbf {\bibinfo {volume} {32}} (\bibinfo {year} {2019})}\BibitemShut {NoStop}%
\bibitem [{\citenamefont {{Lemos}}\ \emph {et~al.}(2023)\citenamefont {{Lemos}}, \citenamefont {{Cranmer}}, \citenamefont {{Abidi}}, \citenamefont {{Hahn}}, \citenamefont {{Eickenberg}}, \citenamefont {{Massara}}, \citenamefont {{Yallup}},\ and\ \citenamefont {{Ho}}}]{Lemos2023}%
  \BibitemOpen
  \bibfield  {author} {\bibinfo {author} {\bibfnamefont {P.}~\bibnamefont {{Lemos}}}, \bibinfo {author} {\bibfnamefont {M.}~\bibnamefont {{Cranmer}}}, \bibinfo {author} {\bibfnamefont {M.}~\bibnamefont {{Abidi}}}, \bibinfo {author} {\bibfnamefont {C.}~\bibnamefont {{Hahn}}}, \bibinfo {author} {\bibfnamefont {M.}~\bibnamefont {{Eickenberg}}}, \bibinfo {author} {\bibfnamefont {E.}~\bibnamefont {{Massara}}}, \bibinfo {author} {\bibfnamefont {D.}~\bibnamefont {{Yallup}}}, \ and\ \bibinfo {author} {\bibfnamefont {S.}~\bibnamefont {{Ho}}},\ }\href {\doibase 10.1088/2632-2153/acbb53} {\bibfield  {journal} {\bibinfo  {journal} {Machine Learning: Science and Technology}\ }\textbf {\bibinfo {volume} {4}},\ \bibinfo {eid} {01LT01} (\bibinfo {year} {2023})},\ \Eprint {http://arxiv.org/abs/2207.08435} {arXiv:2207.08435 [astro-ph.CO]} \BibitemShut {NoStop}%
\bibitem [{\citenamefont {Jin}\ \emph {et~al.}(2019)\citenamefont {Jin}, \citenamefont {Song},\ and\ \citenamefont {Hu}}]{jin2019auto}%
  \BibitemOpen
  \bibfield  {author} {\bibinfo {author} {\bibfnamefont {H.}~\bibnamefont {Jin}}, \bibinfo {author} {\bibfnamefont {Q.}~\bibnamefont {Song}}, \ and\ \bibinfo {author} {\bibfnamefont {X.}~\bibnamefont {Hu}},\ }in\ \href@noop {} {\emph {\bibinfo {booktitle} {Proceedings of the 25th ACM SIGKDD international conference on knowledge discovery \& data mining}}}\ (\bibinfo {year} {2019})\ pp.\ \bibinfo {pages} {1946--1956}\BibitemShut {NoStop}%
\bibitem [{\citenamefont {White}\ \emph {et~al.}(2021)\citenamefont {White}, \citenamefont {Neiswanger},\ and\ \citenamefont {Savani}}]{white2021bananas}%
  \BibitemOpen
  \bibfield  {author} {\bibinfo {author} {\bibfnamefont {C.}~\bibnamefont {White}}, \bibinfo {author} {\bibfnamefont {W.}~\bibnamefont {Neiswanger}}, \ and\ \bibinfo {author} {\bibfnamefont {Y.}~\bibnamefont {Savani}},\ }in\ \href@noop {} {\emph {\bibinfo {booktitle} {Proceedings of the AAAI Conference on Artificial Intelligence}}}\ (\bibinfo {year} {2021})\BibitemShut {NoStop}%
\bibitem [{\citenamefont {{Wang}}\ \emph {et~al.}(2023)\citenamefont {{Wang}}, \citenamefont {{Leja}}, \citenamefont {{Villar}},\ and\ \citenamefont {{Speagle}}}]{sbipp2023}%
  \BibitemOpen
  \bibfield  {author} {\bibinfo {author} {\bibfnamefont {B.}~\bibnamefont {{Wang}}}, \bibinfo {author} {\bibfnamefont {J.}~\bibnamefont {{Leja}}}, \bibinfo {author} {\bibfnamefont {V.~A.}\ \bibnamefont {{Villar}}}, \ and\ \bibinfo {author} {\bibfnamefont {J.~S.}\ \bibnamefont {{Speagle}}},\ }\href {\doibase 10.3847/2041-8213/ace361} {\bibfield  {journal} {\bibinfo  {journal} {\apjl}\ }\textbf {\bibinfo {volume} {952}},\ \bibinfo {eid} {L10} (\bibinfo {year} {2023})},\ \Eprint {http://arxiv.org/abs/2304.05281} {arXiv:2304.05281 [astro-ph.IM]} \BibitemShut {NoStop}%
\bibitem [{\citenamefont {Modi}\ \emph {et~al.}(2023)\citenamefont {Modi}, \citenamefont {Pandey}, \citenamefont {Ho}, \citenamefont {Hahn}, \citenamefont {Blancard},\ and\ \citenamefont {Wandelt}}]{modi2023sensitivity}%
  \BibitemOpen
  \bibfield  {author} {\bibinfo {author} {\bibfnamefont {C.}~\bibnamefont {Modi}}, \bibinfo {author} {\bibfnamefont {S.}~\bibnamefont {Pandey}}, \bibinfo {author} {\bibfnamefont {M.}~\bibnamefont {Ho}}, \bibinfo {author} {\bibfnamefont {C.}~\bibnamefont {Hahn}}, \bibinfo {author} {\bibfnamefont {B.}~\bibnamefont {Blancard}}, \ and\ \bibinfo {author} {\bibfnamefont {B.}~\bibnamefont {Wandelt}},\ }\href@noop {} {\bibfield  {journal} {\bibinfo  {journal} {arXiv preprint arXiv:2309.15071}\ } (\bibinfo {year} {2023})}\BibitemShut {NoStop}%
\bibitem [{\citenamefont {Tejero-Cantero}\ \emph {et~al.}(2020)\citenamefont {Tejero-Cantero}, \citenamefont {Boelts}, \citenamefont {Deistler}, \citenamefont {Lueckmann}, \citenamefont {Durkan}, \citenamefont {Gonçalves}, \citenamefont {Greenberg},\ and\ \citenamefont {Macke}}]{tejero-cantero2020sbi}%
  \BibitemOpen
  \bibfield  {author} {\bibinfo {author} {\bibfnamefont {A.}~\bibnamefont {Tejero-Cantero}}, \bibinfo {author} {\bibfnamefont {J.}~\bibnamefont {Boelts}}, \bibinfo {author} {\bibfnamefont {M.}~\bibnamefont {Deistler}}, \bibinfo {author} {\bibfnamefont {J.-M.}\ \bibnamefont {Lueckmann}}, \bibinfo {author} {\bibfnamefont {C.}~\bibnamefont {Durkan}}, \bibinfo {author} {\bibfnamefont {P.~J.}\ \bibnamefont {Gonçalves}}, \bibinfo {author} {\bibfnamefont {D.~S.}\ \bibnamefont {Greenberg}}, \ and\ \bibinfo {author} {\bibfnamefont {J.~H.}\ \bibnamefont {Macke}},\ }\href {\doibase 10.21105/joss.02505} {\bibfield  {journal} {\bibinfo  {journal} {Journal of Open Source Software}\ }\textbf {\bibinfo {volume} {5}},\ \bibinfo {pages} {2505} (\bibinfo {year} {2020})}\BibitemShut {NoStop}%
\bibitem [{\citenamefont {Alsing}\ \emph {et~al.}(2019)\citenamefont {Alsing}, \citenamefont {Charnock}, \citenamefont {Feeney},\ and\ \citenamefont {Wandelt}}]{alsing2019fast}%
  \BibitemOpen
  \bibfield  {author} {\bibinfo {author} {\bibfnamefont {J.}~\bibnamefont {Alsing}}, \bibinfo {author} {\bibfnamefont {T.}~\bibnamefont {Charnock}}, \bibinfo {author} {\bibfnamefont {S.}~\bibnamefont {Feeney}}, \ and\ \bibinfo {author} {\bibfnamefont {B.}~\bibnamefont {Wandelt}},\ }\href@noop {} {\bibfield  {journal} {\bibinfo  {journal} {Monthly Notices of the Royal Astronomical Society}\ }\textbf {\bibinfo {volume} {488}},\ \bibinfo {pages} {4440} (\bibinfo {year} {2019})}\BibitemShut {NoStop}%
\bibitem [{\citenamefont {Rozet}\ \emph {et~al.}(2021)\citenamefont {Rozet}, \citenamefont {Delaunoy}, \citenamefont {Miller} \emph {et~al.}}]{rozet2021lampe}%
  \BibitemOpen
  \bibfield  {author} {\bibinfo {author} {\bibfnamefont {F.}~\bibnamefont {Rozet}}, \bibinfo {author} {\bibfnamefont {A.}~\bibnamefont {Delaunoy}}, \bibinfo {author} {\bibfnamefont {B.}~\bibnamefont {Miller}},  \emph {et~al.},\ }\href {\doibase 10.5281/zenodo.8405782} {\enquote {\bibinfo {title} {{LAMPE}: Likelihood-free amortized posterior estimation},}\ } (\bibinfo {year} {2021})\BibitemShut {NoStop}%
\bibitem [{\citenamefont {Lemos}\ \emph {et~al.}(2023)\citenamefont {Lemos}, \citenamefont {Coogan}, \citenamefont {Hezaveh},\ and\ \citenamefont {Levasseur}}]{tarp}%
  \BibitemOpen
  \bibfield  {author} {\bibinfo {author} {\bibfnamefont {P.}~\bibnamefont {Lemos}}, \bibinfo {author} {\bibfnamefont {A.}~\bibnamefont {Coogan}}, \bibinfo {author} {\bibfnamefont {Y.}~\bibnamefont {Hezaveh}}, \ and\ \bibinfo {author} {\bibfnamefont {L.~P.}\ \bibnamefont {Levasseur}},\ }in\ \href {https://proceedings.mlr.press/v202/lemos23a.html} {\emph {\bibinfo {booktitle} {International Conference on Machine Learning, {ICML} 2023, 23-29 July 2023, Honolulu, Hawaii, {USA}}}},\ \bibinfo {series} {Proceedings of Machine Learning Research}, Vol.\ \bibinfo {volume} {202},\ \bibinfo {editor} {edited by\ \bibinfo {editor} {\bibfnamefont {A.}~\bibnamefont {Krause}}, \bibinfo {editor} {\bibfnamefont {E.}~\bibnamefont {Brunskill}}, \bibinfo {editor} {\bibfnamefont {K.}~\bibnamefont {Cho}}, \bibinfo {editor} {\bibfnamefont {B.}~\bibnamefont {Engelhardt}}, \bibinfo {editor} {\bibfnamefont {S.}~\bibnamefont {Sabato}}, \ and\ \bibinfo {editor} {\bibfnamefont {J.}~\bibnamefont {Scarlett}}}\ (\bibinfo  {publisher}
  {{PMLR}},\ \bibinfo {year} {2023})\ pp.\ \bibinfo {pages} {19256--19273}\BibitemShut {NoStop}%
\bibitem [{\citenamefont {{Tegmark}}\ \emph {et~al.}(2004)\citenamefont {{Tegmark}}, \citenamefont {{Blanton}}, \citenamefont {{Strauss}}, \citenamefont {{Hoyle}}, \citenamefont {{Schlegel}}, \citenamefont {{Scoccimarro}}, \citenamefont {{Vogeley}}, \citenamefont {{Weinberg}}, \citenamefont {{Zehavi}}, \citenamefont {{Berlind}}, \citenamefont {{Budavari}}, \citenamefont {{Connolly}}, \citenamefont {{Eisenstein}}, \citenamefont {{Finkbeiner}}, \citenamefont {{Frieman}}, \citenamefont {{Gunn}}, \citenamefont {{Hamilton}}, \citenamefont {{Hui}}, \citenamefont {{Jain}}, \citenamefont {{Johnston}}, \citenamefont {{Kent}}, \citenamefont {{Lin}}, \citenamefont {{Nakajima}}, \citenamefont {{Nichol}}, \citenamefont {{Ostriker}}, \citenamefont {{Pope}}, \citenamefont {{Scranton}}, \citenamefont {{Seljak}}, \citenamefont {{Sheth}}, \citenamefont {{Stebbins}}, \citenamefont {{Szalay}}, \citenamefont {{Szapudi}}, \citenamefont {{Verde}}, \citenamefont {{Xu}}, \citenamefont {{Annis}}, \citenamefont {{Bahcall}},
  \citenamefont {{Brinkmann}}, \citenamefont {{Burles}}, \citenamefont {{Castander}}, \citenamefont {{Csabai}}, \citenamefont {{Loveday}}, \citenamefont {{Doi}}, \citenamefont {{Fukugita}}, \citenamefont {{Gott}}, \citenamefont {{Hennessy}}, \citenamefont {{Hogg}}, \citenamefont {{Ivezi{\'c}}}, \citenamefont {{Knapp}}, \citenamefont {{Lamb}}, \citenamefont {{Lee}}, \citenamefont {{Lupton}}, \citenamefont {{McKay}}, \citenamefont {{Kunszt}}, \citenamefont {{Munn}}, \citenamefont {{O'Connell}}, \citenamefont {{Peoples}}, \citenamefont {{Pier}}, \citenamefont {{Richmond}}, \citenamefont {{Rockosi}}, \citenamefont {{Schneider}}, \citenamefont {{Stoughton}}, \citenamefont {{Tucker}}, \citenamefont {{Vanden Berk}}, \citenamefont {{Yanny}}, \citenamefont {{York}},\ and\ \citenamefont {{SDSS Collaboration}}}]{sdssPk2004}%
  \BibitemOpen
  \bibfield  {author} {\bibinfo {author} {\bibfnamefont {M.}~\bibnamefont {{Tegmark}}}, \bibinfo {author} {\bibfnamefont {M.~R.}\ \bibnamefont {{Blanton}}}, \bibinfo {author} {\bibfnamefont {M.~A.}\ \bibnamefont {{Strauss}}}, \bibinfo {author} {\bibfnamefont {F.}~\bibnamefont {{Hoyle}}}, \bibinfo {author} {\bibfnamefont {D.}~\bibnamefont {{Schlegel}}}, \bibinfo {author} {\bibfnamefont {R.}~\bibnamefont {{Scoccimarro}}}, \bibinfo {author} {\bibfnamefont {M.~S.}\ \bibnamefont {{Vogeley}}}, \bibinfo {author} {\bibfnamefont {D.~H.}\ \bibnamefont {{Weinberg}}}, \bibinfo {author} {\bibfnamefont {I.}~\bibnamefont {{Zehavi}}}, \bibinfo {author} {\bibfnamefont {A.}~\bibnamefont {{Berlind}}}, \bibinfo {author} {\bibfnamefont {T.}~\bibnamefont {{Budavari}}}, \bibinfo {author} {\bibfnamefont {A.}~\bibnamefont {{Connolly}}}, \bibinfo {author} {\bibfnamefont {D.~J.}\ \bibnamefont {{Eisenstein}}}, \bibinfo {author} {\bibfnamefont {D.}~\bibnamefont {{Finkbeiner}}}, \bibinfo {author} {\bibfnamefont {J.~A.}\ \bibnamefont
  {{Frieman}}}, \bibinfo {author} {\bibfnamefont {J.~E.}\ \bibnamefont {{Gunn}}}, \bibinfo {author} {\bibfnamefont {A.~J.~S.}\ \bibnamefont {{Hamilton}}}, \bibinfo {author} {\bibfnamefont {L.}~\bibnamefont {{Hui}}}, \bibinfo {author} {\bibfnamefont {B.}~\bibnamefont {{Jain}}}, \bibinfo {author} {\bibfnamefont {D.}~\bibnamefont {{Johnston}}}, \bibinfo {author} {\bibfnamefont {S.}~\bibnamefont {{Kent}}}, \bibinfo {author} {\bibfnamefont {H.}~\bibnamefont {{Lin}}}, \bibinfo {author} {\bibfnamefont {R.}~\bibnamefont {{Nakajima}}}, \bibinfo {author} {\bibfnamefont {R.~C.}\ \bibnamefont {{Nichol}}}, \bibinfo {author} {\bibfnamefont {J.~P.}\ \bibnamefont {{Ostriker}}}, \bibinfo {author} {\bibfnamefont {A.}~\bibnamefont {{Pope}}}, \bibinfo {author} {\bibfnamefont {R.}~\bibnamefont {{Scranton}}}, \bibinfo {author} {\bibfnamefont {U.}~\bibnamefont {{Seljak}}}, \bibinfo {author} {\bibfnamefont {R.~K.}\ \bibnamefont {{Sheth}}}, \bibinfo {author} {\bibfnamefont {A.}~\bibnamefont {{Stebbins}}}, \bibinfo {author}
  {\bibfnamefont {A.~S.}\ \bibnamefont {{Szalay}}}, \bibinfo {author} {\bibfnamefont {I.}~\bibnamefont {{Szapudi}}}, \bibinfo {author} {\bibfnamefont {L.}~\bibnamefont {{Verde}}}, \bibinfo {author} {\bibfnamefont {Y.}~\bibnamefont {{Xu}}}, \bibinfo {author} {\bibfnamefont {J.}~\bibnamefont {{Annis}}}, \bibinfo {author} {\bibfnamefont {N.~A.}\ \bibnamefont {{Bahcall}}}, \bibinfo {author} {\bibfnamefont {J.}~\bibnamefont {{Brinkmann}}}, \bibinfo {author} {\bibfnamefont {S.}~\bibnamefont {{Burles}}}, \bibinfo {author} {\bibfnamefont {F.~J.}\ \bibnamefont {{Castander}}}, \bibinfo {author} {\bibfnamefont {I.}~\bibnamefont {{Csabai}}}, \bibinfo {author} {\bibfnamefont {J.}~\bibnamefont {{Loveday}}}, \bibinfo {author} {\bibfnamefont {M.}~\bibnamefont {{Doi}}}, \bibinfo {author} {\bibfnamefont {M.}~\bibnamefont {{Fukugita}}}, \bibinfo {author} {\bibfnamefont {I.}~\bibnamefont {{Gott}}, \bibfnamefont {J.~Richard}}, \bibinfo {author} {\bibfnamefont {G.}~\bibnamefont {{Hennessy}}}, \bibinfo {author} {\bibfnamefont
  {D.~W.}\ \bibnamefont {{Hogg}}}, \bibinfo {author} {\bibfnamefont {{\v{Z}}.}~\bibnamefont {{Ivezi{\'c}}}}, \bibinfo {author} {\bibfnamefont {G.~R.}\ \bibnamefont {{Knapp}}}, \bibinfo {author} {\bibfnamefont {D.~Q.}\ \bibnamefont {{Lamb}}}, \bibinfo {author} {\bibfnamefont {B.~C.}\ \bibnamefont {{Lee}}}, \bibinfo {author} {\bibfnamefont {R.~H.}\ \bibnamefont {{Lupton}}}, \bibinfo {author} {\bibfnamefont {T.~A.}\ \bibnamefont {{McKay}}}, \bibinfo {author} {\bibfnamefont {P.}~\bibnamefont {{Kunszt}}}, \bibinfo {author} {\bibfnamefont {J.~A.}\ \bibnamefont {{Munn}}}, \bibinfo {author} {\bibfnamefont {L.}~\bibnamefont {{O'Connell}}}, \bibinfo {author} {\bibfnamefont {J.}~\bibnamefont {{Peoples}}}, \bibinfo {author} {\bibfnamefont {J.~R.}\ \bibnamefont {{Pier}}}, \bibinfo {author} {\bibfnamefont {M.}~\bibnamefont {{Richmond}}}, \bibinfo {author} {\bibfnamefont {C.}~\bibnamefont {{Rockosi}}}, \bibinfo {author} {\bibfnamefont {D.~P.}\ \bibnamefont {{Schneider}}}, \bibinfo {author} {\bibfnamefont {C.}~\bibnamefont
  {{Stoughton}}}, \bibinfo {author} {\bibfnamefont {D.~L.}\ \bibnamefont {{Tucker}}}, \bibinfo {author} {\bibfnamefont {D.~E.}\ \bibnamefont {{Vanden Berk}}}, \bibinfo {author} {\bibfnamefont {B.}~\bibnamefont {{Yanny}}}, \bibinfo {author} {\bibfnamefont {D.~G.}\ \bibnamefont {{York}}}, \ and\ \bibinfo {author} {\bibnamefont {{SDSS Collaboration}}},\ }\href {\doibase 10.1086/382125} {\bibfield  {journal} {\bibinfo  {journal} {\apj}\ }\textbf {\bibinfo {volume} {606}},\ \bibinfo {pages} {702} (\bibinfo {year} {2004})},\ \Eprint {http://arxiv.org/abs/astro-ph/0310725} {arXiv:astro-ph/0310725 [astro-ph]} \BibitemShut {NoStop}%
\bibitem [{\citenamefont {Christensen}\ and\ \citenamefont {Meyer}(2022)}]{christensen2022parameter}%
  \BibitemOpen
  \bibfield  {author} {\bibinfo {author} {\bibfnamefont {N.}~\bibnamefont {Christensen}}\ and\ \bibinfo {author} {\bibfnamefont {R.}~\bibnamefont {Meyer}},\ }\href@noop {} {\bibfield  {journal} {\bibinfo  {journal} {Reviews of Modern Physics}\ }\textbf {\bibinfo {volume} {94}},\ \bibinfo {pages} {025001} (\bibinfo {year} {2022})}\BibitemShut {NoStop}%
\bibitem [{\citenamefont {Porqueres}\ \emph {et~al.}(2022)\citenamefont {Porqueres}, \citenamefont {Heavens}, \citenamefont {Mortlock},\ and\ \citenamefont {Lavaux}}]{porqueres2022lifting}%
  \BibitemOpen
  \bibfield  {author} {\bibinfo {author} {\bibfnamefont {N.}~\bibnamefont {Porqueres}}, \bibinfo {author} {\bibfnamefont {A.}~\bibnamefont {Heavens}}, \bibinfo {author} {\bibfnamefont {D.}~\bibnamefont {Mortlock}}, \ and\ \bibinfo {author} {\bibfnamefont {G.}~\bibnamefont {Lavaux}},\ }\href@noop {} {\bibfield  {journal} {\bibinfo  {journal} {Monthly Notices of the Royal Astronomical Society}\ }\textbf {\bibinfo {volume} {509}},\ \bibinfo {pages} {3194} (\bibinfo {year} {2022})}\BibitemShut {NoStop}%
\bibitem [{\citenamefont {Boese}\ and\ \citenamefont {Doebereiner}(2001)}]{boese2001maximum}%
  \BibitemOpen
  \bibfield  {author} {\bibinfo {author} {\bibfnamefont {F.}~\bibnamefont {Boese}}\ and\ \bibinfo {author} {\bibfnamefont {S.}~\bibnamefont {Doebereiner}},\ }\href@noop {} {\bibfield  {journal} {\bibinfo  {journal} {Astronomy \& Astrophysics}\ }\textbf {\bibinfo {volume} {370}},\ \bibinfo {pages} {649} (\bibinfo {year} {2001})}\BibitemShut {NoStop}%
\bibitem [{\citenamefont {Braun}\ \emph {et~al.}(2008)\citenamefont {Braun}, \citenamefont {Dumm}, \citenamefont {De~Palma}, \citenamefont {Finley}, \citenamefont {Karle},\ and\ \citenamefont {Montaruli}}]{braun2008methods}%
  \BibitemOpen
  \bibfield  {author} {\bibinfo {author} {\bibfnamefont {J.}~\bibnamefont {Braun}}, \bibinfo {author} {\bibfnamefont {J.}~\bibnamefont {Dumm}}, \bibinfo {author} {\bibfnamefont {F.}~\bibnamefont {De~Palma}}, \bibinfo {author} {\bibfnamefont {C.}~\bibnamefont {Finley}}, \bibinfo {author} {\bibfnamefont {A.}~\bibnamefont {Karle}}, \ and\ \bibinfo {author} {\bibfnamefont {T.}~\bibnamefont {Montaruli}},\ }\href@noop {} {\bibfield  {journal} {\bibinfo  {journal} {Astroparticle Physics}\ }\textbf {\bibinfo {volume} {29}},\ \bibinfo {pages} {299} (\bibinfo {year} {2008})}\BibitemShut {NoStop}%
\bibitem [{\citenamefont {Tsaprazi}\ \emph {et~al.}(2023)\citenamefont {Tsaprazi}, \citenamefont {Jasche}, \citenamefont {Lavaux},\ and\ \citenamefont {Leclercq}}]{tsaprazi2023higher}%
  \BibitemOpen
  \bibfield  {author} {\bibinfo {author} {\bibfnamefont {E.}~\bibnamefont {Tsaprazi}}, \bibinfo {author} {\bibfnamefont {J.}~\bibnamefont {Jasche}}, \bibinfo {author} {\bibfnamefont {G.}~\bibnamefont {Lavaux}}, \ and\ \bibinfo {author} {\bibfnamefont {F.}~\bibnamefont {Leclercq}},\ }\href@noop {} {\bibfield  {journal} {\bibinfo  {journal} {arXiv preprint arXiv:2301.03581}\ } (\bibinfo {year} {2023})}\BibitemShut {NoStop}%
\bibitem [{\citenamefont {Dingeldein}\ \emph {et~al.}(2023)\citenamefont {Dingeldein}, \citenamefont {Cossio},\ and\ \citenamefont {Covino}}]{dingeldein2023simulation}%
  \BibitemOpen
  \bibfield  {author} {\bibinfo {author} {\bibfnamefont {L.}~\bibnamefont {Dingeldein}}, \bibinfo {author} {\bibfnamefont {P.}~\bibnamefont {Cossio}}, \ and\ \bibinfo {author} {\bibfnamefont {R.}~\bibnamefont {Covino}},\ }\href@noop {} {\bibfield  {journal} {\bibinfo  {journal} {Machine Learning: Science and Technology}\ }\textbf {\bibinfo {volume} {4}},\ \bibinfo {pages} {025009} (\bibinfo {year} {2023})}\BibitemShut {NoStop}%
\bibitem [{\citenamefont {Ntampaka}\ \emph {et~al.}(2019)\citenamefont {Ntampaka}, \citenamefont {ZuHone}, \citenamefont {Eisenstein}, \citenamefont {Nagai}, \citenamefont {Vikhlinin}, \citenamefont {Hernquist}, \citenamefont {Marinacci}, \citenamefont {Nelson}, \citenamefont {Pakmor}, \citenamefont {Pillepich} \emph {et~al.}}]{ntampaka2019deep}%
  \BibitemOpen
  \bibfield  {author} {\bibinfo {author} {\bibfnamefont {M.}~\bibnamefont {Ntampaka}}, \bibinfo {author} {\bibfnamefont {J.}~\bibnamefont {ZuHone}}, \bibinfo {author} {\bibfnamefont {D.}~\bibnamefont {Eisenstein}}, \bibinfo {author} {\bibfnamefont {D.}~\bibnamefont {Nagai}}, \bibinfo {author} {\bibfnamefont {A.}~\bibnamefont {Vikhlinin}}, \bibinfo {author} {\bibfnamefont {L.}~\bibnamefont {Hernquist}}, \bibinfo {author} {\bibfnamefont {F.}~\bibnamefont {Marinacci}}, \bibinfo {author} {\bibfnamefont {D.}~\bibnamefont {Nelson}}, \bibinfo {author} {\bibfnamefont {R.}~\bibnamefont {Pakmor}}, \bibinfo {author} {\bibfnamefont {A.}~\bibnamefont {Pillepich}},  \emph {et~al.},\ }\href@noop {} {\bibfield  {journal} {\bibinfo  {journal} {The Astrophysical Journal}\ }\textbf {\bibinfo {volume} {876}},\ \bibinfo {pages} {82} (\bibinfo {year} {2019})}\BibitemShut {NoStop}%
\bibitem [{\citenamefont {Fluri}\ \emph {et~al.}(2022)\citenamefont {Fluri}, \citenamefont {Kacprzak}, \citenamefont {Lucchi}, \citenamefont {Schneider}, \citenamefont {Refregier},\ and\ \citenamefont {Hofmann}}]{fluriWL2022}%
  \BibitemOpen
  \bibfield  {author} {\bibinfo {author} {\bibfnamefont {J.}~\bibnamefont {Fluri}}, \bibinfo {author} {\bibfnamefont {T.}~\bibnamefont {Kacprzak}}, \bibinfo {author} {\bibfnamefont {A.}~\bibnamefont {Lucchi}}, \bibinfo {author} {\bibfnamefont {A.}~\bibnamefont {Schneider}}, \bibinfo {author} {\bibfnamefont {A.}~\bibnamefont {Refregier}}, \ and\ \bibinfo {author} {\bibfnamefont {T.}~\bibnamefont {Hofmann}},\ }\href {\doibase 10.1103/PhysRevD.105.083518} {\bibfield  {journal} {\bibinfo  {journal} {Phys. Rev. D}\ }\textbf {\bibinfo {volume} {105}},\ \bibinfo {pages} {083518} (\bibinfo {year} {2022})}\BibitemShut {NoStop}%
\bibitem [{\citenamefont {Lanusse}\ \emph {et~al.}(2018)\citenamefont {Lanusse}, \citenamefont {Ma}, \citenamefont {Li}, \citenamefont {Collett}, \citenamefont {Li}, \citenamefont {Ravanbakhsh}, \citenamefont {Mandelbaum},\ and\ \citenamefont {P{\'o}czos}}]{lanusse2018cmu}%
  \BibitemOpen
  \bibfield  {author} {\bibinfo {author} {\bibfnamefont {F.}~\bibnamefont {Lanusse}}, \bibinfo {author} {\bibfnamefont {Q.}~\bibnamefont {Ma}}, \bibinfo {author} {\bibfnamefont {N.}~\bibnamefont {Li}}, \bibinfo {author} {\bibfnamefont {T.~E.}\ \bibnamefont {Collett}}, \bibinfo {author} {\bibfnamefont {C.-L.}\ \bibnamefont {Li}}, \bibinfo {author} {\bibfnamefont {S.}~\bibnamefont {Ravanbakhsh}}, \bibinfo {author} {\bibfnamefont {R.}~\bibnamefont {Mandelbaum}}, \ and\ \bibinfo {author} {\bibfnamefont {B.}~\bibnamefont {P{\'o}czos}},\ }\href@noop {} {\bibfield  {journal} {\bibinfo  {journal} {Monthly Notices of the Royal Astronomical Society}\ }\textbf {\bibinfo {volume} {473}},\ \bibinfo {pages} {3895} (\bibinfo {year} {2018})}\BibitemShut {NoStop}%
\bibitem [{\citenamefont {Rubin}(1984)}]{rubinABC}%
  \BibitemOpen
  \bibfield  {author} {\bibinfo {author} {\bibfnamefont {D.~B.}\ \bibnamefont {Rubin}},\ }\href {\doibase 10.1214/aos/1176346785} {\bibfield  {journal} {\bibinfo  {journal} {The Annals of Statistics}\ }\textbf {\bibinfo {volume} {12}},\ \bibinfo {pages} {1151 } (\bibinfo {year} {1984})}\BibitemShut {NoStop}%
\bibitem [{\citenamefont {Schafer}\ and\ \citenamefont {Freeman}(2012)}]{schafer2012likelihood}%
  \BibitemOpen
  \bibfield  {author} {\bibinfo {author} {\bibfnamefont {C.~M.}\ \bibnamefont {Schafer}}\ and\ \bibinfo {author} {\bibfnamefont {P.~E.}\ \bibnamefont {Freeman}},\ }in\ \href@noop {} {\emph {\bibinfo {booktitle} {Statistical Challenges in Modern Astronomy V}}}\ (\bibinfo  {publisher} {Springer},\ \bibinfo {year} {2012})\ pp.\ \bibinfo {pages} {3--19}\BibitemShut {NoStop}%
\bibitem [{\citenamefont {Cameron}\ and\ \citenamefont {Pettitt}(2012)}]{cameron2012approximate}%
  \BibitemOpen
  \bibfield  {author} {\bibinfo {author} {\bibfnamefont {E.}~\bibnamefont {Cameron}}\ and\ \bibinfo {author} {\bibfnamefont {A.}~\bibnamefont {Pettitt}},\ }\href@noop {} {\bibfield  {journal} {\bibinfo  {journal} {Monthly Notices of the Royal Astronomical Society}\ }\textbf {\bibinfo {volume} {425}},\ \bibinfo {pages} {44} (\bibinfo {year} {2012})}\BibitemShut {NoStop}%
\bibitem [{\citenamefont {Hahn}\ \emph {et~al.}(2017)\citenamefont {Hahn}, \citenamefont {Vakili}, \citenamefont {Walsh}, \citenamefont {Hearin}, \citenamefont {Hogg},\ and\ \citenamefont {Campbell}}]{hahn2017approximate}%
  \BibitemOpen
  \bibfield  {author} {\bibinfo {author} {\bibfnamefont {C.}~\bibnamefont {Hahn}}, \bibinfo {author} {\bibfnamefont {M.}~\bibnamefont {Vakili}}, \bibinfo {author} {\bibfnamefont {K.}~\bibnamefont {Walsh}}, \bibinfo {author} {\bibfnamefont {A.~P.}\ \bibnamefont {Hearin}}, \bibinfo {author} {\bibfnamefont {D.~W.}\ \bibnamefont {Hogg}}, \ and\ \bibinfo {author} {\bibfnamefont {D.}~\bibnamefont {Campbell}},\ }\href@noop {} {\bibfield  {journal} {\bibinfo  {journal} {Monthly Notices of the Royal Astronomical Society}\ }\textbf {\bibinfo {volume} {469}},\ \bibinfo {pages} {2791} (\bibinfo {year} {2017})}\BibitemShut {NoStop}%
\bibitem [{\citenamefont {{Alsing}}\ \emph {et~al.}(2018)\citenamefont {{Alsing}}, \citenamefont {{Wandelt}},\ and\ \citenamefont {{Feeney}}}]{Alsing_2018}%
  \BibitemOpen
  \bibfield  {author} {\bibinfo {author} {\bibfnamefont {J.}~\bibnamefont {{Alsing}}}, \bibinfo {author} {\bibfnamefont {B.}~\bibnamefont {{Wandelt}}}, \ and\ \bibinfo {author} {\bibfnamefont {S.}~\bibnamefont {{Feeney}}},\ }\href {\doibase 10.1093/mnras/sty819} {\bibfield  {journal} {\bibinfo  {journal} {\mnras}\ }\textbf {\bibinfo {volume} {477}},\ \bibinfo {pages} {2874} (\bibinfo {year} {2018})},\ \Eprint {http://arxiv.org/abs/1801.01497} {arXiv:1801.01497 [astro-ph.CO]} \BibitemShut {NoStop}%
\bibitem [{\citenamefont {Papamakarios}(2019)}]{papamakarios2019neural}%
  \BibitemOpen
  \bibfield  {author} {\bibinfo {author} {\bibfnamefont {G.}~\bibnamefont {Papamakarios}},\ }\href@noop {} {\bibfield  {journal} {\bibinfo  {journal} {arXiv preprint arXiv:1910.13233}\ } (\bibinfo {year} {2019})}\BibitemShut {NoStop}%
\bibitem [{\citenamefont {Bishop}(1994)}]{bishop1994mixture}%
  \BibitemOpen
  \bibfield  {author} {\bibinfo {author} {\bibfnamefont {C.~M.}\ \bibnamefont {Bishop}},\ }\href@noop {} {\emph {\bibinfo {title} {Mixture density networks}}},\ \bibinfo {type} {WorkingPaper}\ (\bibinfo  {institution} {Aston University},\ \bibinfo {year} {1994})\BibitemShut {NoStop}%
\bibitem [{\citenamefont {Papamakarios}\ and\ \citenamefont {Murray}(2016)}]{papamakarios2016fast}%
  \BibitemOpen
  \bibfield  {author} {\bibinfo {author} {\bibfnamefont {G.}~\bibnamefont {Papamakarios}}\ and\ \bibinfo {author} {\bibfnamefont {I.}~\bibnamefont {Murray}},\ }\href@noop {} {\bibfield  {journal} {\bibinfo  {journal} {Advances in neural information processing systems}\ }\textbf {\bibinfo {volume} {29}} (\bibinfo {year} {2016})}\BibitemShut {NoStop}%
\bibitem [{\citenamefont {Papamakarios}\ \emph {et~al.}(2021)\citenamefont {Papamakarios}, \citenamefont {Nalisnick}, \citenamefont {Rezende}, \citenamefont {Mohamed},\ and\ \citenamefont {Lakshminarayanan}}]{papamakarios2021normalizing}%
  \BibitemOpen
  \bibfield  {author} {\bibinfo {author} {\bibfnamefont {G.}~\bibnamefont {Papamakarios}}, \bibinfo {author} {\bibfnamefont {E.}~\bibnamefont {Nalisnick}}, \bibinfo {author} {\bibfnamefont {D.~J.}\ \bibnamefont {Rezende}}, \bibinfo {author} {\bibfnamefont {S.}~\bibnamefont {Mohamed}}, \ and\ \bibinfo {author} {\bibfnamefont {B.}~\bibnamefont {Lakshminarayanan}},\ }\href@noop {} {\bibfield  {journal} {\bibinfo  {journal} {The Journal of Machine Learning Research}\ }\textbf {\bibinfo {volume} {22}},\ \bibinfo {pages} {2617} (\bibinfo {year} {2021})}\BibitemShut {NoStop}%
\bibitem [{\citenamefont {Greenberg}\ \emph {et~al.}(2019)\citenamefont {Greenberg}, \citenamefont {Nonnenmacher},\ and\ \citenamefont {Macke}}]{greenberg2019automatic}%
  \BibitemOpen
  \bibfield  {author} {\bibinfo {author} {\bibfnamefont {D.}~\bibnamefont {Greenberg}}, \bibinfo {author} {\bibfnamefont {M.}~\bibnamefont {Nonnenmacher}}, \ and\ \bibinfo {author} {\bibfnamefont {J.}~\bibnamefont {Macke}},\ }in\ \href@noop {} {\emph {\bibinfo {booktitle} {International Conference on Machine Learning}}}\ (\bibinfo {organization} {PMLR},\ \bibinfo {year} {2019})\ pp.\ \bibinfo {pages} {2404--2414}\BibitemShut {NoStop}%
\bibitem [{\citenamefont {{Vasist}}\ \emph {et~al.}(2023)\citenamefont {{Vasist}}, \citenamefont {{Rozet}}, \citenamefont {{Absil}}, \citenamefont {{Molli{\`e}re}}, \citenamefont {{Nasedkin}},\ and\ \citenamefont {{Louppe}}}]{vasist2023}%
  \BibitemOpen
  \bibfield  {author} {\bibinfo {author} {\bibfnamefont {M.}~\bibnamefont {{Vasist}}}, \bibinfo {author} {\bibfnamefont {F.}~\bibnamefont {{Rozet}}}, \bibinfo {author} {\bibfnamefont {O.}~\bibnamefont {{Absil}}}, \bibinfo {author} {\bibfnamefont {P.}~\bibnamefont {{Molli{\`e}re}}}, \bibinfo {author} {\bibfnamefont {E.}~\bibnamefont {{Nasedkin}}}, \ and\ \bibinfo {author} {\bibfnamefont {G.}~\bibnamefont {{Louppe}}},\ }\href {\doibase 10.1051/0004-6361/202245263} {\bibfield  {journal} {\bibinfo  {journal} {\aap}\ }\textbf {\bibinfo {volume} {672}},\ \bibinfo {eid} {A147} (\bibinfo {year} {2023})},\ \Eprint {http://arxiv.org/abs/2301.06575} {arXiv:2301.06575 [astro-ph.EP]} \BibitemShut {NoStop}%
\bibitem [{\citenamefont {{Crisostomi}}\ \emph {et~al.}(2023)\citenamefont {{Crisostomi}}, \citenamefont {{Dey}}, \citenamefont {{Barausse}},\ and\ \citenamefont {{Trotta}}}]{cristosomi2023}%
  \BibitemOpen
  \bibfield  {author} {\bibinfo {author} {\bibfnamefont {M.}~\bibnamefont {{Crisostomi}}}, \bibinfo {author} {\bibfnamefont {K.}~\bibnamefont {{Dey}}}, \bibinfo {author} {\bibfnamefont {E.}~\bibnamefont {{Barausse}}}, \ and\ \bibinfo {author} {\bibfnamefont {R.}~\bibnamefont {{Trotta}}},\ }\href {\doibase 10.1103/PhysRevD.108.044029} {\bibfield  {journal} {\bibinfo  {journal} {\prd}\ }\textbf {\bibinfo {volume} {108}},\ \bibinfo {eid} {044029} (\bibinfo {year} {2023})},\ \Eprint {http://arxiv.org/abs/2305.18528} {arXiv:2305.18528 [gr-qc]} \BibitemShut {NoStop}%
\bibitem [{\citenamefont {{Alsing}}\ \emph {et~al.}(2019)\citenamefont {{Alsing}}, \citenamefont {{Charnock}}, \citenamefont {{Feeney}},\ and\ \citenamefont {{Wandelt}}}]{Alsing_2019}%
  \BibitemOpen
  \bibfield  {author} {\bibinfo {author} {\bibfnamefont {J.}~\bibnamefont {{Alsing}}}, \bibinfo {author} {\bibfnamefont {T.}~\bibnamefont {{Charnock}}}, \bibinfo {author} {\bibfnamefont {S.}~\bibnamefont {{Feeney}}}, \ and\ \bibinfo {author} {\bibfnamefont {B.}~\bibnamefont {{Wandelt}}},\ }\href {\doibase 10.1093/mnras/stz1960} {\bibfield  {journal} {\bibinfo  {journal} {\mnras}\ }\textbf {\bibinfo {volume} {488}},\ \bibinfo {pages} {4440} (\bibinfo {year} {2019})},\ \Eprint {http://arxiv.org/abs/1903.00007} {arXiv:1903.00007 [astro-ph.CO]} \BibitemShut {NoStop}%
\bibitem [{\citenamefont {Papamakarios}\ \emph {et~al.}(2019)\citenamefont {Papamakarios}, \citenamefont {Sterratt},\ and\ \citenamefont {Murray}}]{papamakarios2019sequential}%
  \BibitemOpen
  \bibfield  {author} {\bibinfo {author} {\bibfnamefont {G.}~\bibnamefont {Papamakarios}}, \bibinfo {author} {\bibfnamefont {D.}~\bibnamefont {Sterratt}}, \ and\ \bibinfo {author} {\bibfnamefont {I.}~\bibnamefont {Murray}},\ }in\ \href@noop {} {\emph {\bibinfo {booktitle} {The 22nd International Conference on Artificial Intelligence and Statistics}}}\ (\bibinfo {organization} {PMLR},\ \bibinfo {year} {2019})\ pp.\ \bibinfo {pages} {837--848}\BibitemShut {NoStop}%
\bibitem [{\citenamefont {Robert}\ \emph {et~al.}(1999)\citenamefont {Robert}, \citenamefont {Casella},\ and\ \citenamefont {Casella}}]{robert1999monte}%
  \BibitemOpen
  \bibfield  {author} {\bibinfo {author} {\bibfnamefont {C.~P.}\ \bibnamefont {Robert}}, \bibinfo {author} {\bibfnamefont {G.}~\bibnamefont {Casella}}, \ and\ \bibinfo {author} {\bibfnamefont {G.}~\bibnamefont {Casella}},\ }\href@noop {} {\emph {\bibinfo {title} {Monte Carlo statistical methods}}},\ Vol.~\bibinfo {volume} {2}\ (\bibinfo  {publisher} {Springer},\ \bibinfo {year} {1999})\BibitemShut {NoStop}%
\bibitem [{\citenamefont {Blei}\ \emph {et~al.}(2017)\citenamefont {Blei}, \citenamefont {Kucukelbir},\ and\ \citenamefont {McAuliffe}}]{blei2017variational}%
  \BibitemOpen
  \bibfield  {author} {\bibinfo {author} {\bibfnamefont {D.~M.}\ \bibnamefont {Blei}}, \bibinfo {author} {\bibfnamefont {A.}~\bibnamefont {Kucukelbir}}, \ and\ \bibinfo {author} {\bibfnamefont {J.~D.}\ \bibnamefont {McAuliffe}},\ }\href@noop {} {\bibfield  {journal} {\bibinfo  {journal} {Journal of the American statistical Association}\ }\textbf {\bibinfo {volume} {112}},\ \bibinfo {pages} {859} (\bibinfo {year} {2017})}\BibitemShut {NoStop}%
\bibitem [{\citenamefont {Robert}\ \emph {et~al.}(2004)\citenamefont {Robert}, \citenamefont {Casella}, \citenamefont {Robert},\ and\ \citenamefont {Casella}}]{robert2004metropolis}%
  \BibitemOpen
  \bibfield  {author} {\bibinfo {author} {\bibfnamefont {C.~P.}\ \bibnamefont {Robert}}, \bibinfo {author} {\bibfnamefont {G.}~\bibnamefont {Casella}}, \bibinfo {author} {\bibfnamefont {C.~P.}\ \bibnamefont {Robert}}, \ and\ \bibinfo {author} {\bibfnamefont {G.}~\bibnamefont {Casella}},\ }\href@noop {} {\bibfield  {journal} {\bibinfo  {journal} {Monte Carlo statistical methods}\ ,\ \bibinfo {pages} {267}} (\bibinfo {year} {2004})}\BibitemShut {NoStop}%
\bibitem [{\citenamefont {{Jeffrey}}\ \emph {et~al.}(2021)\citenamefont {{Jeffrey}}, \citenamefont {{Alsing}},\ and\ \citenamefont {{Lanusse}}}]{jeffreyDES2021}%
  \BibitemOpen
  \bibfield  {author} {\bibinfo {author} {\bibfnamefont {N.}~\bibnamefont {{Jeffrey}}}, \bibinfo {author} {\bibfnamefont {J.}~\bibnamefont {{Alsing}}}, \ and\ \bibinfo {author} {\bibfnamefont {F.}~\bibnamefont {{Lanusse}}},\ }\href {\doibase 10.1093/mnras/staa3594} {\bibfield  {journal} {\bibinfo  {journal} {\mnras}\ }\textbf {\bibinfo {volume} {501}},\ \bibinfo {pages} {954} (\bibinfo {year} {2021})},\ \Eprint {http://arxiv.org/abs/2009.08459} {arXiv:2009.08459 [astro-ph.CO]} \BibitemShut {NoStop}%
\bibitem [{\citenamefont {Brehmer}\ \emph {et~al.}(2020)\citenamefont {Brehmer}, \citenamefont {Kling}, \citenamefont {Espejo},\ and\ \citenamefont {Cranmer}}]{Brehmer2019xox}%
  \BibitemOpen
  \bibfield  {author} {\bibinfo {author} {\bibfnamefont {J.}~\bibnamefont {Brehmer}}, \bibinfo {author} {\bibfnamefont {F.}~\bibnamefont {Kling}}, \bibinfo {author} {\bibfnamefont {I.}~\bibnamefont {Espejo}}, \ and\ \bibinfo {author} {\bibfnamefont {K.}~\bibnamefont {Cranmer}},\ }\href {\doibase 10.1007/s41781-020-0035-2} {\bibfield  {journal} {\bibinfo  {journal} {Comput. Softw. Big Sci.}\ }\textbf {\bibinfo {volume} {4}},\ \bibinfo {pages} {3} (\bibinfo {year} {2020})},\ \Eprint {http://arxiv.org/abs/1907.10621} {arXiv:1907.10621 [hep-ph]} \BibitemShut {NoStop}%
\bibitem [{\citenamefont {Hermans}\ \emph {et~al.}(2020)\citenamefont {Hermans}, \citenamefont {Begy},\ and\ \citenamefont {Louppe}}]{hermans2020likelihood}%
  \BibitemOpen
  \bibfield  {author} {\bibinfo {author} {\bibfnamefont {J.}~\bibnamefont {Hermans}}, \bibinfo {author} {\bibfnamefont {V.}~\bibnamefont {Begy}}, \ and\ \bibinfo {author} {\bibfnamefont {G.}~\bibnamefont {Louppe}},\ }in\ \href@noop {} {\emph {\bibinfo {booktitle} {International conference on machine learning}}}\ (\bibinfo {organization} {PMLR},\ \bibinfo {year} {2020})\ pp.\ \bibinfo {pages} {4239--4248}\BibitemShut {NoStop}%
\bibitem [{\citenamefont {Cranmer}\ \emph {et~al.}(2015)\citenamefont {Cranmer}, \citenamefont {Pavez},\ and\ \citenamefont {Louppe}}]{cranmer2015approximating}%
  \BibitemOpen
  \bibfield  {author} {\bibinfo {author} {\bibfnamefont {K.}~\bibnamefont {Cranmer}}, \bibinfo {author} {\bibfnamefont {J.}~\bibnamefont {Pavez}}, \ and\ \bibinfo {author} {\bibfnamefont {G.}~\bibnamefont {Louppe}},\ }\href@noop {} {\bibfield  {journal} {\bibinfo  {journal} {arXiv preprint arXiv:1506.02169}\ } (\bibinfo {year} {2015})}\BibitemShut {NoStop}%
\bibitem [{\citenamefont {{Cole}}\ \emph {et~al.}(2022)\citenamefont {{Cole}}, \citenamefont {{Miller}}, \citenamefont {{Witte}}, \citenamefont {{Cai}}, \citenamefont {{Grootes}}, \citenamefont {{Nattino}},\ and\ \citenamefont {{Weniger}}}]{coleTMNRE}%
  \BibitemOpen
  \bibfield  {author} {\bibinfo {author} {\bibfnamefont {A.}~\bibnamefont {{Cole}}}, \bibinfo {author} {\bibfnamefont {B.~K.}\ \bibnamefont {{Miller}}}, \bibinfo {author} {\bibfnamefont {S.~J.}\ \bibnamefont {{Witte}}}, \bibinfo {author} {\bibfnamefont {M.~X.}\ \bibnamefont {{Cai}}}, \bibinfo {author} {\bibfnamefont {M.~W.}\ \bibnamefont {{Grootes}}}, \bibinfo {author} {\bibfnamefont {F.}~\bibnamefont {{Nattino}}}, \ and\ \bibinfo {author} {\bibfnamefont {C.}~\bibnamefont {{Weniger}}},\ }\href {\doibase 10.1088/1475-7516/2022/09/004} {\bibfield  {journal} {\bibinfo  {journal} {\jcap}\ }\textbf {\bibinfo {volume} {2022}},\ \bibinfo {eid} {004} (\bibinfo {year} {2022})},\ \Eprint {http://arxiv.org/abs/2111.08030} {arXiv:2111.08030 [astro-ph.CO]} \BibitemShut {NoStop}%
\bibitem [{\citenamefont {{Karchev}}\ \emph {et~al.}(2023)\citenamefont {{Karchev}}, \citenamefont {{Trotta}},\ and\ \citenamefont {{Weniger}}}]{karchevTMNRE}%
  \BibitemOpen
  \bibfield  {author} {\bibinfo {author} {\bibfnamefont {K.}~\bibnamefont {{Karchev}}}, \bibinfo {author} {\bibfnamefont {R.}~\bibnamefont {{Trotta}}}, \ and\ \bibinfo {author} {\bibfnamefont {C.}~\bibnamefont {{Weniger}}},\ }\href {\doibase 10.1093/mnras/stac3785} {\bibfield  {journal} {\bibinfo  {journal} {\mnras}\ }\textbf {\bibinfo {volume} {520}},\ \bibinfo {pages} {1056} (\bibinfo {year} {2023})},\ \Eprint {http://arxiv.org/abs/2209.06733} {arXiv:2209.06733 [astro-ph.CO]} \BibitemShut {NoStop}%
\bibitem [{\citenamefont {{Miller}}\ \emph {et~al.}(2021)\citenamefont {{Miller}}, \citenamefont {{Cole}}, \citenamefont {{Forr{\'e}}}, \citenamefont {{Louppe}},\ and\ \citenamefont {{Weniger}}}]{millerTMNRE}%
  \BibitemOpen
  \bibfield  {author} {\bibinfo {author} {\bibfnamefont {B.}~\bibnamefont {{Miller}}}, \bibinfo {author} {\bibfnamefont {A.}~\bibnamefont {{Cole}}}, \bibinfo {author} {\bibfnamefont {P.}~\bibnamefont {{Forr{\'e}}}}, \bibinfo {author} {\bibfnamefont {G.}~\bibnamefont {{Louppe}}}, \ and\ \bibinfo {author} {\bibfnamefont {C.}~\bibnamefont {{Weniger}}},\ }\href {\doibase 10.48550/arXiv.2107.01214} {\bibfield  {journal} {\bibinfo  {journal} {Advances in Neural Information Processing Systems}\ }\textbf {\bibinfo {volume} {34}},\ \bibinfo {pages} {129} (\bibinfo {year} {2021})},\ \Eprint {http://arxiv.org/abs/2107.01214} {arXiv:2107.01214 [stat.ML]} \BibitemShut {NoStop}%
\bibitem [{\citenamefont {{Delaunoy}}\ \emph {et~al.}(2020)\citenamefont {{Delaunoy}}, \citenamefont {{Wehenkel}}, \citenamefont {{Hinderer}}, \citenamefont {{Nissanke}}, \citenamefont {{Weniger}}, \citenamefont {{Williamson}},\ and\ \citenamefont {{Louppe}}}]{delaunoy2020}%
  \BibitemOpen
  \bibfield  {author} {\bibinfo {author} {\bibfnamefont {A.}~\bibnamefont {{Delaunoy}}}, \bibinfo {author} {\bibfnamefont {A.}~\bibnamefont {{Wehenkel}}}, \bibinfo {author} {\bibfnamefont {T.}~\bibnamefont {{Hinderer}}}, \bibinfo {author} {\bibfnamefont {S.}~\bibnamefont {{Nissanke}}}, \bibinfo {author} {\bibfnamefont {C.}~\bibnamefont {{Weniger}}}, \bibinfo {author} {\bibfnamefont {A.~R.}\ \bibnamefont {{Williamson}}}, \ and\ \bibinfo {author} {\bibfnamefont {G.}~\bibnamefont {{Louppe}}},\ }\href {\doibase 10.48550/arXiv.2010.12931} {\bibfield  {journal} {\bibinfo  {journal} {arXiv e-prints}\ ,\ \bibinfo {eid} {arXiv:2010.12931}} (\bibinfo {year} {2020})},\ \Eprint {http://arxiv.org/abs/2010.12931} {arXiv:2010.12931 [astro-ph.IM]} \BibitemShut {NoStop}%
\bibitem [{\citenamefont {{Bhardwaj}}\ \emph {et~al.}(2023)\citenamefont {{Bhardwaj}}, \citenamefont {{Alvey}}, \citenamefont {{Miller}}, \citenamefont {{Nissanke}},\ and\ \citenamefont {{Weniger}}}]{peregrine2023}%
  \BibitemOpen
  \bibfield  {author} {\bibinfo {author} {\bibfnamefont {U.}~\bibnamefont {{Bhardwaj}}}, \bibinfo {author} {\bibfnamefont {J.}~\bibnamefont {{Alvey}}}, \bibinfo {author} {\bibfnamefont {B.~K.}\ \bibnamefont {{Miller}}}, \bibinfo {author} {\bibfnamefont {S.}~\bibnamefont {{Nissanke}}}, \ and\ \bibinfo {author} {\bibfnamefont {C.}~\bibnamefont {{Weniger}}},\ }\href {\doibase 10.1103/PhysRevD.108.042004} {\bibfield  {journal} {\bibinfo  {journal} {\prd}\ }\textbf {\bibinfo {volume} {108}},\ \bibinfo {eid} {042004} (\bibinfo {year} {2023})},\ \Eprint {http://arxiv.org/abs/2304.02035} {arXiv:2304.02035 [gr-qc]} \BibitemShut {NoStop}%
\bibitem [{\citenamefont {Miller}\ \emph {et~al.}(2022)\citenamefont {Miller}, \citenamefont {Cole}, \citenamefont {Weniger}, \citenamefont {Nattino}, \citenamefont {Ku},\ and\ \citenamefont {Grootes}}]{Miller:2022shs}%
  \BibitemOpen
  \bibfield  {author} {\bibinfo {author} {\bibfnamefont {B.~K.}\ \bibnamefont {Miller}}, \bibinfo {author} {\bibfnamefont {A.}~\bibnamefont {Cole}}, \bibinfo {author} {\bibfnamefont {C.}~\bibnamefont {Weniger}}, \bibinfo {author} {\bibfnamefont {F.}~\bibnamefont {Nattino}}, \bibinfo {author} {\bibfnamefont {O.}~\bibnamefont {Ku}}, \ and\ \bibinfo {author} {\bibfnamefont {M.~W.}\ \bibnamefont {Grootes}},\ }\href {\doibase 10.21105/joss.04205} {\bibfield  {journal} {\bibinfo  {journal} {J. Open Source Softw.}\ }\textbf {\bibinfo {volume} {7}},\ \bibinfo {pages} {4205} (\bibinfo {year} {2022})}\BibitemShut {NoStop}%
\bibitem [{\citenamefont {Song}\ \emph {et~al.}(2020)\citenamefont {Song}, \citenamefont {Sohl-Dickstein}, \citenamefont {Kingma}, \citenamefont {Kumar}, \citenamefont {Ermon},\ and\ \citenamefont {Poole}}]{song2020score}%
  \BibitemOpen
  \bibfield  {author} {\bibinfo {author} {\bibfnamefont {Y.}~\bibnamefont {Song}}, \bibinfo {author} {\bibfnamefont {J.}~\bibnamefont {Sohl-Dickstein}}, \bibinfo {author} {\bibfnamefont {D.~P.}\ \bibnamefont {Kingma}}, \bibinfo {author} {\bibfnamefont {A.}~\bibnamefont {Kumar}}, \bibinfo {author} {\bibfnamefont {S.}~\bibnamefont {Ermon}}, \ and\ \bibinfo {author} {\bibfnamefont {B.}~\bibnamefont {Poole}},\ }\href@noop {} {\bibfield  {journal} {\bibinfo  {journal} {arXiv preprint arXiv:2011.13456}\ } (\bibinfo {year} {2020})}\BibitemShut {NoStop}%
\bibitem [{\citenamefont {Lueckmann}\ \emph {et~al.}(2021)\citenamefont {Lueckmann}, \citenamefont {Boelts}, \citenamefont {Greenberg}, \citenamefont {Goncalves},\ and\ \citenamefont {Macke}}]{lueckmann2021benchmarking}%
  \BibitemOpen
  \bibfield  {author} {\bibinfo {author} {\bibfnamefont {J.-M.}\ \bibnamefont {Lueckmann}}, \bibinfo {author} {\bibfnamefont {J.}~\bibnamefont {Boelts}}, \bibinfo {author} {\bibfnamefont {D.}~\bibnamefont {Greenberg}}, \bibinfo {author} {\bibfnamefont {P.}~\bibnamefont {Goncalves}}, \ and\ \bibinfo {author} {\bibfnamefont {J.}~\bibnamefont {Macke}},\ }in\ \href@noop {} {\emph {\bibinfo {booktitle} {International conference on artificial intelligence and statistics}}}\ (\bibinfo {organization} {PMLR},\ \bibinfo {year} {2021})\ pp.\ \bibinfo {pages} {343--351}\BibitemShut {NoStop}%
\bibitem [{\citenamefont {Watson-Parris}\ \emph {et~al.}(2021)\citenamefont {Watson-Parris}, \citenamefont {Williams}, \citenamefont {Deaconu},\ and\ \citenamefont {Stier}}]{climateILI}%
  \BibitemOpen
  \bibfield  {author} {\bibinfo {author} {\bibfnamefont {D.}~\bibnamefont {Watson-Parris}}, \bibinfo {author} {\bibfnamefont {A.}~\bibnamefont {Williams}}, \bibinfo {author} {\bibfnamefont {L.}~\bibnamefont {Deaconu}}, \ and\ \bibinfo {author} {\bibfnamefont {P.}~\bibnamefont {Stier}},\ }\href {\doibase 10.5194/gmd-14-7659-2021} {\bibfield  {journal} {\bibinfo  {journal} {Geoscientific Model Development}\ }\textbf {\bibinfo {volume} {14}},\ \bibinfo {pages} {7659} (\bibinfo {year} {2021})}\BibitemShut {NoStop}%
\bibitem [{\citenamefont {Gutmann}\ \emph {et~al.}(2016)\citenamefont {Gutmann}, \citenamefont {Cor} \emph {et~al.}}]{gutmann2016bayesian}%
  \BibitemOpen
  \bibfield  {author} {\bibinfo {author} {\bibfnamefont {M.~U.}\ \bibnamefont {Gutmann}}, \bibinfo {author} {\bibfnamefont {J.}~\bibnamefont {Cor}},  \emph {et~al.},\ }\href@noop {} {\bibfield  {journal} {\bibinfo  {journal} {Journal of Machine Learning Research}\ }\textbf {\bibinfo {volume} {17}},\ \bibinfo {pages} {1} (\bibinfo {year} {2016})}\BibitemShut {NoStop}%
\bibitem [{\citenamefont {{Leclercq}}(2018)}]{Leclercq2018}%
  \BibitemOpen
  \bibfield  {author} {\bibinfo {author} {\bibfnamefont {F.}~\bibnamefont {{Leclercq}}},\ }\href {\doibase 10.1103/PhysRevD.98.063511} {\bibfield  {journal} {\bibinfo  {journal} {\prd}\ }\textbf {\bibinfo {volume} {98}},\ \bibinfo {eid} {063511} (\bibinfo {year} {2018})},\ \Eprint {http://arxiv.org/abs/1805.07152} {arXiv:1805.07152 [astro-ph.CO]} \BibitemShut {NoStop}%
\bibitem [{\citenamefont {Lopez-Paz}\ and\ \citenamefont {Oquab}(2016)}]{lopez2016revisiting}%
  \BibitemOpen
  \bibfield  {author} {\bibinfo {author} {\bibfnamefont {D.}~\bibnamefont {Lopez-Paz}}\ and\ \bibinfo {author} {\bibfnamefont {M.}~\bibnamefont {Oquab}},\ }\href@noop {} {\bibfield  {journal} {\bibinfo  {journal} {arXiv preprint arXiv:1610.06545}\ } (\bibinfo {year} {2016})}\BibitemShut {NoStop}%
\bibitem [{\citenamefont {Cook}\ \emph {et~al.}(2006)\citenamefont {Cook}, \citenamefont {Gelman},\ and\ \citenamefont {Rubin}}]{cook2006}%
  \BibitemOpen
  \bibfield  {author} {\bibinfo {author} {\bibfnamefont {S.~R.}\ \bibnamefont {Cook}}, \bibinfo {author} {\bibfnamefont {A.}~\bibnamefont {Gelman}}, \ and\ \bibinfo {author} {\bibfnamefont {D.~B.}\ \bibnamefont {Rubin}},\ }\href {http://www.jstor.org/stable/27594203} {\bibfield  {journal} {\bibinfo  {journal} {Journal of Computational and Graphical Statistics}\ }\textbf {\bibinfo {volume} {15}},\ \bibinfo {pages} {675} (\bibinfo {year} {2006})}\BibitemShut {NoStop}%
\bibitem [{\citenamefont {Zhao}\ \emph {et~al.}(2021)\citenamefont {Zhao}, \citenamefont {Dalmasso}, \citenamefont {Izbicki},\ and\ \citenamefont {Lee}}]{zhao2021diagnostics}%
  \BibitemOpen
  \bibfield  {author} {\bibinfo {author} {\bibfnamefont {D.}~\bibnamefont {Zhao}}, \bibinfo {author} {\bibfnamefont {N.}~\bibnamefont {Dalmasso}}, \bibinfo {author} {\bibfnamefont {R.}~\bibnamefont {Izbicki}}, \ and\ \bibinfo {author} {\bibfnamefont {A.~B.}\ \bibnamefont {Lee}},\ }in\ \href@noop {} {\emph {\bibinfo {booktitle} {Uncertainty in Artificial Intelligence}}}\ (\bibinfo {organization} {PMLR},\ \bibinfo {year} {2021})\ pp.\ \bibinfo {pages} {1830--1840}\BibitemShut {NoStop}%
\bibitem [{\citenamefont {Bordoloi}\ \emph {et~al.}(2010)\citenamefont {Bordoloi}, \citenamefont {Lilly},\ and\ \citenamefont {Amara}}]{bordoloi2010photo}%
  \BibitemOpen
  \bibfield  {author} {\bibinfo {author} {\bibfnamefont {R.}~\bibnamefont {Bordoloi}}, \bibinfo {author} {\bibfnamefont {S.~J.}\ \bibnamefont {Lilly}}, \ and\ \bibinfo {author} {\bibfnamefont {A.}~\bibnamefont {Amara}},\ }\href@noop {} {\bibfield  {journal} {\bibinfo  {journal} {Monthly Notices of the Royal Astronomical Society}\ }\textbf {\bibinfo {volume} {406}},\ \bibinfo {pages} {881} (\bibinfo {year} {2010})}\BibitemShut {NoStop}%
\bibitem [{\citenamefont {Tanaka}\ \emph {et~al.}(2018)\citenamefont {Tanaka}, \citenamefont {Coupon}, \citenamefont {Hsieh}, \citenamefont {Mineo}, \citenamefont {Nishizawa}, \citenamefont {Speagle}, \citenamefont {Furusawa}, \citenamefont {Miyazaki},\ and\ \citenamefont {Murayama}}]{tanaka2018photometric}%
  \BibitemOpen
  \bibfield  {author} {\bibinfo {author} {\bibfnamefont {M.}~\bibnamefont {Tanaka}}, \bibinfo {author} {\bibfnamefont {J.}~\bibnamefont {Coupon}}, \bibinfo {author} {\bibfnamefont {B.-C.}\ \bibnamefont {Hsieh}}, \bibinfo {author} {\bibfnamefont {S.}~\bibnamefont {Mineo}}, \bibinfo {author} {\bibfnamefont {A.~J.}\ \bibnamefont {Nishizawa}}, \bibinfo {author} {\bibfnamefont {J.}~\bibnamefont {Speagle}}, \bibinfo {author} {\bibfnamefont {H.}~\bibnamefont {Furusawa}}, \bibinfo {author} {\bibfnamefont {S.}~\bibnamefont {Miyazaki}}, \ and\ \bibinfo {author} {\bibfnamefont {H.}~\bibnamefont {Murayama}},\ }\href@noop {} {\bibfield  {journal} {\bibinfo  {journal} {Publications of the Astronomical Society of Japan}\ }\textbf {\bibinfo {volume} {70}},\ \bibinfo {pages} {S9} (\bibinfo {year} {2018})}\BibitemShut {NoStop}%
\bibitem [{\citenamefont {{Talts}}\ \emph {et~al.}(2018)\citenamefont {{Talts}}, \citenamefont {{Betancourt}}, \citenamefont {{Simpson}}, \citenamefont {{Vehtari}},\ and\ \citenamefont {{Gelman}}}]{talts2018}%
  \BibitemOpen
  \bibfield  {author} {\bibinfo {author} {\bibfnamefont {S.}~\bibnamefont {{Talts}}}, \bibinfo {author} {\bibfnamefont {M.}~\bibnamefont {{Betancourt}}}, \bibinfo {author} {\bibfnamefont {D.}~\bibnamefont {{Simpson}}}, \bibinfo {author} {\bibfnamefont {A.}~\bibnamefont {{Vehtari}}}, \ and\ \bibinfo {author} {\bibfnamefont {A.}~\bibnamefont {{Gelman}}},\ }\href {\doibase 10.48550/arXiv.1804.06788} {\bibfield  {journal} {\bibinfo  {journal} {arXiv e-prints}\ ,\ \bibinfo {eid} {arXiv:1804.06788}} (\bibinfo {year} {2018})},\ \Eprint {http://arxiv.org/abs/1804.06788} {arXiv:1804.06788 [stat.ME]} \BibitemShut {NoStop}%
\bibitem [{\citenamefont {Wang}\ \emph {et~al.}(2019)\citenamefont {Wang}, \citenamefont {Mittal}, \citenamefont {Brooks},\ and\ \citenamefont {Oney}}]{wang2019data}%
  \BibitemOpen
  \bibfield  {author} {\bibinfo {author} {\bibfnamefont {A.~Y.}\ \bibnamefont {Wang}}, \bibinfo {author} {\bibfnamefont {A.}~\bibnamefont {Mittal}}, \bibinfo {author} {\bibfnamefont {C.}~\bibnamefont {Brooks}}, \ and\ \bibinfo {author} {\bibfnamefont {S.}~\bibnamefont {Oney}},\ }\href@noop {} {\bibfield  {journal} {\bibinfo  {journal} {Proceedings of the ACM on Human-Computer Interaction}\ }\textbf {\bibinfo {volume} {3}},\ \bibinfo {pages} {1} (\bibinfo {year} {2019})}\BibitemShut {NoStop}%
\bibitem [{\citenamefont {Singh}\ and\ \citenamefont {Singh}(2020)}]{singh2020investigating}%
  \BibitemOpen
  \bibfield  {author} {\bibinfo {author} {\bibfnamefont {D.}~\bibnamefont {Singh}}\ and\ \bibinfo {author} {\bibfnamefont {B.}~\bibnamefont {Singh}},\ }\href@noop {} {\bibfield  {journal} {\bibinfo  {journal} {Applied Soft Computing}\ }\textbf {\bibinfo {volume} {97}},\ \bibinfo {pages} {105524} (\bibinfo {year} {2020})}\BibitemShut {NoStop}%
\bibitem [{\citenamefont {Kingma}\ and\ \citenamefont {Ba}(2014)}]{kingma2014adam}%
  \BibitemOpen
  \bibfield  {author} {\bibinfo {author} {\bibfnamefont {D.~P.}\ \bibnamefont {Kingma}}\ and\ \bibinfo {author} {\bibfnamefont {J.}~\bibnamefont {Ba}},\ }\href@noop {} {\bibfield  {journal} {\bibinfo  {journal} {arXiv preprint arXiv:1412.6980}\ } (\bibinfo {year} {2014})}\BibitemShut {NoStop}%
\bibitem [{\citenamefont {Bai}\ \emph {et~al.}(2021)\citenamefont {Bai}, \citenamefont {Yang}, \citenamefont {Han}, \citenamefont {Yang}, \citenamefont {Li}, \citenamefont {Mao}, \citenamefont {Niu},\ and\ \citenamefont {Liu}}]{bai2021understanding}%
  \BibitemOpen
  \bibfield  {author} {\bibinfo {author} {\bibfnamefont {Y.}~\bibnamefont {Bai}}, \bibinfo {author} {\bibfnamefont {E.}~\bibnamefont {Yang}}, \bibinfo {author} {\bibfnamefont {B.}~\bibnamefont {Han}}, \bibinfo {author} {\bibfnamefont {Y.}~\bibnamefont {Yang}}, \bibinfo {author} {\bibfnamefont {J.}~\bibnamefont {Li}}, \bibinfo {author} {\bibfnamefont {Y.}~\bibnamefont {Mao}}, \bibinfo {author} {\bibfnamefont {G.}~\bibnamefont {Niu}}, \ and\ \bibinfo {author} {\bibfnamefont {T.}~\bibnamefont {Liu}},\ }\href@noop {} {\bibfield  {journal} {\bibinfo  {journal} {Advances in Neural Information Processing Systems}\ }\textbf {\bibinfo {volume} {34}},\ \bibinfo {pages} {24392} (\bibinfo {year} {2021})}\BibitemShut {NoStop}%
\bibitem [{\citenamefont {Abadi}\ \emph {et~al.}(2016)\citenamefont {Abadi}, \citenamefont {Barham}, \citenamefont {Chen}, \citenamefont {Chen}, \citenamefont {Davis}, \citenamefont {Dean}, \citenamefont {Devin}, \citenamefont {Ghemawat}, \citenamefont {Irving}, \citenamefont {Isard} \emph {et~al.}}]{abadi2016tensorflow}%
  \BibitemOpen
  \bibfield  {author} {\bibinfo {author} {\bibfnamefont {M.}~\bibnamefont {Abadi}}, \bibinfo {author} {\bibfnamefont {P.}~\bibnamefont {Barham}}, \bibinfo {author} {\bibfnamefont {J.}~\bibnamefont {Chen}}, \bibinfo {author} {\bibfnamefont {Z.}~\bibnamefont {Chen}}, \bibinfo {author} {\bibfnamefont {A.}~\bibnamefont {Davis}}, \bibinfo {author} {\bibfnamefont {J.}~\bibnamefont {Dean}}, \bibinfo {author} {\bibfnamefont {M.}~\bibnamefont {Devin}}, \bibinfo {author} {\bibfnamefont {S.}~\bibnamefont {Ghemawat}}, \bibinfo {author} {\bibfnamefont {G.}~\bibnamefont {Irving}}, \bibinfo {author} {\bibfnamefont {M.}~\bibnamefont {Isard}},  \emph {et~al.},\ }in\ \href@noop {} {\emph {\bibinfo {booktitle} {12th USENIX symposium on operating systems design and implementation (OSDI 16)}}}\ (\bibinfo {year} {2016})\ pp.\ \bibinfo {pages} {265--283}\BibitemShut {NoStop}%
\bibitem [{\citenamefont {Paszke}\ \emph {et~al.}(2019)\citenamefont {Paszke}, \citenamefont {Gross}, \citenamefont {Massa}, \citenamefont {Lerer}, \citenamefont {Bradbury}, \citenamefont {Chanan}, \citenamefont {Killeen}, \citenamefont {Lin}, \citenamefont {Gimelshein}, \citenamefont {Antiga} \emph {et~al.}}]{paszke2019pytorch}%
  \BibitemOpen
  \bibfield  {author} {\bibinfo {author} {\bibfnamefont {A.}~\bibnamefont {Paszke}}, \bibinfo {author} {\bibfnamefont {S.}~\bibnamefont {Gross}}, \bibinfo {author} {\bibfnamefont {F.}~\bibnamefont {Massa}}, \bibinfo {author} {\bibfnamefont {A.}~\bibnamefont {Lerer}}, \bibinfo {author} {\bibfnamefont {J.}~\bibnamefont {Bradbury}}, \bibinfo {author} {\bibfnamefont {G.}~\bibnamefont {Chanan}}, \bibinfo {author} {\bibfnamefont {T.}~\bibnamefont {Killeen}}, \bibinfo {author} {\bibfnamefont {Z.}~\bibnamefont {Lin}}, \bibinfo {author} {\bibfnamefont {N.}~\bibnamefont {Gimelshein}}, \bibinfo {author} {\bibfnamefont {L.}~\bibnamefont {Antiga}},  \emph {et~al.},\ }\href@noop {} {\bibfield  {journal} {\bibinfo  {journal} {Advances in neural information processing systems}\ }\textbf {\bibinfo {volume} {32}} (\bibinfo {year} {2019})}\BibitemShut {NoStop}%
\bibitem [{\citenamefont {Charnock}\ \emph {et~al.}(2018)\citenamefont {Charnock}, \citenamefont {Lavaux},\ and\ \citenamefont {Wandelt}}]{charnock2018automatic}%
  \BibitemOpen
  \bibfield  {author} {\bibinfo {author} {\bibfnamefont {T.}~\bibnamefont {Charnock}}, \bibinfo {author} {\bibfnamefont {G.}~\bibnamefont {Lavaux}}, \ and\ \bibinfo {author} {\bibfnamefont {B.~D.}\ \bibnamefont {Wandelt}},\ }\href@noop {} {\bibfield  {journal} {\bibinfo  {journal} {Physical Review D}\ }\textbf {\bibinfo {volume} {97}},\ \bibinfo {pages} {083004} (\bibinfo {year} {2018})}\BibitemShut {NoStop}%
\bibitem [{\citenamefont {Alsing}\ and\ \citenamefont {Wandelt}(2018)}]{alsing2018generalized}%
  \BibitemOpen
  \bibfield  {author} {\bibinfo {author} {\bibfnamefont {J.}~\bibnamefont {Alsing}}\ and\ \bibinfo {author} {\bibfnamefont {B.}~\bibnamefont {Wandelt}},\ }\href@noop {} {\bibfield  {journal} {\bibinfo  {journal} {Monthly Notices of the Royal Astronomical Society: Letters}\ }\textbf {\bibinfo {volume} {476}},\ \bibinfo {pages} {L60} (\bibinfo {year} {2018})}\BibitemShut {NoStop}%
\bibitem [{\citenamefont {Alsing}\ \emph {et~al.}(2018)\citenamefont {Alsing}, \citenamefont {Wandelt},\ and\ \citenamefont {Feeney}}]{alsing2018massive}%
  \BibitemOpen
  \bibfield  {author} {\bibinfo {author} {\bibfnamefont {J.}~\bibnamefont {Alsing}}, \bibinfo {author} {\bibfnamefont {B.}~\bibnamefont {Wandelt}}, \ and\ \bibinfo {author} {\bibfnamefont {S.}~\bibnamefont {Feeney}},\ }\href@noop {} {\bibfield  {journal} {\bibinfo  {journal} {Monthly Notices of the Royal Astronomical Society}\ }\textbf {\bibinfo {volume} {477}},\ \bibinfo {pages} {2874} (\bibinfo {year} {2018})}\BibitemShut {NoStop}%
\bibitem [{\citenamefont {Lueckmann}\ \emph {et~al.}(2017)\citenamefont {Lueckmann}, \citenamefont {Goncalves}, \citenamefont {Bassetto}, \citenamefont {{\"O}cal}, \citenamefont {Nonnenmacher},\ and\ \citenamefont {Macke}}]{lueckmann2017flexible}%
  \BibitemOpen
  \bibfield  {author} {\bibinfo {author} {\bibfnamefont {J.-M.}\ \bibnamefont {Lueckmann}}, \bibinfo {author} {\bibfnamefont {P.~J.}\ \bibnamefont {Goncalves}}, \bibinfo {author} {\bibfnamefont {G.}~\bibnamefont {Bassetto}}, \bibinfo {author} {\bibfnamefont {K.}~\bibnamefont {{\"O}cal}}, \bibinfo {author} {\bibfnamefont {M.}~\bibnamefont {Nonnenmacher}}, \ and\ \bibinfo {author} {\bibfnamefont {J.~H.}\ \bibnamefont {Macke}},\ }\href@noop {} {\bibfield  {journal} {\bibinfo  {journal} {Advances in neural information processing systems}\ }\textbf {\bibinfo {volume} {30}} (\bibinfo {year} {2017})}\BibitemShut {NoStop}%
\bibitem [{\citenamefont {Durkan}\ \emph {et~al.}(2020)\citenamefont {Durkan}, \citenamefont {Bekasov}, \citenamefont {Murray},\ and\ \citenamefont {Papamakarios}}]{nflows}%
  \BibitemOpen
  \bibfield  {author} {\bibinfo {author} {\bibfnamefont {C.}~\bibnamefont {Durkan}}, \bibinfo {author} {\bibfnamefont {A.}~\bibnamefont {Bekasov}}, \bibinfo {author} {\bibfnamefont {I.}~\bibnamefont {Murray}}, \ and\ \bibinfo {author} {\bibfnamefont {G.}~\bibnamefont {Papamakarios}},\ }\href {\doibase 10.5281/zenodo.4296287} {\enquote {\bibinfo {title} {{nflows}: normalizing flows in {PyTorch}},}\ } (\bibinfo {year} {2020})\BibitemShut {NoStop}%
\bibitem [{\citenamefont {Rozet}\ \emph {et~al.}(2022)\citenamefont {Rozet} \emph {et~al.}}]{rozet2022zuko}%
  \BibitemOpen
  \bibfield  {author} {\bibinfo {author} {\bibfnamefont {F.}~\bibnamefont {Rozet}} \emph {et~al.},\ }\href {\doibase 10.5281/zenodo.7625672} {\emph {\bibinfo {title} {{Zuko}: Normalizing flows in PyTorch}}} (\bibinfo {year} {2022})\BibitemShut {NoStop}%
\bibitem [{\citenamefont {Hermans}\ \emph {et~al.}(2022)\citenamefont {Hermans}, \citenamefont {Delaunoy}, \citenamefont {Rozet}, \citenamefont {Wehenkel}, \citenamefont {Begy},\ and\ \citenamefont {Louppe}}]{hermans2022trust}%
  \BibitemOpen
  \bibfield  {author} {\bibinfo {author} {\bibfnamefont {J.}~\bibnamefont {Hermans}}, \bibinfo {author} {\bibfnamefont {A.}~\bibnamefont {Delaunoy}}, \bibinfo {author} {\bibfnamefont {F.}~\bibnamefont {Rozet}}, \bibinfo {author} {\bibfnamefont {A.}~\bibnamefont {Wehenkel}}, \bibinfo {author} {\bibfnamefont {V.}~\bibnamefont {Begy}}, \ and\ \bibinfo {author} {\bibfnamefont {G.}~\bibnamefont {Louppe}},\ }\href@noop {} {\bibfield  {journal} {\bibinfo  {journal} {stat.}\ }\textbf {\bibinfo {volume} {1050}} (\bibinfo {year} {2022})}\BibitemShut {NoStop}%
\bibitem [{\citenamefont {Blundell}\ \emph {et~al.}(2015)\citenamefont {Blundell}, \citenamefont {Cornebise}, \citenamefont {Kavukcuoglu},\ and\ \citenamefont {Wierstra}}]{blundell2015weight}%
  \BibitemOpen
  \bibfield  {author} {\bibinfo {author} {\bibfnamefont {C.}~\bibnamefont {Blundell}}, \bibinfo {author} {\bibfnamefont {J.}~\bibnamefont {Cornebise}}, \bibinfo {author} {\bibfnamefont {K.}~\bibnamefont {Kavukcuoglu}}, \ and\ \bibinfo {author} {\bibfnamefont {D.}~\bibnamefont {Wierstra}},\ }in\ \href@noop {} {\emph {\bibinfo {booktitle} {International conference on machine learning}}}\ (\bibinfo {organization} {PMLR},\ \bibinfo {year} {2015})\ pp.\ \bibinfo {pages} {1613--1622}\BibitemShut {NoStop}%
\bibitem [{\citenamefont {Cobb}\ and\ \citenamefont {Jalaian}(2021)}]{cobb2021scaling}%
  \BibitemOpen
  \bibfield  {author} {\bibinfo {author} {\bibfnamefont {A.~D.}\ \bibnamefont {Cobb}}\ and\ \bibinfo {author} {\bibfnamefont {B.}~\bibnamefont {Jalaian}},\ }in\ \href@noop {} {\emph {\bibinfo {booktitle} {Uncertainty in Artificial Intelligence}}}\ (\bibinfo {organization} {PMLR},\ \bibinfo {year} {2021})\ pp.\ \bibinfo {pages} {675--685}\BibitemShut {NoStop}%
\bibitem [{\citenamefont {Chollet}\ \emph {et~al.}(2015)\citenamefont {Chollet} \emph {et~al.}}]{chollet2015keras}%
  \BibitemOpen
  \bibfield  {author} {\bibinfo {author} {\bibfnamefont {F.}~\bibnamefont {Chollet}} \emph {et~al.},\ }\href@noop {} {\enquote {\bibinfo {title} {Keras},}\ }\bibinfo {howpublished} {\url{https://github.com/fchollet/keras}} (\bibinfo {year} {2015})\BibitemShut {NoStop}%
\bibitem [{\citenamefont {LeCun}\ \emph {et~al.}(1998)\citenamefont {LeCun}, \citenamefont {Bottou}, \citenamefont {Bengio},\ and\ \citenamefont {Haffner}}]{lecun1998gradient}%
  \BibitemOpen
  \bibfield  {author} {\bibinfo {author} {\bibfnamefont {Y.}~\bibnamefont {LeCun}}, \bibinfo {author} {\bibfnamefont {L.}~\bibnamefont {Bottou}}, \bibinfo {author} {\bibfnamefont {Y.}~\bibnamefont {Bengio}}, \ and\ \bibinfo {author} {\bibfnamefont {P.}~\bibnamefont {Haffner}},\ }\href@noop {} {\bibfield  {journal} {\bibinfo  {journal} {Proceedings of the IEEE}\ }\textbf {\bibinfo {volume} {86}},\ \bibinfo {pages} {2278} (\bibinfo {year} {1998})}\BibitemShut {NoStop}%
\bibitem [{\citenamefont {Kipf}\ and\ \citenamefont {Welling}(2016)}]{kipf2016semi}%
  \BibitemOpen
  \bibfield  {author} {\bibinfo {author} {\bibfnamefont {T.~N.}\ \bibnamefont {Kipf}}\ and\ \bibinfo {author} {\bibfnamefont {M.}~\bibnamefont {Welling}},\ }\href@noop {} {\bibfield  {journal} {\bibinfo  {journal} {arXiv preprint arXiv:1609.02907}\ } (\bibinfo {year} {2016})}\BibitemShut {NoStop}%
\bibitem [{\citenamefont {Foreman-Mackey}\ \emph {et~al.}(2013)\citenamefont {Foreman-Mackey}, \citenamefont {Hogg}, \citenamefont {Lang},\ and\ \citenamefont {Goodman}}]{foreman2013emcee}%
  \BibitemOpen
  \bibfield  {author} {\bibinfo {author} {\bibfnamefont {D.}~\bibnamefont {Foreman-Mackey}}, \bibinfo {author} {\bibfnamefont {D.~W.}\ \bibnamefont {Hogg}}, \bibinfo {author} {\bibfnamefont {D.}~\bibnamefont {Lang}}, \ and\ \bibinfo {author} {\bibfnamefont {J.}~\bibnamefont {Goodman}},\ }\href@noop {} {\bibfield  {journal} {\bibinfo  {journal} {Publications of the Astronomical Society of the Pacific}\ }\textbf {\bibinfo {volume} {125}},\ \bibinfo {pages} {306} (\bibinfo {year} {2013})}\BibitemShut {NoStop}%
\bibitem [{\citenamefont {Goodman}\ and\ \citenamefont {Weare}(2010)}]{goodman2010ensemble}%
  \BibitemOpen
  \bibfield  {author} {\bibinfo {author} {\bibfnamefont {J.}~\bibnamefont {Goodman}}\ and\ \bibinfo {author} {\bibfnamefont {J.}~\bibnamefont {Weare}},\ }\href@noop {} {\bibfield  {journal} {\bibinfo  {journal} {Communications in applied mathematics and computational science}\ }\textbf {\bibinfo {volume} {5}},\ \bibinfo {pages} {65} (\bibinfo {year} {2010})}\BibitemShut {NoStop}%
\bibitem [{\citenamefont {Bingham}\ \emph {et~al.}(2019)\citenamefont {Bingham}, \citenamefont {Chen}, \citenamefont {Jankowiak}, \citenamefont {Obermeyer}, \citenamefont {Pradhan}, \citenamefont {Karaletsos}, \citenamefont {Singh}, \citenamefont {Szerlip}, \citenamefont {Horsfall},\ and\ \citenamefont {Goodman}}]{bingham2019pyro}%
  \BibitemOpen
  \bibfield  {author} {\bibinfo {author} {\bibfnamefont {E.}~\bibnamefont {Bingham}}, \bibinfo {author} {\bibfnamefont {J.~P.}\ \bibnamefont {Chen}}, \bibinfo {author} {\bibfnamefont {M.}~\bibnamefont {Jankowiak}}, \bibinfo {author} {\bibfnamefont {F.}~\bibnamefont {Obermeyer}}, \bibinfo {author} {\bibfnamefont {N.}~\bibnamefont {Pradhan}}, \bibinfo {author} {\bibfnamefont {T.}~\bibnamefont {Karaletsos}}, \bibinfo {author} {\bibfnamefont {R.}~\bibnamefont {Singh}}, \bibinfo {author} {\bibfnamefont {P.~A.}\ \bibnamefont {Szerlip}}, \bibinfo {author} {\bibfnamefont {P.}~\bibnamefont {Horsfall}}, \ and\ \bibinfo {author} {\bibfnamefont {N.~D.}\ \bibnamefont {Goodman}},\ }\href {http://jmlr.org/papers/v20/18-403.html} {\bibfield  {journal} {\bibinfo  {journal} {J. Mach. Learn. Res.}\ }\textbf {\bibinfo {volume} {20}},\ \bibinfo {pages} {28:1} (\bibinfo {year} {2019})}\BibitemShut {NoStop}%
\bibitem [{\citenamefont {Neal}(2003)}]{neal2003slice}%
  \BibitemOpen
  \bibfield  {author} {\bibinfo {author} {\bibfnamefont {R.~M.}\ \bibnamefont {Neal}},\ }\href@noop {} {\bibfield  {journal} {\bibinfo  {journal} {The annals of statistics}\ }\textbf {\bibinfo {volume} {31}},\ \bibinfo {pages} {705} (\bibinfo {year} {2003})}\BibitemShut {NoStop}%
\bibitem [{\citenamefont {Neal}\ \emph {et~al.}(2011)\citenamefont {Neal} \emph {et~al.}}]{neal2011mcmc}%
  \BibitemOpen
  \bibfield  {author} {\bibinfo {author} {\bibfnamefont {R.~M.}\ \bibnamefont {Neal}} \emph {et~al.},\ }\href@noop {} {\bibfield  {journal} {\bibinfo  {journal} {Handbook of markov chain monte carlo}\ }\textbf {\bibinfo {volume} {2}},\ \bibinfo {pages} {2} (\bibinfo {year} {2011})}\BibitemShut {NoStop}%
\bibitem [{\citenamefont {Hoffman}\ \emph {et~al.}(2014)\citenamefont {Hoffman}, \citenamefont {Gelman} \emph {et~al.}}]{hoffman2014no}%
  \BibitemOpen
  \bibfield  {author} {\bibinfo {author} {\bibfnamefont {M.~D.}\ \bibnamefont {Hoffman}}, \bibinfo {author} {\bibfnamefont {A.}~\bibnamefont {Gelman}},  \emph {et~al.},\ }\href@noop {} {\bibfield  {journal} {\bibinfo  {journal} {J. Mach. Learn. Res.}\ }\textbf {\bibinfo {volume} {15}},\ \bibinfo {pages} {1593} (\bibinfo {year} {2014})}\BibitemShut {NoStop}%
\bibitem [{\citenamefont {Graves}(2011)}]{graves2011practical}%
  \BibitemOpen
  \bibfield  {author} {\bibinfo {author} {\bibfnamefont {A.}~\bibnamefont {Graves}},\ }\href@noop {} {\bibfield  {journal} {\bibinfo  {journal} {Advances in neural information processing systems}\ }\textbf {\bibinfo {volume} {24}} (\bibinfo {year} {2011})}\BibitemShut {NoStop}%
\bibitem [{\citenamefont {Bartlett}\ \emph {et~al.}(2023)\citenamefont {Bartlett}, \citenamefont {Kammerer}, \citenamefont {Kronberger}, \citenamefont {Desmond}, \citenamefont {Ferreira}, \citenamefont {Wandelt}, \citenamefont {Burlacu}, \citenamefont {Alonso},\ and\ \citenamefont {Zennaro}}]{bartlett2023precise}%
  \BibitemOpen
  \bibfield  {author} {\bibinfo {author} {\bibfnamefont {D.~J.}\ \bibnamefont {Bartlett}}, \bibinfo {author} {\bibfnamefont {L.}~\bibnamefont {Kammerer}}, \bibinfo {author} {\bibfnamefont {G.}~\bibnamefont {Kronberger}}, \bibinfo {author} {\bibfnamefont {H.}~\bibnamefont {Desmond}}, \bibinfo {author} {\bibfnamefont {P.~G.}\ \bibnamefont {Ferreira}}, \bibinfo {author} {\bibfnamefont {B.~D.}\ \bibnamefont {Wandelt}}, \bibinfo {author} {\bibfnamefont {B.}~\bibnamefont {Burlacu}}, \bibinfo {author} {\bibfnamefont {D.}~\bibnamefont {Alonso}}, \ and\ \bibinfo {author} {\bibfnamefont {M.}~\bibnamefont {Zennaro}},\ }\href@noop {} {\bibfield  {journal} {\bibinfo  {journal} {arXiv preprint arXiv:2311.15865}\ } (\bibinfo {year} {2023})}\BibitemShut {NoStop}%
\bibitem [{\citenamefont {{Greenberg}}\ \emph {et~al.}(2019)\citenamefont {{Greenberg}}, \citenamefont {{Nonnenmacher}},\ and\ \citenamefont {{Macke}}}]{Greenberg_2019}%
  \BibitemOpen
  \bibfield  {author} {\bibinfo {author} {\bibfnamefont {D.~S.}\ \bibnamefont {{Greenberg}}}, \bibinfo {author} {\bibfnamefont {M.}~\bibnamefont {{Nonnenmacher}}}, \ and\ \bibinfo {author} {\bibfnamefont {J.~H.}\ \bibnamefont {{Macke}}},\ }\href {\doibase 10.48550/arXiv.1905.07488} {\bibfield  {journal} {\bibinfo  {journal} {arXiv e-prints}\ ,\ \bibinfo {eid} {arXiv:1905.07488}} (\bibinfo {year} {2019})},\ \Eprint {http://arxiv.org/abs/1905.07488} {arXiv:1905.07488 [cs.LG]} \BibitemShut {NoStop}%
\bibitem [{\citenamefont {Ramesh}\ \emph {et~al.}(2022)\citenamefont {Ramesh}, \citenamefont {Lueckmann}, \citenamefont {Boelts}, \citenamefont {Tejero-Cantero}, \citenamefont {Greenberg}, \citenamefont {Gon{\c{c}}alves},\ and\ \citenamefont {Macke}}]{ramesh2022gatsbi}%
  \BibitemOpen
  \bibfield  {author} {\bibinfo {author} {\bibfnamefont {P.}~\bibnamefont {Ramesh}}, \bibinfo {author} {\bibfnamefont {J.-M.}\ \bibnamefont {Lueckmann}}, \bibinfo {author} {\bibfnamefont {J.}~\bibnamefont {Boelts}}, \bibinfo {author} {\bibfnamefont {{\'A}.}~\bibnamefont {Tejero-Cantero}}, \bibinfo {author} {\bibfnamefont {D.~S.}\ \bibnamefont {Greenberg}}, \bibinfo {author} {\bibfnamefont {P.~J.}\ \bibnamefont {Gon{\c{c}}alves}}, \ and\ \bibinfo {author} {\bibfnamefont {J.~H.}\ \bibnamefont {Macke}},\ }\href@noop {} {\bibfield  {journal} {\bibinfo  {journal} {arXiv preprint arXiv:2203.06481}\ } (\bibinfo {year} {2022})}\BibitemShut {NoStop}%
\bibitem [{\citenamefont {Ho}\ \emph {et~al.}(2023)\citenamefont {Ho}, \citenamefont {Soltis}, \citenamefont {Farahi}, \citenamefont {Nagai}, \citenamefont {Evrard},\ and\ \citenamefont {Ntampaka}}]{ho2023benchmarks}%
  \BibitemOpen
  \bibfield  {author} {\bibinfo {author} {\bibfnamefont {M.}~\bibnamefont {Ho}}, \bibinfo {author} {\bibfnamefont {J.}~\bibnamefont {Soltis}}, \bibinfo {author} {\bibfnamefont {A.}~\bibnamefont {Farahi}}, \bibinfo {author} {\bibfnamefont {D.}~\bibnamefont {Nagai}}, \bibinfo {author} {\bibfnamefont {A.}~\bibnamefont {Evrard}}, \ and\ \bibinfo {author} {\bibfnamefont {M.}~\bibnamefont {Ntampaka}},\ }\href@noop {} {\bibfield  {journal} {\bibinfo  {journal} {arXiv preprint arXiv:2303.00005}\ } (\bibinfo {year} {2023})}\BibitemShut {NoStop}%
\bibitem [{\citenamefont {Dolag}\ \emph {et~al.}(2016)\citenamefont {Dolag}, \citenamefont {Komatsu},\ and\ \citenamefont {Sunyaev}}]{dolag2016sz}%
  \BibitemOpen
  \bibfield  {author} {\bibinfo {author} {\bibfnamefont {K.}~\bibnamefont {Dolag}}, \bibinfo {author} {\bibfnamefont {E.}~\bibnamefont {Komatsu}}, \ and\ \bibinfo {author} {\bibfnamefont {R.}~\bibnamefont {Sunyaev}},\ }\href@noop {} {\bibfield  {journal} {\bibinfo  {journal} {Monthly Notices of the Royal Astronomical Society}\ }\textbf {\bibinfo {volume} {463}},\ \bibinfo {pages} {1797} (\bibinfo {year} {2016})}\BibitemShut {NoStop}%
\bibitem [{\citenamefont {Soltis}\ \emph {et~al.}(2022)\citenamefont {Soltis}, \citenamefont {Ntampaka}, \citenamefont {Wu}, \citenamefont {ZuHone}, \citenamefont {Evrard}, \citenamefont {Farahi}, \citenamefont {Ho},\ and\ \citenamefont {Nagai}}]{soltis2022machine}%
  \BibitemOpen
  \bibfield  {author} {\bibinfo {author} {\bibfnamefont {J.}~\bibnamefont {Soltis}}, \bibinfo {author} {\bibfnamefont {M.}~\bibnamefont {Ntampaka}}, \bibinfo {author} {\bibfnamefont {J.~F.}\ \bibnamefont {Wu}}, \bibinfo {author} {\bibfnamefont {J.}~\bibnamefont {ZuHone}}, \bibinfo {author} {\bibfnamefont {A.}~\bibnamefont {Evrard}}, \bibinfo {author} {\bibfnamefont {A.}~\bibnamefont {Farahi}}, \bibinfo {author} {\bibfnamefont {M.}~\bibnamefont {Ho}}, \ and\ \bibinfo {author} {\bibfnamefont {D.}~\bibnamefont {Nagai}},\ }\href@noop {} {\bibfield  {journal} {\bibinfo  {journal} {The Astrophysical Journal}\ }\textbf {\bibinfo {volume} {940}},\ \bibinfo {pages} {60} (\bibinfo {year} {2022})}\BibitemShut {NoStop}%
\bibitem [{\citenamefont {Papamakarios}\ \emph {et~al.}(2017)\citenamefont {Papamakarios}, \citenamefont {Pavlakou},\ and\ \citenamefont {Murray}}]{papamakarios2017masked}%
  \BibitemOpen
  \bibfield  {author} {\bibinfo {author} {\bibfnamefont {G.}~\bibnamefont {Papamakarios}}, \bibinfo {author} {\bibfnamefont {T.}~\bibnamefont {Pavlakou}}, \ and\ \bibinfo {author} {\bibfnamefont {I.}~\bibnamefont {Murray}},\ }\href@noop {} {\bibfield  {journal} {\bibinfo  {journal} {Advances in neural information processing systems}\ }\textbf {\bibinfo {volume} {30}} (\bibinfo {year} {2017})}\BibitemShut {NoStop}%
\bibitem [{\citenamefont {Cooray}\ and\ \citenamefont {Sheth}(2002)}]{cooray2002halo}%
  \BibitemOpen
  \bibfield  {author} {\bibinfo {author} {\bibfnamefont {A.}~\bibnamefont {Cooray}}\ and\ \bibinfo {author} {\bibfnamefont {R.}~\bibnamefont {Sheth}},\ }\href@noop {} {\bibfield  {journal} {\bibinfo  {journal} {Physics reports}\ }\textbf {\bibinfo {volume} {372}},\ \bibinfo {pages} {1} (\bibinfo {year} {2002})}\BibitemShut {NoStop}%
\bibitem [{\citenamefont {Spurio~Mancini}\ \emph {et~al.}(2022)\citenamefont {Spurio~Mancini}, \citenamefont {Piras}, \citenamefont {Alsing}, \citenamefont {Joachimi},\ and\ \citenamefont {Hobson}}]{spurio2022cosmopower}%
  \BibitemOpen
  \bibfield  {author} {\bibinfo {author} {\bibfnamefont {A.}~\bibnamefont {Spurio~Mancini}}, \bibinfo {author} {\bibfnamefont {D.}~\bibnamefont {Piras}}, \bibinfo {author} {\bibfnamefont {J.}~\bibnamefont {Alsing}}, \bibinfo {author} {\bibfnamefont {B.}~\bibnamefont {Joachimi}}, \ and\ \bibinfo {author} {\bibfnamefont {M.~P.}\ \bibnamefont {Hobson}},\ }\href@noop {} {\bibfield  {journal} {\bibinfo  {journal} {Monthly Notices of the Royal Astronomical Society}\ }\textbf {\bibinfo {volume} {511}},\ \bibinfo {pages} {1771} (\bibinfo {year} {2022})}\BibitemShut {NoStop}%
\bibitem [{\citenamefont {Hahn}\ \emph {et~al.}(2023{\natexlab{b}})\citenamefont {Hahn}, \citenamefont {Eickenberg}, \citenamefont {Ho}, \citenamefont {Hou}, \citenamefont {Lemos}, \citenamefont {Massara}, \citenamefont {Modi}, \citenamefont {Dizgah}, \citenamefont {R{\'e}galdo-Saint~Blancard},\ and\ \citenamefont {Abidi}}]{hahn2023simbig}%
  \BibitemOpen
  \bibfield  {author} {\bibinfo {author} {\bibfnamefont {C.}~\bibnamefont {Hahn}}, \bibinfo {author} {\bibfnamefont {M.}~\bibnamefont {Eickenberg}}, \bibinfo {author} {\bibfnamefont {S.}~\bibnamefont {Ho}}, \bibinfo {author} {\bibfnamefont {J.}~\bibnamefont {Hou}}, \bibinfo {author} {\bibfnamefont {P.}~\bibnamefont {Lemos}}, \bibinfo {author} {\bibfnamefont {E.}~\bibnamefont {Massara}}, \bibinfo {author} {\bibfnamefont {C.}~\bibnamefont {Modi}}, \bibinfo {author} {\bibfnamefont {A.~M.}\ \bibnamefont {Dizgah}}, \bibinfo {author} {\bibfnamefont {B.}~\bibnamefont {R{\'e}galdo-Saint~Blancard}}, \ and\ \bibinfo {author} {\bibfnamefont {M.~M.}\ \bibnamefont {Abidi}},\ }\href@noop {} {\bibfield  {journal} {\bibinfo  {journal} {Journal of Cosmology and Astroparticle Physics}\ }\textbf {\bibinfo {volume} {2023}},\ \bibinfo {pages} {010} (\bibinfo {year} {2023}{\natexlab{b}})}\BibitemShut {NoStop}%
\bibitem [{\citenamefont {{Villaescusa-Navarro}}\ \emph {et~al.}(2020)\citenamefont {{Villaescusa-Navarro}}, \citenamefont {{Hahn}}, \citenamefont {{Massara}}, \citenamefont {{Banerjee}}, \citenamefont {{Delgado}}, \citenamefont {{Ramanah}}, \citenamefont {{Charnock}}, \citenamefont {{Giusarma}}, \citenamefont {{Li}}, \citenamefont {{Allys}}, \citenamefont {{Brochard}}, \citenamefont {{Uhlemann}}, \citenamefont {{Chiang}}, \citenamefont {{He}}, \citenamefont {{Pisani}}, \citenamefont {{Obuljen}}, \citenamefont {{Feng}}, \citenamefont {{Castorina}}, \citenamefont {{Contardo}}, \citenamefont {{Kreisch}}, \citenamefont {{Nicola}}, \citenamefont {{Alsing}}, \citenamefont {{Scoccimarro}}, \citenamefont {{Verde}}, \citenamefont {{Viel}}, \citenamefont {{Ho}}, \citenamefont {{Mallat}}, \citenamefont {{Wandelt}},\ and\ \citenamefont {{Spergel}}}]{Quijote_sims}%
  \BibitemOpen
  \bibfield  {author} {\bibinfo {author} {\bibfnamefont {F.}~\bibnamefont {{Villaescusa-Navarro}}}, \bibinfo {author} {\bibfnamefont {C.}~\bibnamefont {{Hahn}}}, \bibinfo {author} {\bibfnamefont {E.}~\bibnamefont {{Massara}}}, \bibinfo {author} {\bibfnamefont {A.}~\bibnamefont {{Banerjee}}}, \bibinfo {author} {\bibfnamefont {A.~M.}\ \bibnamefont {{Delgado}}}, \bibinfo {author} {\bibfnamefont {D.~K.}\ \bibnamefont {{Ramanah}}}, \bibinfo {author} {\bibfnamefont {T.}~\bibnamefont {{Charnock}}}, \bibinfo {author} {\bibfnamefont {E.}~\bibnamefont {{Giusarma}}}, \bibinfo {author} {\bibfnamefont {Y.}~\bibnamefont {{Li}}}, \bibinfo {author} {\bibfnamefont {E.}~\bibnamefont {{Allys}}}, \bibinfo {author} {\bibfnamefont {A.}~\bibnamefont {{Brochard}}}, \bibinfo {author} {\bibfnamefont {C.}~\bibnamefont {{Uhlemann}}}, \bibinfo {author} {\bibfnamefont {C.-T.}\ \bibnamefont {{Chiang}}}, \bibinfo {author} {\bibfnamefont {S.}~\bibnamefont {{He}}}, \bibinfo {author} {\bibfnamefont {A.}~\bibnamefont {{Pisani}}}, \bibinfo
  {author} {\bibfnamefont {A.}~\bibnamefont {{Obuljen}}}, \bibinfo {author} {\bibfnamefont {Y.}~\bibnamefont {{Feng}}}, \bibinfo {author} {\bibfnamefont {E.}~\bibnamefont {{Castorina}}}, \bibinfo {author} {\bibfnamefont {G.}~\bibnamefont {{Contardo}}}, \bibinfo {author} {\bibfnamefont {C.~D.}\ \bibnamefont {{Kreisch}}}, \bibinfo {author} {\bibfnamefont {A.}~\bibnamefont {{Nicola}}}, \bibinfo {author} {\bibfnamefont {J.}~\bibnamefont {{Alsing}}}, \bibinfo {author} {\bibfnamefont {R.}~\bibnamefont {{Scoccimarro}}}, \bibinfo {author} {\bibfnamefont {L.}~\bibnamefont {{Verde}}}, \bibinfo {author} {\bibfnamefont {M.}~\bibnamefont {{Viel}}}, \bibinfo {author} {\bibfnamefont {S.}~\bibnamefont {{Ho}}}, \bibinfo {author} {\bibfnamefont {S.}~\bibnamefont {{Mallat}}}, \bibinfo {author} {\bibfnamefont {B.}~\bibnamefont {{Wandelt}}}, \ and\ \bibinfo {author} {\bibfnamefont {D.~N.}\ \bibnamefont {{Spergel}}},\ }\href {\doibase 10.3847/1538-4365/ab9d82} {\bibfield  {journal} {\bibinfo  {journal} {\apjs}\ }\textbf {\bibinfo
  {volume} {250}},\ \bibinfo {eid} {2} (\bibinfo {year} {2020})},\ \Eprint {http://arxiv.org/abs/1909.05273} {arXiv:1909.05273 [astro-ph.CO]} \BibitemShut {NoStop}%
\bibitem [{\citenamefont {Makinen}\ \emph {et~al.}(2022{\natexlab{b}})\citenamefont {Makinen}, \citenamefont {Charnock}, \citenamefont {Lemos}, \citenamefont {Porqueres}, \citenamefont {Heavens},\ and\ \citenamefont {Wandelt}}]{Makinen2022Cosmic}%
  \BibitemOpen
  \bibfield  {author} {\bibinfo {author} {\bibfnamefont {T.~L.}\ \bibnamefont {Makinen}}, \bibinfo {author} {\bibfnamefont {T.}~\bibnamefont {Charnock}}, \bibinfo {author} {\bibfnamefont {P.}~\bibnamefont {Lemos}}, \bibinfo {author} {\bibfnamefont {N.}~\bibnamefont {Porqueres}}, \bibinfo {author} {\bibfnamefont {A.}~\bibnamefont {Heavens}}, \ and\ \bibinfo {author} {\bibfnamefont {B.~D.}\ \bibnamefont {Wandelt}},\ }\href {\doibase 10.21105/astro.2207.05202} {\bibfield  {journal} {\bibinfo  {journal} {The Open Journal of Astrophysics}\ }\textbf {\bibinfo {volume} {5}} (\bibinfo {year} {2022}{\natexlab{b}}),\ 10.21105/astro.2207.05202}\BibitemShut {NoStop}%
\bibitem [{\citenamefont {Cuesta-Lazaro}\ and\ \citenamefont {Mishra-Sharma}(2023)}]{cuesta2023point}%
  \BibitemOpen
  \bibfield  {author} {\bibinfo {author} {\bibfnamefont {C.}~\bibnamefont {Cuesta-Lazaro}}\ and\ \bibinfo {author} {\bibfnamefont {S.}~\bibnamefont {Mishra-Sharma}},\ }\href@noop {} {\bibfield  {journal} {\bibinfo  {journal} {arXiv preprint arXiv:2311.17141}\ } (\bibinfo {year} {2023})}\BibitemShut {NoStop}%
\bibitem [{\citenamefont {de~Santi}\ \emph {et~al.}(2023{\natexlab{b}})\citenamefont {de~Santi}, \citenamefont {Shao}, \citenamefont {Villaescusa-Navarro}, \citenamefont {Abramo}, \citenamefont {Teyssier}, \citenamefont {Villanueva-Domingo}, \citenamefont {Ni}, \citenamefont {Anglés-Alcázar}, \citenamefont {Genel}, \citenamefont {Hernández-Martínez}, \citenamefont {Steinwandel}, \citenamefont {Lovell}, \citenamefont {Dolag}, \citenamefont {Castro},\ and\ \citenamefont {Vogelsberger}}]{deSanti_2023}%
  \BibitemOpen
  \bibfield  {author} {\bibinfo {author} {\bibfnamefont {N.~S.~M.}\ \bibnamefont {de~Santi}}, \bibinfo {author} {\bibfnamefont {H.}~\bibnamefont {Shao}}, \bibinfo {author} {\bibfnamefont {F.}~\bibnamefont {Villaescusa-Navarro}}, \bibinfo {author} {\bibfnamefont {L.~R.}\ \bibnamefont {Abramo}}, \bibinfo {author} {\bibfnamefont {R.}~\bibnamefont {Teyssier}}, \bibinfo {author} {\bibfnamefont {P.}~\bibnamefont {Villanueva-Domingo}}, \bibinfo {author} {\bibfnamefont {Y.}~\bibnamefont {Ni}}, \bibinfo {author} {\bibfnamefont {D.}~\bibnamefont {Anglés-Alcázar}}, \bibinfo {author} {\bibfnamefont {S.}~\bibnamefont {Genel}}, \bibinfo {author} {\bibfnamefont {E.}~\bibnamefont {Hernández-Martínez}}, \bibinfo {author} {\bibfnamefont {U.~P.}\ \bibnamefont {Steinwandel}}, \bibinfo {author} {\bibfnamefont {C.~C.}\ \bibnamefont {Lovell}}, \bibinfo {author} {\bibfnamefont {K.}~\bibnamefont {Dolag}}, \bibinfo {author} {\bibfnamefont {T.}~\bibnamefont {Castro}}, \ and\ \bibinfo {author} {\bibfnamefont {M.}~\bibnamefont
  {Vogelsberger}},\ }\href {\doibase 10.3847/1538-4357/acd1e2} {\bibfield  {journal} {\bibinfo  {journal} {The Astrophysical Journal}\ }\textbf {\bibinfo {volume} {952}},\ \bibinfo {pages} {69} (\bibinfo {year} {2023}{\natexlab{b}})}\BibitemShut {NoStop}%
\bibitem [{\citenamefont {Gilmer}\ \emph {et~al.}(2020)\citenamefont {Gilmer}, \citenamefont {Schoenholz}, \citenamefont {Riley}, \citenamefont {Vinyals},\ and\ \citenamefont {Dahl}}]{gilmer2020message}%
  \BibitemOpen
  \bibfield  {author} {\bibinfo {author} {\bibfnamefont {J.}~\bibnamefont {Gilmer}}, \bibinfo {author} {\bibfnamefont {S.~S.}\ \bibnamefont {Schoenholz}}, \bibinfo {author} {\bibfnamefont {P.~F.}\ \bibnamefont {Riley}}, \bibinfo {author} {\bibfnamefont {O.}~\bibnamefont {Vinyals}}, \ and\ \bibinfo {author} {\bibfnamefont {G.~E.}\ \bibnamefont {Dahl}},\ }\href@noop {} {\bibfield  {journal} {\bibinfo  {journal} {Machine learning meets quantum physics}\ ,\ \bibinfo {pages} {199}} (\bibinfo {year} {2020})}\BibitemShut {NoStop}%
\bibitem [{\citenamefont {{Cutler}}\ and\ \citenamefont {{Flanagan}}(1994)}]{cutler1994}%
  \BibitemOpen
  \bibfield  {author} {\bibinfo {author} {\bibfnamefont {C.}~\bibnamefont {{Cutler}}}\ and\ \bibinfo {author} {\bibfnamefont {{\'E}.~E.}\ \bibnamefont {{Flanagan}}},\ }\href {\doibase 10.1103/PhysRevD.49.2658} {\bibfield  {journal} {\bibinfo  {journal} {\prd}\ }\textbf {\bibinfo {volume} {49}},\ \bibinfo {pages} {2658} (\bibinfo {year} {1994})},\ \Eprint {http://arxiv.org/abs/gr-qc/9402014} {arXiv:gr-qc/9402014 [gr-qc]} \BibitemShut {NoStop}%
\bibitem [{\citenamefont {{Thrane}}\ and\ \citenamefont {{Talbot}}(2019)}]{gw_inference}%
  \BibitemOpen
  \bibfield  {author} {\bibinfo {author} {\bibfnamefont {E.}~\bibnamefont {{Thrane}}}\ and\ \bibinfo {author} {\bibfnamefont {C.}~\bibnamefont {{Talbot}}},\ }\href {\doibase 10.1017/pasa.2019.2} {\bibfield  {journal} {\bibinfo  {journal} {\pasa}\ }\textbf {\bibinfo {volume} {36}},\ \bibinfo {eid} {e010} (\bibinfo {year} {2019})},\ \Eprint {http://arxiv.org/abs/1809.02293} {arXiv:1809.02293 [astro-ph.IM]} \BibitemShut {NoStop}%
\bibitem [{\citenamefont {{Green}}\ \emph {et~al.}(2020)\citenamefont {{Green}}, \citenamefont {{Simpson}},\ and\ \citenamefont {{Gair}}}]{green_gwflow}%
  \BibitemOpen
  \bibfield  {author} {\bibinfo {author} {\bibfnamefont {S.~R.}\ \bibnamefont {{Green}}}, \bibinfo {author} {\bibfnamefont {C.}~\bibnamefont {{Simpson}}}, \ and\ \bibinfo {author} {\bibfnamefont {J.}~\bibnamefont {{Gair}}},\ }\href {\doibase 10.1103/PhysRevD.102.104057} {\bibfield  {journal} {\bibinfo  {journal} {\prd}\ }\textbf {\bibinfo {volume} {102}},\ \bibinfo {eid} {104057} (\bibinfo {year} {2020})},\ \Eprint {http://arxiv.org/abs/2002.07656} {arXiv:2002.07656 [astro-ph.IM]} \BibitemShut {NoStop}%
\bibitem [{\citenamefont {{Green}}\ and\ \citenamefont {{Gair}}(2020)}]{green_gw150914}%
  \BibitemOpen
  \bibfield  {author} {\bibinfo {author} {\bibfnamefont {S.~R.}\ \bibnamefont {{Green}}}\ and\ \bibinfo {author} {\bibfnamefont {J.}~\bibnamefont {{Gair}}},\ }\href {\doibase 10.48550/arXiv.2008.03312} {\bibfield  {journal} {\bibinfo  {journal} {arXiv e-prints}\ ,\ \bibinfo {eid} {arXiv:2008.03312}} (\bibinfo {year} {2020})},\ \Eprint {http://arxiv.org/abs/2008.03312} {arXiv:2008.03312 [astro-ph.IM]} \BibitemShut {NoStop}%
\bibitem [{\citenamefont {{Dax}}\ \emph {et~al.}(2021{\natexlab{b}})\citenamefont {{Dax}}, \citenamefont {{Green}}, \citenamefont {{Gair}}, \citenamefont {{Deistler}}, \citenamefont {{Sch{\"o}lkopf}},\ and\ \citenamefont {{Macke}}}]{dax_gnpe}%
  \BibitemOpen
  \bibfield  {author} {\bibinfo {author} {\bibfnamefont {M.}~\bibnamefont {{Dax}}}, \bibinfo {author} {\bibfnamefont {S.~R.}\ \bibnamefont {{Green}}}, \bibinfo {author} {\bibfnamefont {J.}~\bibnamefont {{Gair}}}, \bibinfo {author} {\bibfnamefont {M.}~\bibnamefont {{Deistler}}}, \bibinfo {author} {\bibfnamefont {B.}~\bibnamefont {{Sch{\"o}lkopf}}}, \ and\ \bibinfo {author} {\bibfnamefont {J.~H.}\ \bibnamefont {{Macke}}},\ }\href {\doibase 10.48550/arXiv.2111.13139} {\bibfield  {journal} {\bibinfo  {journal} {arXiv e-prints}\ ,\ \bibinfo {eid} {arXiv:2111.13139}} (\bibinfo {year} {2021}{\natexlab{b}})},\ \Eprint {http://arxiv.org/abs/2111.13139} {arXiv:2111.13139 [cs.LG]} \BibitemShut {NoStop}%
\bibitem [{\citenamefont {{Wildberger}}\ \emph {et~al.}(2023)\citenamefont {{Wildberger}}, \citenamefont {{Dax}}, \citenamefont {{Buchholz}}, \citenamefont {{Green}}, \citenamefont {{Macke}},\ and\ \citenamefont {{Sch{\"o}lkopf}}}]{fmpe_gw}%
  \BibitemOpen
  \bibfield  {author} {\bibinfo {author} {\bibfnamefont {J.~B.}\ \bibnamefont {{Wildberger}}}, \bibinfo {author} {\bibfnamefont {M.}~\bibnamefont {{Dax}}}, \bibinfo {author} {\bibfnamefont {S.}~\bibnamefont {{Buchholz}}}, \bibinfo {author} {\bibfnamefont {S.~R.}\ \bibnamefont {{Green}}}, \bibinfo {author} {\bibfnamefont {J.}~\bibnamefont {{Macke}}}, \ and\ \bibinfo {author} {\bibfnamefont {B.}~\bibnamefont {{Sch{\"o}lkopf}}},\ }in\ \href {\doibase 10.48550/arXiv.2305.17161} {\emph {\bibinfo {booktitle} {Machine Learning for Astrophysics}}}\ (\bibinfo {year} {2023})\ p.~\bibinfo {pages} {34},\ \Eprint {http://arxiv.org/abs/2305.17161} {arXiv:2305.17161 [cs.LG]} \BibitemShut {NoStop}%
\bibitem [{\citenamefont {{Khan}}\ \emph {et~al.}(2016)\citenamefont {{Khan}}, \citenamefont {{Husa}}, \citenamefont {{Hannam}}, \citenamefont {{Ohme}}, \citenamefont {{P{\"u}rrer}}, \citenamefont {{Forteza}},\ and\ \citenamefont {{Boh{\'e}}}}]{khan_imr}%
  \BibitemOpen
  \bibfield  {author} {\bibinfo {author} {\bibfnamefont {S.}~\bibnamefont {{Khan}}}, \bibinfo {author} {\bibfnamefont {S.}~\bibnamefont {{Husa}}}, \bibinfo {author} {\bibfnamefont {M.}~\bibnamefont {{Hannam}}}, \bibinfo {author} {\bibfnamefont {F.}~\bibnamefont {{Ohme}}}, \bibinfo {author} {\bibfnamefont {M.}~\bibnamefont {{P{\"u}rrer}}}, \bibinfo {author} {\bibfnamefont {X.~J.}\ \bibnamefont {{Forteza}}}, \ and\ \bibinfo {author} {\bibfnamefont {A.}~\bibnamefont {{Boh{\'e}}}},\ }\href {\doibase 10.1103/PhysRevD.93.044007} {\bibfield  {journal} {\bibinfo  {journal} {\prd}\ }\textbf {\bibinfo {volume} {93}},\ \bibinfo {eid} {044007} (\bibinfo {year} {2016})},\ \Eprint {http://arxiv.org/abs/1508.07253} {arXiv:1508.07253 [gr-qc]} \BibitemShut {NoStop}%
\bibitem [{\citenamefont {{Ashton}}\ \emph {et~al.}(2019)\citenamefont {{Ashton}}, \citenamefont {{H{\"u}bner}}, \citenamefont {{Lasky}}, \citenamefont {{Talbot}}, \citenamefont {{Ackley}}, \citenamefont {{Biscoveanu}}, \citenamefont {{Chu}}, \citenamefont {{Divakarla}}, \citenamefont {{Easter}}, \citenamefont {{Goncharov}}, \citenamefont {{Hernandez Vivanco}}, \citenamefont {{Harms}}, \citenamefont {{Lower}}, \citenamefont {{Meadors}}, \citenamefont {{Melchor}}, \citenamefont {{Payne}}, \citenamefont {{Pitkin}}, \citenamefont {{Powell}}, \citenamefont {{Sarin}}, \citenamefont {{Smith}},\ and\ \citenamefont {{Thrane}}}]{bibly_ashton}%
  \BibitemOpen
  \bibfield  {author} {\bibinfo {author} {\bibfnamefont {G.}~\bibnamefont {{Ashton}}}, \bibinfo {author} {\bibfnamefont {M.}~\bibnamefont {{H{\"u}bner}}}, \bibinfo {author} {\bibfnamefont {P.~D.}\ \bibnamefont {{Lasky}}}, \bibinfo {author} {\bibfnamefont {C.}~\bibnamefont {{Talbot}}}, \bibinfo {author} {\bibfnamefont {K.}~\bibnamefont {{Ackley}}}, \bibinfo {author} {\bibfnamefont {S.}~\bibnamefont {{Biscoveanu}}}, \bibinfo {author} {\bibfnamefont {Q.}~\bibnamefont {{Chu}}}, \bibinfo {author} {\bibfnamefont {A.}~\bibnamefont {{Divakarla}}}, \bibinfo {author} {\bibfnamefont {P.~J.}\ \bibnamefont {{Easter}}}, \bibinfo {author} {\bibfnamefont {B.}~\bibnamefont {{Goncharov}}}, \bibinfo {author} {\bibfnamefont {F.}~\bibnamefont {{Hernandez Vivanco}}}, \bibinfo {author} {\bibfnamefont {J.}~\bibnamefont {{Harms}}}, \bibinfo {author} {\bibfnamefont {M.~E.}\ \bibnamefont {{Lower}}}, \bibinfo {author} {\bibfnamefont {G.~D.}\ \bibnamefont {{Meadors}}}, \bibinfo {author} {\bibfnamefont {D.}~\bibnamefont {{Melchor}}},
  \bibinfo {author} {\bibfnamefont {E.}~\bibnamefont {{Payne}}}, \bibinfo {author} {\bibfnamefont {M.~D.}\ \bibnamefont {{Pitkin}}}, \bibinfo {author} {\bibfnamefont {J.}~\bibnamefont {{Powell}}}, \bibinfo {author} {\bibfnamefont {N.}~\bibnamefont {{Sarin}}}, \bibinfo {author} {\bibfnamefont {R.~J.~E.}\ \bibnamefont {{Smith}}}, \ and\ \bibinfo {author} {\bibfnamefont {E.}~\bibnamefont {{Thrane}}},\ }\href {\doibase 10.3847/1538-4365/ab06fc} {\bibfield  {journal} {\bibinfo  {journal} {\apjs}\ }\textbf {\bibinfo {volume} {241}},\ \bibinfo {eid} {27} (\bibinfo {year} {2019})},\ \Eprint {http://arxiv.org/abs/1811.02042} {arXiv:1811.02042 [astro-ph.IM]} \BibitemShut {NoStop}%
\bibitem [{\citenamefont {{Romero-Shaw}}\ \emph {et~al.}(2020)\citenamefont {{Romero-Shaw}}, \citenamefont {{Talbot}}, \citenamefont {{Biscoveanu}}, \citenamefont {{D'Emilio}}, \citenamefont {{Ashton}}, \citenamefont {{Berry}}, \citenamefont {{Coughlin}}, \citenamefont {{Galaudage}}, \citenamefont {{Hoy}}, \citenamefont {{H{\"u}bner}}, \citenamefont {{Phukon}}, \citenamefont {{Pitkin}}, \citenamefont {{Rizzo}}, \citenamefont {{Sarin}}, \citenamefont {{Smith}}, \citenamefont {{Stevenson}}, \citenamefont {{Vajpeyi}}, \citenamefont {{Ar{\`e}ne}}, \citenamefont {{Athar}}, \citenamefont {{Banagiri}}, \citenamefont {{Bose}}, \citenamefont {{Carney}}, \citenamefont {{Chatziioannou}}, \citenamefont {{Clark}}, \citenamefont {{Colleoni}}, \citenamefont {{Cotesta}}, \citenamefont {{Edelman}}, \citenamefont {{Estell{\'e}s}}, \citenamefont {{Garc{\'\i}a-Quir{\'o}s}}, \citenamefont {{Ghosh}}, \citenamefont {{Green}}, \citenamefont {{Haster}}, \citenamefont {{Husa}}, \citenamefont {{Keitel}}, \citenamefont {{Kim}},
  \citenamefont {{Hernandez-Vivanco}}, \citenamefont {{Maga{\~n}a Hernandez}}, \citenamefont {{Karathanasis}}, \citenamefont {{Lasky}}, \citenamefont {{De Lillo}}, \citenamefont {{Lower}}, \citenamefont {{Macleod}}, \citenamefont {{Mateu-Lucena}}, \citenamefont {{Miller}}, \citenamefont {{Millhouse}}, \citenamefont {{Morisaki}}, \citenamefont {{Oh}}, \citenamefont {{Ossokine}}, \citenamefont {{Payne}}, \citenamefont {{Powell}}, \citenamefont {{Pratten}}, \citenamefont {{P{\"u}rrer}}, \citenamefont {{Ramos-Buades}}, \citenamefont {{Raymond}}, \citenamefont {{Thrane}}, \citenamefont {{Veitch}}, \citenamefont {{Williams}}, \citenamefont {{Williams}},\ and\ \citenamefont {{Xiao}}}]{bilby_shaw}%
  \BibitemOpen
  \bibfield  {author} {\bibinfo {author} {\bibfnamefont {I.~M.}\ \bibnamefont {{Romero-Shaw}}}, \bibinfo {author} {\bibfnamefont {C.}~\bibnamefont {{Talbot}}}, \bibinfo {author} {\bibfnamefont {S.}~\bibnamefont {{Biscoveanu}}}, \bibinfo {author} {\bibfnamefont {V.}~\bibnamefont {{D'Emilio}}}, \bibinfo {author} {\bibfnamefont {G.}~\bibnamefont {{Ashton}}}, \bibinfo {author} {\bibfnamefont {C.~P.~L.}\ \bibnamefont {{Berry}}}, \bibinfo {author} {\bibfnamefont {S.}~\bibnamefont {{Coughlin}}}, \bibinfo {author} {\bibfnamefont {S.}~\bibnamefont {{Galaudage}}}, \bibinfo {author} {\bibfnamefont {C.}~\bibnamefont {{Hoy}}}, \bibinfo {author} {\bibfnamefont {M.}~\bibnamefont {{H{\"u}bner}}}, \bibinfo {author} {\bibfnamefont {K.~S.}\ \bibnamefont {{Phukon}}}, \bibinfo {author} {\bibfnamefont {M.}~\bibnamefont {{Pitkin}}}, \bibinfo {author} {\bibfnamefont {M.}~\bibnamefont {{Rizzo}}}, \bibinfo {author} {\bibfnamefont {N.}~\bibnamefont {{Sarin}}}, \bibinfo {author} {\bibfnamefont {R.}~\bibnamefont {{Smith}}}, \bibinfo
  {author} {\bibfnamefont {S.}~\bibnamefont {{Stevenson}}}, \bibinfo {author} {\bibfnamefont {A.}~\bibnamefont {{Vajpeyi}}}, \bibinfo {author} {\bibfnamefont {M.}~\bibnamefont {{Ar{\`e}ne}}}, \bibinfo {author} {\bibfnamefont {K.}~\bibnamefont {{Athar}}}, \bibinfo {author} {\bibfnamefont {S.}~\bibnamefont {{Banagiri}}}, \bibinfo {author} {\bibfnamefont {N.}~\bibnamefont {{Bose}}}, \bibinfo {author} {\bibfnamefont {M.}~\bibnamefont {{Carney}}}, \bibinfo {author} {\bibfnamefont {K.}~\bibnamefont {{Chatziioannou}}}, \bibinfo {author} {\bibfnamefont {J.~A.}\ \bibnamefont {{Clark}}}, \bibinfo {author} {\bibfnamefont {M.}~\bibnamefont {{Colleoni}}}, \bibinfo {author} {\bibfnamefont {R.}~\bibnamefont {{Cotesta}}}, \bibinfo {author} {\bibfnamefont {B.}~\bibnamefont {{Edelman}}}, \bibinfo {author} {\bibfnamefont {H.}~\bibnamefont {{Estell{\'e}s}}}, \bibinfo {author} {\bibfnamefont {C.}~\bibnamefont {{Garc{\'\i}a-Quir{\'o}s}}}, \bibinfo {author} {\bibfnamefont {A.}~\bibnamefont {{Ghosh}}}, \bibinfo {author}
  {\bibfnamefont {R.}~\bibnamefont {{Green}}}, \bibinfo {author} {\bibfnamefont {C.~J.}\ \bibnamefont {{Haster}}}, \bibinfo {author} {\bibfnamefont {S.}~\bibnamefont {{Husa}}}, \bibinfo {author} {\bibfnamefont {D.}~\bibnamefont {{Keitel}}}, \bibinfo {author} {\bibfnamefont {A.~X.}\ \bibnamefont {{Kim}}}, \bibinfo {author} {\bibfnamefont {F.}~\bibnamefont {{Hernandez-Vivanco}}}, \bibinfo {author} {\bibfnamefont {I.}~\bibnamefont {{Maga{\~n}a Hernandez}}}, \bibinfo {author} {\bibfnamefont {C.}~\bibnamefont {{Karathanasis}}}, \bibinfo {author} {\bibfnamefont {P.~D.}\ \bibnamefont {{Lasky}}}, \bibinfo {author} {\bibfnamefont {N.}~\bibnamefont {{De Lillo}}}, \bibinfo {author} {\bibfnamefont {M.~E.}\ \bibnamefont {{Lower}}}, \bibinfo {author} {\bibfnamefont {D.}~\bibnamefont {{Macleod}}}, \bibinfo {author} {\bibfnamefont {M.}~\bibnamefont {{Mateu-Lucena}}}, \bibinfo {author} {\bibfnamefont {A.}~\bibnamefont {{Miller}}}, \bibinfo {author} {\bibfnamefont {M.}~\bibnamefont {{Millhouse}}}, \bibinfo {author}
  {\bibfnamefont {S.}~\bibnamefont {{Morisaki}}}, \bibinfo {author} {\bibfnamefont {S.~H.}\ \bibnamefont {{Oh}}}, \bibinfo {author} {\bibfnamefont {S.}~\bibnamefont {{Ossokine}}}, \bibinfo {author} {\bibfnamefont {E.}~\bibnamefont {{Payne}}}, \bibinfo {author} {\bibfnamefont {J.}~\bibnamefont {{Powell}}}, \bibinfo {author} {\bibfnamefont {G.}~\bibnamefont {{Pratten}}}, \bibinfo {author} {\bibfnamefont {M.}~\bibnamefont {{P{\"u}rrer}}}, \bibinfo {author} {\bibfnamefont {A.}~\bibnamefont {{Ramos-Buades}}}, \bibinfo {author} {\bibfnamefont {V.}~\bibnamefont {{Raymond}}}, \bibinfo {author} {\bibfnamefont {E.}~\bibnamefont {{Thrane}}}, \bibinfo {author} {\bibfnamefont {J.}~\bibnamefont {{Veitch}}}, \bibinfo {author} {\bibfnamefont {D.}~\bibnamefont {{Williams}}}, \bibinfo {author} {\bibfnamefont {M.~J.}\ \bibnamefont {{Williams}}}, \ and\ \bibinfo {author} {\bibfnamefont {L.}~\bibnamefont {{Xiao}}},\ }\href {\doibase 10.1093/mnras/staa2850} {\bibfield  {journal} {\bibinfo  {journal} {\mnras}\ }\textbf {\bibinfo
  {volume} {499}},\ \bibinfo {pages} {3295} (\bibinfo {year} {2020})},\ \Eprint {http://arxiv.org/abs/2006.00714} {arXiv:2006.00714 [astro-ph.IM]} \BibitemShut {NoStop}%
\bibitem [{\citenamefont {{Veitch}}\ \emph {et~al.}(2015)\citenamefont {{Veitch}}, \citenamefont {{Raymond}}, \citenamefont {{Farr}}, \citenamefont {{Farr}}, \citenamefont {{Graff}}, \citenamefont {{Vitale}}, \citenamefont {{Aylott}}, \citenamefont {{Blackburn}}, \citenamefont {{Christensen}}, \citenamefont {{Coughlin}}, \citenamefont {{Del Pozzo}}, \citenamefont {{Feroz}}, \citenamefont {{Gair}}, \citenamefont {{Haster}}, \citenamefont {{Kalogera}}, \citenamefont {{Littenberg}}, \citenamefont {{Mandel}}, \citenamefont {{O'Shaughnessy}}, \citenamefont {{Pitkin}}, \citenamefont {{Rodriguez}}, \citenamefont {{R{\"o}ver}}, \citenamefont {{Sidery}}, \citenamefont {{Smith}}, \citenamefont {{Van Der Sluys}}, \citenamefont {{Vecchio}}, \citenamefont {{Vousden}},\ and\ \citenamefont {{Wade}}}]{veitch_2015}%
  \BibitemOpen
  \bibfield  {author} {\bibinfo {author} {\bibfnamefont {J.}~\bibnamefont {{Veitch}}}, \bibinfo {author} {\bibfnamefont {V.}~\bibnamefont {{Raymond}}}, \bibinfo {author} {\bibfnamefont {B.}~\bibnamefont {{Farr}}}, \bibinfo {author} {\bibfnamefont {W.}~\bibnamefont {{Farr}}}, \bibinfo {author} {\bibfnamefont {P.}~\bibnamefont {{Graff}}}, \bibinfo {author} {\bibfnamefont {S.}~\bibnamefont {{Vitale}}}, \bibinfo {author} {\bibfnamefont {B.}~\bibnamefont {{Aylott}}}, \bibinfo {author} {\bibfnamefont {K.}~\bibnamefont {{Blackburn}}}, \bibinfo {author} {\bibfnamefont {N.}~\bibnamefont {{Christensen}}}, \bibinfo {author} {\bibfnamefont {M.}~\bibnamefont {{Coughlin}}}, \bibinfo {author} {\bibfnamefont {W.}~\bibnamefont {{Del Pozzo}}}, \bibinfo {author} {\bibfnamefont {F.}~\bibnamefont {{Feroz}}}, \bibinfo {author} {\bibfnamefont {J.}~\bibnamefont {{Gair}}}, \bibinfo {author} {\bibfnamefont {C.~J.}\ \bibnamefont {{Haster}}}, \bibinfo {author} {\bibfnamefont {V.}~\bibnamefont {{Kalogera}}}, \bibinfo {author}
  {\bibfnamefont {T.}~\bibnamefont {{Littenberg}}}, \bibinfo {author} {\bibfnamefont {I.}~\bibnamefont {{Mandel}}}, \bibinfo {author} {\bibfnamefont {R.}~\bibnamefont {{O'Shaughnessy}}}, \bibinfo {author} {\bibfnamefont {M.}~\bibnamefont {{Pitkin}}}, \bibinfo {author} {\bibfnamefont {C.}~\bibnamefont {{Rodriguez}}}, \bibinfo {author} {\bibfnamefont {C.}~\bibnamefont {{R{\"o}ver}}}, \bibinfo {author} {\bibfnamefont {T.}~\bibnamefont {{Sidery}}}, \bibinfo {author} {\bibfnamefont {R.}~\bibnamefont {{Smith}}}, \bibinfo {author} {\bibfnamefont {M.}~\bibnamefont {{Van Der Sluys}}}, \bibinfo {author} {\bibfnamefont {A.}~\bibnamefont {{Vecchio}}}, \bibinfo {author} {\bibfnamefont {W.}~\bibnamefont {{Vousden}}}, \ and\ \bibinfo {author} {\bibfnamefont {L.}~\bibnamefont {{Wade}}},\ }\href {\doibase 10.1103/PhysRevD.91.042003} {\bibfield  {journal} {\bibinfo  {journal} {\prd}\ }\textbf {\bibinfo {volume} {91}},\ \bibinfo {eid} {042003} (\bibinfo {year} {2015})},\ \Eprint {http://arxiv.org/abs/1409.7215} {arXiv:1409.7215
  [gr-qc]} \BibitemShut {NoStop}%
\bibitem [{\citenamefont {Vitale}\ \emph {et~al.}(2017)\citenamefont {Vitale}, \citenamefont {Lynch}, \citenamefont {Raymond}, \citenamefont {Sturani}, \citenamefont {Veitch},\ and\ \citenamefont {Graff}}]{Vitale:2016avz}%
  \BibitemOpen
  \bibfield  {author} {\bibinfo {author} {\bibfnamefont {S.}~\bibnamefont {Vitale}}, \bibinfo {author} {\bibfnamefont {R.}~\bibnamefont {Lynch}}, \bibinfo {author} {\bibfnamefont {V.}~\bibnamefont {Raymond}}, \bibinfo {author} {\bibfnamefont {R.}~\bibnamefont {Sturani}}, \bibinfo {author} {\bibfnamefont {J.}~\bibnamefont {Veitch}}, \ and\ \bibinfo {author} {\bibfnamefont {P.}~\bibnamefont {Graff}},\ }\href {\doibase 10.1103/PhysRevD.95.064053} {\bibfield  {journal} {\bibinfo  {journal} {Phys. Rev. D}\ }\textbf {\bibinfo {volume} {95}},\ \bibinfo {pages} {064053} (\bibinfo {year} {2017})},\ \Eprint {http://arxiv.org/abs/1611.01122} {arXiv:1611.01122 [gr-qc]} \BibitemShut {NoStop}%
\bibitem [{\citenamefont {{Racine}}(2008)}]{racine_spins}%
  \BibitemOpen
  \bibfield  {author} {\bibinfo {author} {\bibfnamefont {{\'E}.}~\bibnamefont {{Racine}}},\ }\href {\doibase 10.1103/PhysRevD.78.044021} {\bibfield  {journal} {\bibinfo  {journal} {\prd}\ }\textbf {\bibinfo {volume} {78}},\ \bibinfo {eid} {044021} (\bibinfo {year} {2008})},\ \Eprint {http://arxiv.org/abs/0803.1820} {arXiv:0803.1820 [gr-qc]} \BibitemShut {NoStop}%
\bibitem [{\citenamefont {Abbott}\ \emph {et~al.}(2016)\citenamefont {Abbott} \emph {et~al.}}]{LIGOScientific:2016vlm}%
  \BibitemOpen
  \bibfield  {author} {\bibinfo {author} {\bibfnamefont {B.~P.}\ \bibnamefont {Abbott}} \emph {et~al.} (\bibinfo {collaboration} {LIGO Scientific, Virgo}),\ }\href {\doibase 10.1103/PhysRevLett.116.241102} {\bibfield  {journal} {\bibinfo  {journal} {Phys. Rev. Lett.}\ }\textbf {\bibinfo {volume} {116}},\ \bibinfo {pages} {241102} (\bibinfo {year} {2016})},\ \Eprint {http://arxiv.org/abs/1602.03840} {arXiv:1602.03840 [gr-qc]} \BibitemShut {NoStop}%
\bibitem [{\citenamefont {Karathanasis}\ \emph {et~al.}(2023)\citenamefont {Karathanasis}, \citenamefont {Revenu}, \citenamefont {Mukherjee},\ and\ \citenamefont {Stachurski}}]{gwsim}%
  \BibitemOpen
  \bibfield  {author} {\bibinfo {author} {\bibfnamefont {C.}~\bibnamefont {Karathanasis}}, \bibinfo {author} {\bibfnamefont {B.}~\bibnamefont {Revenu}}, \bibinfo {author} {\bibfnamefont {S.}~\bibnamefont {Mukherjee}}, \ and\ \bibinfo {author} {\bibfnamefont {F.}~\bibnamefont {Stachurski}},\ }\href {\doibase 10.1051/0004-6361/202245216} {\bibfield  {journal} {\bibinfo  {journal} {Astron. Astrophys.}\ }\textbf {\bibinfo {volume} {677}},\ \bibinfo {pages} {A124} (\bibinfo {year} {2023})},\ \Eprint {http://arxiv.org/abs/2210.05724} {arXiv:2210.05724 [astro-ph.CO]} \BibitemShut {NoStop}%
\bibitem [{\citenamefont {{Silva}}\ \emph {et~al.}(1998)\citenamefont {{Silva}}, \citenamefont {{Granato}}, \citenamefont {{Bressan}},\ and\ \citenamefont {{Danese}}}]{silva1998}%
  \BibitemOpen
  \bibfield  {author} {\bibinfo {author} {\bibfnamefont {L.}~\bibnamefont {{Silva}}}, \bibinfo {author} {\bibfnamefont {G.~L.}\ \bibnamefont {{Granato}}}, \bibinfo {author} {\bibfnamefont {A.}~\bibnamefont {{Bressan}}}, \ and\ \bibinfo {author} {\bibfnamefont {L.}~\bibnamefont {{Danese}}},\ }\href {\doibase 10.1086/306476} {\bibfield  {journal} {\bibinfo  {journal} {\apj}\ }\textbf {\bibinfo {volume} {509}},\ \bibinfo {pages} {103} (\bibinfo {year} {1998})}\BibitemShut {NoStop}%
\bibitem [{\citenamefont {Crocce}\ \emph {et~al.}(2016)\citenamefont {Crocce}, \citenamefont {Carretero}, \citenamefont {Bauer}, \citenamefont {Ross}, \citenamefont {Sevilla-Noarbe}, \citenamefont {Giannantonio}, \citenamefont {Sobreira}, \citenamefont {Sanchez}, \citenamefont {Gaztanaga}, \citenamefont {Kind} \emph {et~al.}}]{crocce2016galaxy}%
  \BibitemOpen
  \bibfield  {author} {\bibinfo {author} {\bibfnamefont {M.}~\bibnamefont {Crocce}}, \bibinfo {author} {\bibfnamefont {J.}~\bibnamefont {Carretero}}, \bibinfo {author} {\bibfnamefont {A.~H.}\ \bibnamefont {Bauer}}, \bibinfo {author} {\bibfnamefont {A.}~\bibnamefont {Ross}}, \bibinfo {author} {\bibfnamefont {I.}~\bibnamefont {Sevilla-Noarbe}}, \bibinfo {author} {\bibfnamefont {T.}~\bibnamefont {Giannantonio}}, \bibinfo {author} {\bibfnamefont {F.}~\bibnamefont {Sobreira}}, \bibinfo {author} {\bibfnamefont {J.}~\bibnamefont {Sanchez}}, \bibinfo {author} {\bibfnamefont {E.}~\bibnamefont {Gaztanaga}}, \bibinfo {author} {\bibfnamefont {M.~C.}\ \bibnamefont {Kind}},  \emph {et~al.},\ }\href@noop {} {\bibfield  {journal} {\bibinfo  {journal} {Monthly Notices of the Royal Astronomical Society}\ }\textbf {\bibinfo {volume} {455}},\ \bibinfo {pages} {4301} (\bibinfo {year} {2016})}\BibitemShut {NoStop}%
\bibitem [{\citenamefont {Trayford}\ \emph {et~al.}(2015)\citenamefont {Trayford}, \citenamefont {Theuns}, \citenamefont {Bower}, \citenamefont {Schaye}, \citenamefont {Furlong}, \citenamefont {Schaller}, \citenamefont {Frenk}, \citenamefont {Crain}, \citenamefont {Vecchia},\ and\ \citenamefont {McCarthy}}]{trayford2015}%
  \BibitemOpen
  \bibfield  {author} {\bibinfo {author} {\bibfnamefont {J.~W.}\ \bibnamefont {Trayford}}, \bibinfo {author} {\bibfnamefont {T.}~\bibnamefont {Theuns}}, \bibinfo {author} {\bibfnamefont {R.~G.}\ \bibnamefont {Bower}}, \bibinfo {author} {\bibfnamefont {J.}~\bibnamefont {Schaye}}, \bibinfo {author} {\bibfnamefont {M.}~\bibnamefont {Furlong}}, \bibinfo {author} {\bibfnamefont {M.}~\bibnamefont {Schaller}}, \bibinfo {author} {\bibfnamefont {C.~S.}\ \bibnamefont {Frenk}}, \bibinfo {author} {\bibfnamefont {R.~A.}\ \bibnamefont {Crain}}, \bibinfo {author} {\bibfnamefont {C.~D.}\ \bibnamefont {Vecchia}}, \ and\ \bibinfo {author} {\bibfnamefont {I.~G.}\ \bibnamefont {McCarthy}},\ }\href {\doibase 10.1093/mnras/stv1461} {\bibfield  {journal} {\bibinfo  {journal} {Monthly Notices of the Royal Astronomical Society}\ }\textbf {\bibinfo {volume} {452}},\ \bibinfo {pages} {2879} (\bibinfo {year} {2015})},\ \Eprint {http://arxiv.org/abs/https://academic.oup.com/mnras/article-pdf/452/3/2879/4922335/stv1461.pdf}
  {https://academic.oup.com/mnras/article-pdf/452/3/2879/4922335/stv1461.pdf} \BibitemShut {NoStop}%
\bibitem [{\citenamefont {{Dav{\'e}}}\ \emph {et~al.}(2019)\citenamefont {{Dav{\'e}}}, \citenamefont {{Angl{\'e}s-Alc{\'a}zar}}, \citenamefont {{Narayanan}}, \citenamefont {{Li}}, \citenamefont {{Rafieferantsoa}},\ and\ \citenamefont {{Appleby}}}]{dave_simba_2019}%
  \BibitemOpen
  \bibfield  {author} {\bibinfo {author} {\bibfnamefont {R.}~\bibnamefont {{Dav{\'e}}}}, \bibinfo {author} {\bibfnamefont {D.}~\bibnamefont {{Angl{\'e}s-Alc{\'a}zar}}}, \bibinfo {author} {\bibfnamefont {D.}~\bibnamefont {{Narayanan}}}, \bibinfo {author} {\bibfnamefont {Q.}~\bibnamefont {{Li}}}, \bibinfo {author} {\bibfnamefont {M.~H.}\ \bibnamefont {{Rafieferantsoa}}}, \ and\ \bibinfo {author} {\bibfnamefont {S.}~\bibnamefont {{Appleby}}},\ }\href {\doibase 10.1093/mnras/stz937} {\bibfield  {journal} {\bibinfo  {journal} {\mnras}\ }\textbf {\bibinfo {volume} {486}},\ \bibinfo {pages} {2827} (\bibinfo {year} {2019})},\ \Eprint {http://arxiv.org/abs/1901.10203} {arXiv:1901.10203 [astro-ph.GA]} \BibitemShut {NoStop}%
\bibitem [{\citenamefont {{Bruzual}}\ and\ \citenamefont {{Charlot}}(2003)}]{bruzual_charlot_2003}%
  \BibitemOpen
  \bibfield  {author} {\bibinfo {author} {\bibfnamefont {G.}~\bibnamefont {{Bruzual}}}\ and\ \bibinfo {author} {\bibfnamefont {S.}~\bibnamefont {{Charlot}}},\ }\href {\doibase 10.1046/j.1365-8711.2003.06897.x} {\bibfield  {journal} {\bibinfo  {journal} {\mnras}\ }\textbf {\bibinfo {volume} {344}},\ \bibinfo {pages} {1000} (\bibinfo {year} {2003})},\ \Eprint {http://arxiv.org/abs/astro-ph/0309134} {arXiv:astro-ph/0309134 [astro-ph]} \BibitemShut {NoStop}%
\bibitem [{\citenamefont {{Carr}}\ \emph {et~al.}(2023)\citenamefont {{Carr}}, \citenamefont {{Bryan}}, \citenamefont {{Fielding}}, \citenamefont {{Pandya}},\ and\ \citenamefont {{Somerville}}}]{carr2023}%
  \BibitemOpen
  \bibfield  {author} {\bibinfo {author} {\bibfnamefont {C.}~\bibnamefont {{Carr}}}, \bibinfo {author} {\bibfnamefont {G.~L.}\ \bibnamefont {{Bryan}}}, \bibinfo {author} {\bibfnamefont {D.~B.}\ \bibnamefont {{Fielding}}}, \bibinfo {author} {\bibfnamefont {V.}~\bibnamefont {{Pandya}}}, \ and\ \bibinfo {author} {\bibfnamefont {R.~S.}\ \bibnamefont {{Somerville}}},\ }\href {\doibase 10.3847/1538-4357/acc4c7} {\bibfield  {journal} {\bibinfo  {journal} {\apj}\ }\textbf {\bibinfo {volume} {949}},\ \bibinfo {eid} {21} (\bibinfo {year} {2023})},\ \Eprint {http://arxiv.org/abs/2211.05115} {arXiv:2211.05115 [astro-ph.GA]} \BibitemShut {NoStop}%
\bibitem [{\citenamefont {{Pandya}}\ \emph {et~al.}(2023)\citenamefont {{Pandya}}, \citenamefont {{Fielding}}, \citenamefont {{Bryan}}, \citenamefont {{Carr}}, \citenamefont {{Somerville}}, \citenamefont {{Stern}}, \citenamefont {{Faucher-Gigu{\`e}re}}, \citenamefont {{Hafen}}, \citenamefont {{Angl{\'e}s-Alc{\'a}zar}},\ and\ \citenamefont {{Forbes}}}]{pandya23_unified}%
  \BibitemOpen
  \bibfield  {author} {\bibinfo {author} {\bibfnamefont {V.}~\bibnamefont {{Pandya}}}, \bibinfo {author} {\bibfnamefont {D.~B.}\ \bibnamefont {{Fielding}}}, \bibinfo {author} {\bibfnamefont {G.~L.}\ \bibnamefont {{Bryan}}}, \bibinfo {author} {\bibfnamefont {C.}~\bibnamefont {{Carr}}}, \bibinfo {author} {\bibfnamefont {R.~S.}\ \bibnamefont {{Somerville}}}, \bibinfo {author} {\bibfnamefont {J.}~\bibnamefont {{Stern}}}, \bibinfo {author} {\bibfnamefont {C.-A.}\ \bibnamefont {{Faucher-Gigu{\`e}re}}}, \bibinfo {author} {\bibfnamefont {Z.}~\bibnamefont {{Hafen}}}, \bibinfo {author} {\bibfnamefont {D.}~\bibnamefont {{Angl{\'e}s-Alc{\'a}zar}}}, \ and\ \bibinfo {author} {\bibfnamefont {J.~C.}\ \bibnamefont {{Forbes}}},\ }\href {\doibase 10.3847/1538-4357/acf3ea} {\bibfield  {journal} {\bibinfo  {journal} {\apj}\ }\textbf {\bibinfo {volume} {956}},\ \bibinfo {eid} {118} (\bibinfo {year} {2023})},\ \Eprint {http://arxiv.org/abs/2211.09755} {arXiv:2211.09755 [astro-ph.GA]} \BibitemShut {NoStop}%
\bibitem [{\citenamefont {{Pillepich}}\ \emph {et~al.}(2018)\citenamefont {{Pillepich}}, \citenamefont {{Springel}}, \citenamefont {{Nelson}}, \citenamefont {{Genel}}, \citenamefont {{Naiman}}, \citenamefont {{Pakmor}}, \citenamefont {{Hernquist}}, \citenamefont {{Torrey}}, \citenamefont {{Vogelsberger}}, \citenamefont {{Weinberger}},\ and\ \citenamefont {{Marinacci}}}]{pillepich2018one}%
  \BibitemOpen
  \bibfield  {author} {\bibinfo {author} {\bibfnamefont {A.}~\bibnamefont {{Pillepich}}}, \bibinfo {author} {\bibfnamefont {V.}~\bibnamefont {{Springel}}}, \bibinfo {author} {\bibfnamefont {D.}~\bibnamefont {{Nelson}}}, \bibinfo {author} {\bibfnamefont {S.}~\bibnamefont {{Genel}}}, \bibinfo {author} {\bibfnamefont {J.}~\bibnamefont {{Naiman}}}, \bibinfo {author} {\bibfnamefont {R.}~\bibnamefont {{Pakmor}}}, \bibinfo {author} {\bibfnamefont {L.}~\bibnamefont {{Hernquist}}}, \bibinfo {author} {\bibfnamefont {P.}~\bibnamefont {{Torrey}}}, \bibinfo {author} {\bibfnamefont {M.}~\bibnamefont {{Vogelsberger}}}, \bibinfo {author} {\bibfnamefont {R.}~\bibnamefont {{Weinberger}}}, \ and\ \bibinfo {author} {\bibfnamefont {F.}~\bibnamefont {{Marinacci}}},\ }\href {\doibase 10.1093/mnras/stx2656} {\bibfield  {journal} {\bibinfo  {journal} {\mnras}\ }\textbf {\bibinfo {volume} {473}},\ \bibinfo {pages} {4077} (\bibinfo {year} {2018})},\ \Eprint {http://arxiv.org/abs/1703.02970} {arXiv:1703.02970 [astro-ph.GA]} \BibitemShut
  {NoStop}%
\bibitem [{\citenamefont {Makinen}\ \emph {et~al.}(2023)\citenamefont {Makinen}, \citenamefont {Alsing},\ and\ \citenamefont {Wandelt}}]{makinen2023fishnets}%
  \BibitemOpen
  \bibfield  {author} {\bibinfo {author} {\bibfnamefont {T.~L.}\ \bibnamefont {Makinen}}, \bibinfo {author} {\bibfnamefont {J.}~\bibnamefont {Alsing}}, \ and\ \bibinfo {author} {\bibfnamefont {B.~D.}\ \bibnamefont {Wandelt}},\ }\href@noop {} {\enquote {\bibinfo {title} {Fishnets: Information-optimal, scalable aggregation for sets and graphs},}\ } (\bibinfo {year} {2023}),\ \Eprint {http://arxiv.org/abs/2310.03812} {arXiv:2310.03812 [cs.LG]} \BibitemShut {NoStop}%
\bibitem [{\citenamefont {Durkan}\ \emph {et~al.}(2019)\citenamefont {Durkan}, \citenamefont {Bekasov}, \citenamefont {Murray},\ and\ \citenamefont {Papamakarios}}]{durkan2019neural}%
  \BibitemOpen
  \bibfield  {author} {\bibinfo {author} {\bibfnamefont {C.}~\bibnamefont {Durkan}}, \bibinfo {author} {\bibfnamefont {A.}~\bibnamefont {Bekasov}}, \bibinfo {author} {\bibfnamefont {I.}~\bibnamefont {Murray}}, \ and\ \bibinfo {author} {\bibfnamefont {G.}~\bibnamefont {Papamakarios}},\ }\href@noop {} {\bibfield  {journal} {\bibinfo  {journal} {Advances in neural information processing systems}\ }\textbf {\bibinfo {volume} {32}} (\bibinfo {year} {2019})}\BibitemShut {NoStop}%
\bibitem [{\citenamefont {Meng}\ \emph {et~al.}(2020)\citenamefont {Meng}, \citenamefont {Song}, \citenamefont {Song},\ and\ \citenamefont {Ermon}}]{meng2020gaussianization}%
  \BibitemOpen
  \bibfield  {author} {\bibinfo {author} {\bibfnamefont {C.}~\bibnamefont {Meng}}, \bibinfo {author} {\bibfnamefont {Y.}~\bibnamefont {Song}}, \bibinfo {author} {\bibfnamefont {J.}~\bibnamefont {Song}}, \ and\ \bibinfo {author} {\bibfnamefont {S.}~\bibnamefont {Ermon}},\ }in\ \href@noop {} {\emph {\bibinfo {booktitle} {International Conference on Artificial Intelligence and Statistics}}}\ (\bibinfo {organization} {PMLR},\ \bibinfo {year} {2020})\ pp.\ \bibinfo {pages} {4336--4345}\BibitemShut {NoStop}%
\bibitem [{\citenamefont {Cannon}\ \emph {et~al.}(2022)\citenamefont {Cannon}, \citenamefont {Ward},\ and\ \citenamefont {Schmon}}]{cannon2022investigating}%
  \BibitemOpen
  \bibfield  {author} {\bibinfo {author} {\bibfnamefont {P.}~\bibnamefont {Cannon}}, \bibinfo {author} {\bibfnamefont {D.}~\bibnamefont {Ward}}, \ and\ \bibinfo {author} {\bibfnamefont {S.~M.}\ \bibnamefont {Schmon}},\ }\href@noop {} {\bibfield  {journal} {\bibinfo  {journal} {arXiv preprint arXiv:2209.01845}\ } (\bibinfo {year} {2022})}\BibitemShut {NoStop}%
\bibitem [{\citenamefont {Kelly}\ \emph {et~al.}(2023)\citenamefont {Kelly}, \citenamefont {Nott}, \citenamefont {Frazier}, \citenamefont {Warne},\ and\ \citenamefont {Drovandi}}]{kelly2023misspecification}%
  \BibitemOpen
  \bibfield  {author} {\bibinfo {author} {\bibfnamefont {R.~P.}\ \bibnamefont {Kelly}}, \bibinfo {author} {\bibfnamefont {D.~J.}\ \bibnamefont {Nott}}, \bibinfo {author} {\bibfnamefont {D.~T.}\ \bibnamefont {Frazier}}, \bibinfo {author} {\bibfnamefont {D.~J.}\ \bibnamefont {Warne}}, \ and\ \bibinfo {author} {\bibfnamefont {C.}~\bibnamefont {Drovandi}},\ }\href@noop {} {\bibfield  {journal} {\bibinfo  {journal} {arXiv preprint arXiv:2301.13368}\ } (\bibinfo {year} {2023})}\BibitemShut {NoStop}%
\bibitem [{\citenamefont {Huang}\ \emph {et~al.}(2023)\citenamefont {Huang}, \citenamefont {Bharti}, \citenamefont {Souza}, \citenamefont {Acerbi},\ and\ \citenamefont {Kaski}}]{huang2023learning}%
  \BibitemOpen
  \bibfield  {author} {\bibinfo {author} {\bibfnamefont {D.}~\bibnamefont {Huang}}, \bibinfo {author} {\bibfnamefont {A.}~\bibnamefont {Bharti}}, \bibinfo {author} {\bibfnamefont {A.}~\bibnamefont {Souza}}, \bibinfo {author} {\bibfnamefont {L.}~\bibnamefont {Acerbi}}, \ and\ \bibinfo {author} {\bibfnamefont {S.}~\bibnamefont {Kaski}},\ }\href@noop {} {\bibfield  {journal} {\bibinfo  {journal} {arXiv preprint arXiv:2305.15871}\ } (\bibinfo {year} {2023})}\BibitemShut {NoStop}%
\bibitem [{\citenamefont {{Leclercq}}(2022)}]{Leclercq2022}%
  \BibitemOpen
  \bibfield  {author} {\bibinfo {author} {\bibfnamefont {F.}~\bibnamefont {{Leclercq}}},\ }\href {\doibase 10.48550/arXiv.2209.11057} {\bibfield  {journal} {\bibinfo  {journal} {arXiv e-prints}\ ,\ \bibinfo {eid} {arXiv:2209.11057}} (\bibinfo {year} {2022})},\ \Eprint {http://arxiv.org/abs/2209.11057} {arXiv:2209.11057 [stat.ME]} \BibitemShut {NoStop}%
\bibitem [{\citenamefont {Handley}\ \emph {et~al.}(2015)\citenamefont {Handley}, \citenamefont {Hobson},\ and\ \citenamefont {Lasenby}}]{handley2015polychord}%
  \BibitemOpen
  \bibfield  {author} {\bibinfo {author} {\bibfnamefont {W.}~\bibnamefont {Handley}}, \bibinfo {author} {\bibfnamefont {M.}~\bibnamefont {Hobson}}, \ and\ \bibinfo {author} {\bibfnamefont {A.}~\bibnamefont {Lasenby}},\ }\href@noop {} {\bibfield  {journal} {\bibinfo  {journal} {Monthly Notices of the Royal Astronomical Society: Letters}\ }\textbf {\bibinfo {volume} {450}},\ \bibinfo {pages} {L61} (\bibinfo {year} {2015})}\BibitemShut {NoStop}%
\bibitem [{\citenamefont {McEwen}\ \emph {et~al.}(2021)\citenamefont {McEwen}, \citenamefont {Wallis}, \citenamefont {Price},\ and\ \citenamefont {Docherty}}]{harmonic}%
  \BibitemOpen
  \bibfield  {author} {\bibinfo {author} {\bibfnamefont {J.~D.}\ \bibnamefont {McEwen}}, \bibinfo {author} {\bibfnamefont {C.~G.~R.}\ \bibnamefont {Wallis}}, \bibinfo {author} {\bibfnamefont {M.~A.}\ \bibnamefont {Price}}, \ and\ \bibinfo {author} {\bibfnamefont {M.~M.}\ \bibnamefont {Docherty}},\ }\href@noop {} {\bibfield  {journal} {\bibinfo  {journal} {ArXiv}\ } (\bibinfo {year} {2021})},\ \Eprint {http://arxiv.org/abs/arXiv:2111.12720} {arXiv:2111.12720} \BibitemShut {NoStop}%
\bibitem [{\citenamefont {Spurio~Mancini}\ \emph {et~al.}(2023)\citenamefont {Spurio~Mancini}, \citenamefont {Docherty}, \citenamefont {Price},\ and\ \citenamefont {McEwen}}]{spurio2023bayesian}%
  \BibitemOpen
  \bibfield  {author} {\bibinfo {author} {\bibfnamefont {A.}~\bibnamefont {Spurio~Mancini}}, \bibinfo {author} {\bibfnamefont {M.}~\bibnamefont {Docherty}}, \bibinfo {author} {\bibfnamefont {M.}~\bibnamefont {Price}}, \ and\ \bibinfo {author} {\bibfnamefont {J.}~\bibnamefont {McEwen}},\ }\href@noop {} {\bibfield  {journal} {\bibinfo  {journal} {RAS Techniques and Instruments}\ }\textbf {\bibinfo {volume} {2}},\ \bibinfo {pages} {710} (\bibinfo {year} {2023})}\BibitemShut {NoStop}%
\bibitem [{\citenamefont {Jeffrey}\ and\ \citenamefont {Wandelt}(2024)}]{jeffrey2024evidence}%
  \BibitemOpen
  \bibfield  {author} {\bibinfo {author} {\bibfnamefont {N.}~\bibnamefont {Jeffrey}}\ and\ \bibinfo {author} {\bibfnamefont {B.~D.}\ \bibnamefont {Wandelt}},\ }\href@noop {} {\bibfield  {journal} {\bibinfo  {journal} {Machine Learning: Science and Technology}\ }\textbf {\bibinfo {volume} {5}},\ \bibinfo {pages} {015008} (\bibinfo {year} {2024})}\BibitemShut {NoStop}%
\bibitem [{\citenamefont {Masserano}\ \emph {et~al.}(2022)\citenamefont {Masserano}, \citenamefont {Dorigo}, \citenamefont {Izbicki}, \citenamefont {Kuusela},\ and\ \citenamefont {Lee}}]{masserano2022simulation}%
  \BibitemOpen
  \bibfield  {author} {\bibinfo {author} {\bibfnamefont {L.}~\bibnamefont {Masserano}}, \bibinfo {author} {\bibfnamefont {T.}~\bibnamefont {Dorigo}}, \bibinfo {author} {\bibfnamefont {R.}~\bibnamefont {Izbicki}}, \bibinfo {author} {\bibfnamefont {M.}~\bibnamefont {Kuusela}}, \ and\ \bibinfo {author} {\bibfnamefont {A.~B.}\ \bibnamefont {Lee}},\ }\href@noop {} {\bibfield  {journal} {\bibinfo  {journal} {arXiv preprint arXiv:2205.15680}\ } (\bibinfo {year} {2022})}\BibitemShut {NoStop}%
\bibitem [{\citenamefont {Belkin}\ \emph {et~al.}(2019)\citenamefont {Belkin}, \citenamefont {Hsu}, \citenamefont {Ma},\ and\ \citenamefont {Mandal}}]{belkin2019reconciling}%
  \BibitemOpen
  \bibfield  {author} {\bibinfo {author} {\bibfnamefont {M.}~\bibnamefont {Belkin}}, \bibinfo {author} {\bibfnamefont {D.}~\bibnamefont {Hsu}}, \bibinfo {author} {\bibfnamefont {S.}~\bibnamefont {Ma}}, \ and\ \bibinfo {author} {\bibfnamefont {S.}~\bibnamefont {Mandal}},\ }\href@noop {} {\bibfield  {journal} {\bibinfo  {journal} {Proceedings of the National Academy of Sciences}\ }\textbf {\bibinfo {volume} {116}},\ \bibinfo {pages} {15849} (\bibinfo {year} {2019})}\BibitemShut {NoStop}%
\bibitem [{\citenamefont {Akiba}\ \emph {et~al.}(2019)\citenamefont {Akiba}, \citenamefont {Sano}, \citenamefont {Yanase}, \citenamefont {Ohta},\ and\ \citenamefont {Koyama}}]{akiba2019optuna}%
  \BibitemOpen
  \bibfield  {author} {\bibinfo {author} {\bibfnamefont {T.}~\bibnamefont {Akiba}}, \bibinfo {author} {\bibfnamefont {S.}~\bibnamefont {Sano}}, \bibinfo {author} {\bibfnamefont {T.}~\bibnamefont {Yanase}}, \bibinfo {author} {\bibfnamefont {T.}~\bibnamefont {Ohta}}, \ and\ \bibinfo {author} {\bibfnamefont {M.}~\bibnamefont {Koyama}},\ }in\ \href@noop {} {\emph {\bibinfo {booktitle} {Proceedings of the 25th ACM SIGKDD international conference on knowledge discovery \& data mining}}}\ (\bibinfo {year} {2019})\ pp.\ \bibinfo {pages} {2623--2631}\BibitemShut {NoStop}%
\bibitem [{\citenamefont {Jasche}\ and\ \citenamefont {Wandelt}(2013)}]{jasche2013bayesian}%
  \BibitemOpen
  \bibfield  {author} {\bibinfo {author} {\bibfnamefont {J.}~\bibnamefont {Jasche}}\ and\ \bibinfo {author} {\bibfnamefont {B.~D.}\ \bibnamefont {Wandelt}},\ }\href@noop {} {\bibfield  {journal} {\bibinfo  {journal} {Monthly Notices of the Royal Astronomical Society}\ }\textbf {\bibinfo {volume} {432}},\ \bibinfo {pages} {894} (\bibinfo {year} {2013})}\BibitemShut {NoStop}%
\bibitem [{\citenamefont {de~Andres}\ \emph {et~al.}(2024)\citenamefont {de~Andres}, \citenamefont {Cui}, \citenamefont {Yepes}, \citenamefont {De~Petris}, \citenamefont {Ferragamo}, \citenamefont {De~Luca}, \citenamefont {Aversano},\ and\ \citenamefont {Rennehan}}]{de2024three}%
  \BibitemOpen
  \bibfield  {author} {\bibinfo {author} {\bibfnamefont {D.}~\bibnamefont {de~Andres}}, \bibinfo {author} {\bibfnamefont {W.}~\bibnamefont {Cui}}, \bibinfo {author} {\bibfnamefont {G.}~\bibnamefont {Yepes}}, \bibinfo {author} {\bibfnamefont {M.}~\bibnamefont {De~Petris}}, \bibinfo {author} {\bibfnamefont {A.}~\bibnamefont {Ferragamo}}, \bibinfo {author} {\bibfnamefont {F.}~\bibnamefont {De~Luca}}, \bibinfo {author} {\bibfnamefont {G.}~\bibnamefont {Aversano}}, \ and\ \bibinfo {author} {\bibfnamefont {D.}~\bibnamefont {Rennehan}},\ }\href@noop {} {\bibfield  {journal} {\bibinfo  {journal} {Monthly Notices of the Royal Astronomical Society}\ ,\ \bibinfo {pages} {stae071}} (\bibinfo {year} {2024})}\BibitemShut {NoStop}%
\bibitem [{\citenamefont {Jeffrey}\ and\ \citenamefont {Wandelt}(2020)}]{jeffrey2020solving}%
  \BibitemOpen
  \bibfield  {author} {\bibinfo {author} {\bibfnamefont {N.}~\bibnamefont {Jeffrey}}\ and\ \bibinfo {author} {\bibfnamefont {B.~D.}\ \bibnamefont {Wandelt}},\ }\href@noop {} {\bibfield  {journal} {\bibinfo  {journal} {arXiv preprint arXiv:2011.05991}\ } (\bibinfo {year} {2020})}\BibitemShut {NoStop}%
\bibitem [{\citenamefont {Legin}\ \emph {et~al.}(2024)\citenamefont {Legin}, \citenamefont {Ho}, \citenamefont {Lemos}, \citenamefont {Perreault-Levasseur}, \citenamefont {Ho}, \citenamefont {Hezaveh},\ and\ \citenamefont {Wandelt}}]{legin2024posterior}%
  \BibitemOpen
  \bibfield  {author} {\bibinfo {author} {\bibfnamefont {R.}~\bibnamefont {Legin}}, \bibinfo {author} {\bibfnamefont {M.}~\bibnamefont {Ho}}, \bibinfo {author} {\bibfnamefont {P.}~\bibnamefont {Lemos}}, \bibinfo {author} {\bibfnamefont {L.}~\bibnamefont {Perreault-Levasseur}}, \bibinfo {author} {\bibfnamefont {S.}~\bibnamefont {Ho}}, \bibinfo {author} {\bibfnamefont {Y.}~\bibnamefont {Hezaveh}}, \ and\ \bibinfo {author} {\bibfnamefont {B.}~\bibnamefont {Wandelt}},\ }\href@noop {} {\bibfield  {journal} {\bibinfo  {journal} {Monthly Notices of the Royal Astronomical Society: Letters}\ }\textbf {\bibinfo {volume} {527}},\ \bibinfo {pages} {L173} (\bibinfo {year} {2024})}\BibitemShut {NoStop}%
\bibitem [{\citenamefont {Choustikov}\ \emph {et~al.}(2024)\citenamefont {Choustikov}, \citenamefont {Stiskalek}, \citenamefont {Saxena}, \citenamefont {Katz}, \citenamefont {Devrient},\ and\ \citenamefont {Slyz}}]{choustikov2024inferring}%
  \BibitemOpen
  \bibfield  {author} {\bibinfo {author} {\bibfnamefont {N.}~\bibnamefont {Choustikov}}, \bibinfo {author} {\bibfnamefont {R.}~\bibnamefont {Stiskalek}}, \bibinfo {author} {\bibfnamefont {A.}~\bibnamefont {Saxena}}, \bibinfo {author} {\bibfnamefont {H.}~\bibnamefont {Katz}}, \bibinfo {author} {\bibfnamefont {J.}~\bibnamefont {Devrient}}, \ and\ \bibinfo {author} {\bibfnamefont {A.}~\bibnamefont {Slyz}},\ }\href@noop {} {\bibfield  {journal} {\bibinfo  {journal} {arXiv preprint arXiv:2405.09720}\ } (\bibinfo {year} {2024})}\BibitemShut {NoStop}%
\bibitem [{\citenamefont {Lemos}\ \emph {et~al.}(2024)\citenamefont {Lemos}, \citenamefont {Sharief}, \citenamefont {Malkin}, \citenamefont {Perreault-Levasseur},\ and\ \citenamefont {Hezaveh}}]{lemos2024pqmass}%
  \BibitemOpen
  \bibfield  {author} {\bibinfo {author} {\bibfnamefont {P.}~\bibnamefont {Lemos}}, \bibinfo {author} {\bibfnamefont {S.}~\bibnamefont {Sharief}}, \bibinfo {author} {\bibfnamefont {N.}~\bibnamefont {Malkin}}, \bibinfo {author} {\bibfnamefont {L.}~\bibnamefont {Perreault-Levasseur}}, \ and\ \bibinfo {author} {\bibfnamefont {Y.}~\bibnamefont {Hezaveh}},\ }\href@noop {} {\bibfield  {journal} {\bibinfo  {journal} {arXiv preprint arXiv:2402.04355}\ } (\bibinfo {year} {2024})}\BibitemShut {NoStop}%
\bibitem [{\citenamefont {Miller}\ \emph {et~al.}(2021)\citenamefont {Miller}, \citenamefont {Cole}, \citenamefont {Forr\'e}, \citenamefont {Louppe},\ and\ \citenamefont {Weniger}}]{Miller:2021hys}%
  \BibitemOpen
  \bibfield  {author} {\bibinfo {author} {\bibfnamefont {B.~K.}\ \bibnamefont {Miller}}, \bibinfo {author} {\bibfnamefont {A.}~\bibnamefont {Cole}}, \bibinfo {author} {\bibfnamefont {P.}~\bibnamefont {Forr\'e}}, \bibinfo {author} {\bibfnamefont {G.}~\bibnamefont {Louppe}}, \ and\ \bibinfo {author} {\bibfnamefont {C.}~\bibnamefont {Weniger}},\ }in\ \href {\doibase 10.5281/zenodo.5043706} {\emph {\bibinfo {booktitle} {{35th Conference on Neural Information Processing Systems}}}}\ (\bibinfo {year} {2021})\ \Eprint {http://arxiv.org/abs/2107.01214} {arXiv:2107.01214 [stat.ML]} \BibitemShut {NoStop}%
\bibitem [{\citenamefont {Sharrock}\ \emph {et~al.}(2022)\citenamefont {Sharrock}, \citenamefont {Simons}, \citenamefont {Liu},\ and\ \citenamefont {Beaumont}}]{sharrock2022sequential}%
  \BibitemOpen
  \bibfield  {author} {\bibinfo {author} {\bibfnamefont {L.}~\bibnamefont {Sharrock}}, \bibinfo {author} {\bibfnamefont {J.}~\bibnamefont {Simons}}, \bibinfo {author} {\bibfnamefont {S.}~\bibnamefont {Liu}}, \ and\ \bibinfo {author} {\bibfnamefont {M.}~\bibnamefont {Beaumont}},\ }\href@noop {} {\bibfield  {journal} {\bibinfo  {journal} {arXiv preprint arXiv:2210.04872}\ } (\bibinfo {year} {2022})}\BibitemShut {NoStop}%
\bibitem [{\citenamefont {Dax}\ \emph {et~al.}(2023)\citenamefont {Dax}, \citenamefont {Wildberger}, \citenamefont {Buchholz}, \citenamefont {Green}, \citenamefont {Macke},\ and\ \citenamefont {Sch{\"o}lkopf}}]{dax2023flow}%
  \BibitemOpen
  \bibfield  {author} {\bibinfo {author} {\bibfnamefont {M.}~\bibnamefont {Dax}}, \bibinfo {author} {\bibfnamefont {J.}~\bibnamefont {Wildberger}}, \bibinfo {author} {\bibfnamefont {S.}~\bibnamefont {Buchholz}}, \bibinfo {author} {\bibfnamefont {S.~R.}\ \bibnamefont {Green}}, \bibinfo {author} {\bibfnamefont {J.~H.}\ \bibnamefont {Macke}}, \ and\ \bibinfo {author} {\bibfnamefont {B.}~\bibnamefont {Sch{\"o}lkopf}},\ }\href@noop {} {\bibfield  {journal} {\bibinfo  {journal} {arXiv preprint arXiv:2305.17161}\ } (\bibinfo {year} {2023})}\BibitemShut {NoStop}%
\bibitem [{\citenamefont {Linhart}\ \emph {et~al.}(2024)\citenamefont {Linhart}, \citenamefont {Cardoso}, \citenamefont {Gramfort}, \citenamefont {Corff},\ and\ \citenamefont {Rodrigues}}]{linhart2024diffusion}%
  \BibitemOpen
  \bibfield  {author} {\bibinfo {author} {\bibfnamefont {J.}~\bibnamefont {Linhart}}, \bibinfo {author} {\bibfnamefont {G.~V.}\ \bibnamefont {Cardoso}}, \bibinfo {author} {\bibfnamefont {A.}~\bibnamefont {Gramfort}}, \bibinfo {author} {\bibfnamefont {S.~L.}\ \bibnamefont {Corff}}, \ and\ \bibinfo {author} {\bibfnamefont {P.~L.}\ \bibnamefont {Rodrigues}},\ }\href@noop {} {\bibfield  {journal} {\bibinfo  {journal} {arXiv preprint arXiv:2404.07593}\ } (\bibinfo {year} {2024})}\BibitemShut {NoStop}%
\bibitem [{\citenamefont {Zhang}\ \emph {et~al.}(2023)\citenamefont {Zhang}, \citenamefont {Bloom},\ and\ \citenamefont {Hernitschek}}]{zhang2023nbi}%
  \BibitemOpen
  \bibfield  {author} {\bibinfo {author} {\bibfnamefont {K.}~\bibnamefont {Zhang}}, \bibinfo {author} {\bibfnamefont {J.}~\bibnamefont {Bloom}}, \ and\ \bibinfo {author} {\bibfnamefont {N.}~\bibnamefont {Hernitschek}},\ }in\ \href@noop {} {\emph {\bibinfo {booktitle} {NeurIPS 2023 Workshop on Deep Learning and Inverse Problems}}}\ (\bibinfo {year} {2023})\BibitemShut {NoStop}%
\end{thebibliography}%


\end{document}